\documentclass[sigconf]{acmart}

\AtBeginDocument{%
  }

\usepackage{subcaption}  
\usepackage{graphicx} 
\usepackage{makecell} 

\copyrightyear{2025}
\acmYear{2025}
\setcopyright{acmlicensed}\acmConference[DIS '25]{Designing Interactive Systems Conference}{July 5--9, 2025}{Funchal, Portugal}
\acmBooktitle{Designing Interactive Systems Conference (DIS '25), July 5--9, 2025, Funchal, Portugal}
\acmDOI{10.1145/3715336.3735846}
\acmISBN{979-8-4007-1485-6/2025/07}

\begin{document}
\title[If I were in Space]{"If I were in Space": Understanding and Adapting to Social Isolation through Designing Collaborative Narratives}

\author{Qi Gong}
\email{gongqi_77.0818@sjtu.edu.cn}
\orcid{0000-0002-5621-2673}
\affiliation{
\institution{School of Design\\Shanghai Jiao Tong University}
\city{Shang Hai}
\country{China}}

\author{Ximing Shen}
\email{ximing.shen@kmd.keio.ac.jp}
\orcid{0000-0003-4272-0068}
\affiliation{
\institution{Graduate School of Media Design\\Keio University}
\city{Yokohama}
\country{Japan}}

\author{Ines Ziyou Yin}
\email{ines-z.yin@connect.polyu.hk}
\orcid{0000-0001-8739-1304}
\affiliation{
\institution{School of Design\\The Hong Kong Polytechnic University}
\city{Hong Kong}
\country{China}}

\author{Yaning Li}
\email{yaning.li@mail.polimi.it}
\orcid{0009-0004-6305-5866}
\affiliation{
\institution{Department of Industrial Design\\Xi'an Jiaotong University}
\city{Xi'an}
\country{China}}

\author{Ray Lc}
\authornote{Correspondences should be addressed to LC@raylc.org}
\email{LC@raylc.org}
\orcid{0000-0001-7310-8790}
\affiliation{
\institution{Studio for Narrative Spaces\\City University of Hong Kong}
\city{Hong Kong}
\country{China}}

\renewcommand{\shortauthors}{Gong et al.}

\begin{abstract}
Social isolation can lead to pervasive health issues like anxiety and loneliness. Previous work focused on physical interventions like exercise and teleconferencing, but overlooked the narrative potential of adaptive strategies. To address this, we designed a collaborative online storytelling experience in social VR, enabling participants in isolation to design an imaginary space journey as a metaphor for quarantine, in order to learn about their isolation adaptation strategies in the process. Eighteen individuals participated during real quarantine  undertaken a virtual role-play experience, designing their own spaceship rooms and engaging in collaborative activities that revealed creative adaptative strategies. Qualitative analyses of participant designs, transcripts, and interactions revealed how they coped with isolation, and how the engagement unexpectedly influenced their adaptation process. This study shows how designing playful narrative experiences, rather than solution-driven approaches, can serve as probes to surface how people navigate social isolation. 

\end{abstract}




\begin{teaserfigure}
    \centering
    \includegraphics[width=1\linewidth]{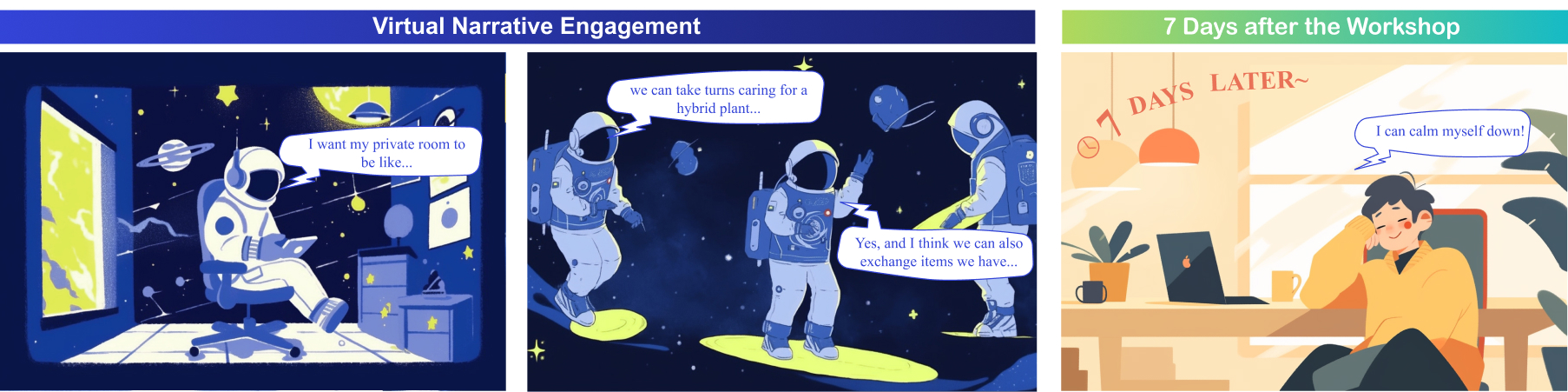}
    \caption{This figure summarizes the process of this study in probing adaptation strategies for social isolation through an immersive virtual experience.\textbf{ (Left) }Participants described how they modified and adapted their isolation private rooms in a virtual spaceflight experience. \textbf{(Center) }Participants engaged in collaborative storytelling, proposing individual or cooperative adaptation methods. \textbf{(Right)} Seven days later, in a follow-up interview, we found that the immersive virtual experience inspired participants' self-regulation and real-world isolation adaptation.}
    \Description{This figure illustrates a three-stage virtual narrative engagement process for studying isolation adaptation strategies. The figure consists of three panels: The left panel shows an astronaut avatar in a personalized virtual space room, depicting how participants designed and described their ideal isolation environments during the virtual spaceflight experience. The astronaut is shown interacting with room elements while expressing preferences through dialogue bubbles. The center panel illustrates the collaborative phase where multiple astronaut avatars engage in shared storytelling activities. Participants are shown floating in space while exchanging ideas about adaptation strategies and resource sharing, represented through their conversation bubbles. The right panel transitions to a post-workshop scene, showing a participant 7 days after the virtual experience. The illustration depicts them implementing learned self-regulation strategies in their actual isolation environment, demonstrating the workshop's real-world impact on adaptation behaviors. Together, these panels demonstrate the progression from individual space design through collaborative engagement to practical application of adaptation strategies.}
    \label{fig:teaser}
\end{teaserfigure}

\maketitle
\section{Introduction}\label{sec:Introduction}
Social isolation has been linked to severe physical and psychological health problems \cite{cacioppo2008loneliness,oluwafemi_review_2021}, with the COVID-19 pandemic amplifying these challenges and exposing gaps in our understanding of how individuals adapt to such conditions \cite{polizzi_stress_2020, saczuk_temporomandibular_2022, yevdokimova_coping_2021}. Beyond pandemics, social isolation spans diverse contexts, including aging populations, extreme work environments, and space exploration, where individuals face extended periods of solitude.
Existing studies have explored coping strategies such as structured routines, mindfulness practices, and digital solutions like teleconferencing \cite{polizzi_stress_2020, korpilahti-leino_resilience_2022}. 
Interventions have included IoT devices for remote connection and virtual platforms to enhance engagement. However, these approaches often focus on specific tools or immediate needs, overlooking systemic patterns and the creative processes through which individuals adapt to the social isolation. This gap underscores the need for research methods that capture the dynamic and nuanced nature of human adaptation during isolation.

Adaptation, defined as a continuous process of modifying one's behavior and strategies to align with changing environments, is critical for managing isolation \cite{Bjorklund_2015}. While prior research has acknowledged this phenomenon \cite{fancourt2021trajectories}, the mechanisms underlying adaptation processes and how individuals refine their coping strategies over time remain poorly understood. Current methodologies—including interviews \cite{harrison_antarctica_2012}, surveys \cite{baloran_knowledge_2020}, and diary studies \cite{peldszus_perfect_2014}—often fail to systematically capture and analyze these adaptive processes as they unfold, limiting our ability to develop sustainable adaptation strategies for future isolation scenarios, particularly in constrained environments.

Research through Design (RtD) offers a promising approach to address these limitations by enabling the study of adaptation as it unfolds. By iteratively creating, testing, and refining designed artifacts, RtD emphasizes emergent insights that develop through design interaction \cite{Desjardins_2020}. 
To operationalize this approach, we leverage virtual narrative engagement - defined as the process of immersing participants in interactive, technology-mediated storytelling to elicit and observe their responses, decision-making, and adaptation strategies in simulated contexts. This method creates a controlled yet flexible environment that projects real isolation conditions while offering greater opportunities for creative expression and reflection.
Building on previous works connecting forced quarantine to spaceflight research \cite{Chouker_2020}, we conducted a study with 18 participants who underwent 1 to 3 weeks of physical social isolation (e.g., hotel quarantine, on-campus isolation, or at-home isolation) due to COVID-19. Through individual and collaborative creative tasks, participants designed their ideal isolation lives on a spaceship, including room layouts and daily routines presented in a storytelling format. Data were collected through interviews conducted immediately after the workshop, and also followed by an email-based interview 7 days later.

Despite our intention to create an assessment probe for understanding those in physical isolation, we observed that our virtual narrative engagement actually led to participant strategic adaptations in the real world. Thus, the engagement served a dual purpose: while designed primarily as a research method to probe adaptation strategies to real-world social isolation, it emerged as a potential intervention that actively affected participants' isolation experiences. Qualitative data revealed reports of emotional relief, renewed routines, and increased motivation through their engagement with the virtual narrative engagement workshop. This duality leads us to explore two interconnected research questions:

    \textbf{RQ1:} \textit{Research Engagement: How may we create a virtual narrative engagement to study people’s adaptation strategies during social isolation?} 

    \textbf{RQ2:} \textit{Potential Intervention: How may this designed virtual narrative engagement actually change people's adaptation strategies during social isolation?} 

By analyzing participants' creative outputs, interactions, and feedback, our study uncovers three dimensions of adaptation strategies and four emergent adaptation patterns. This is because participants are constantly changing their strategies for physical isolation in any case, and must do so for our research probe as well. Our findings reveal that virtual narrative engagement not only uncovers the strategies they are using but also potentially facilitates their developing adaptation. These insights demonstrate how virtual narrative environments can function as both a research probe methodology and, in doing the study, lead to potential interventions that support individuals during real-world social isolation.

Our work makes the following research contributions:  (1) a design framework for virtual narrative engagement to study adaptation strategies during social isolation, and (2) empirical insights about participants' design strategies and behavioral changes during the virtual narrative engagement during social isolation.

\section{Related Works}\label{sec:Related works}
\subsection{Social lsolation Studies}
Social isolation is a persistent problem that exists in many communities such as online networks \cite{Park_2015} , elderly homes \cite{baecker_technology_2014} and people in epidemic like COVID-19 \cite{brooks_psychological_2020,Einav_Margalit_2023, Bonsaksen_2023} . The long-term effects extend to hybrid work \cite{McPhail_2024} and potential future scenarios like space travel \cite{Arone_2021}. Prior works have shown that it has negative impacts on both physiological \cite{Cacioppo_2002, Cacioppo_Hawkley_2003, Uchino_Cacioppo_1996} and psychological \cite{Shankar_2023, Brandt_2022} side of health through social deprivation, monotonous sensory environment, and spatial deprivation \cite{gunderson_individual_1973}. 

Recent research has emphasized adaptation as a key framework for understanding responses to isolation. Adaptation, defined as "a process of deliberate change in anticipation of or in reaction to external stimuli and stress" \cite{nelson2007adaptation}, involves both immediate responses and long-term psychological growth. This process requires adaptive capacity - the preconditions and resources that enable individuals to maintain functionality while developing new capabilities \cite{nelson2007adaptation}.

Studies have documented various adaptation strategies across different contexts. At the individual level, imagination helps mitigate the absence of tangible amenities \cite{oluwafemi_review_2021, peldszus_perfect_2014}, while preparation and proactive planning enhance resilience through activities like meaningful work, exercise, and skill development \cite{baloran_knowledge_2020,farhang_impact_2022,korpilahti-leino_resilience_2022,peldszus_perfect_2014,polizzi_stress_2020,saczuk_temporomandibular_2022,yevdokimova_coping_2021,palinkas_psychosocial_2021,oluwafemi_review_2021,harrison_antarctica_2012}. Mindfulness practices, including meditation, yoga and religious activities serve as effective emotional regulation tools \cite{baloran_knowledge_2020,farhang_impact_2022,sakurai_resilience_2020,harrison_antarctica_2012}.

At the community level, online interactions have become crucial for maintaining social connections. Digital platforms enable various forms of social support, particularly valuable for isolated groups like mariners and astronauts \cite{baloran_knowledge_2020,norton_depression_2007,palinkas_psychosocial_2021,peldszus_perfect_2014,yevdokimova_coping_2021}. HCI and CSCW researchers have extensively studied these online behaviors\cite{Choudhury_Gamon_Counts_Horvitz_2013, Chancellor_Pater_Clear_Choudhury_2016, Chancellor_Lin_Goodman_Zerwas_Choudhury_2016, Newman_Lauterbach_Munson_Resnick_Morris_2011}, focusing especially on marginalized communities \cite{Karimi_Neustaedter_2012, Homan_2014}.

However, two gaps exist in current research. First, studies typically focus on specific isolation contexts, lacking exploration of common adaptation patterns across diverse situations. Second, while these studies identify various coping strategies, they rarely examine how these strategies naturally emerge and evolve over time, limiting our understanding of the adaptation process itself.

\subsection{Design Interventions for social isolation}
In the field of HCI and CSCW, researchers have documented various design interventions to support people during isolation. These interventions primarily focus on two aspects: remote connection technologies and social activity platforms.

For remote connections, researchers have explored novel interaction methods through Internet of Things (IoT) devices that facilitate emotional expression and physical presence, such as systems for conveying mood and enabling pseudo-physical interactions like touching and knocking \cite{gaver_yo_2022}. Studies have also examined specific use cases, such as maintaining parent-child connections across distances \cite{Pan_2013}.

In terms of social activities, research has demonstrated the positive impact of group participation in reducing isolation effects \cite{baloran_knowledge_2020}. For instance, Nardi and Harris \cite{Nardi_Harris_2006} challenged conventional views about online activities causing isolation through their study of World of Warcraft, revealing how lightweight virtual collaborations can foster adaptive learning and social connections. Similarly, Park et al. \cite{Park_Yoo_Choe_Park_Song_2012} showed how converting solitary exercises into social exergames could enhance both physical activity and social interaction, creating relationship-building opportunities comparable to traditional team sports.

Building on these limitations, current design research faces its own challenges. While numerous technological solutions have been developed, they tend to focus on providing immediate functional support rather than facilitating long-term adaptation. Most interventions are designed as standalone solutions, without consideration for how they might integrate with and enhance users' natural adaptation processes. This disconnect between design interventions and natural adaptation capabilities suggests a need to fundamentally rethink how technology can scaffold, rather than substitute, individuals' adaptive capacities in isolation contexts.

\subsection{Research through Design and Emergence Analysis}

Research through Design (RtD) has been widely adopted in HCI as a methodology where design practice and research inquiry are intertwined. While it may appear similar to regular design practice, RtD is distinguished by its systematic approach to interpreting and reinterpreting conventional understanding, and its detailed documentation of design actions and rationale \cite{zimmerman2007research,zimmerman2010analysis,zimmerman2014research}. Indeterminacy is central to this approach, as practitioners must constantly negotiate emerging understandings in their interactions with an unpredictable world - much like how people navigate and adapt to changing circumstances. This inherent uncertainty allows designers to diverge in their approaches to seemingly identical problems \cite{gaver2022emergence}. 

The methodology's value lies in how it produces knowledge through design actions, including novel perspectives on problematic situations, insights about theory operationalization, and artifacts that broaden the space for design action \cite{zimmerman2014research}. Unlike traditional research where theory drives creation, RtD acknowledges that in HCI, artifacts often precede theory, offering a way to speculate about future possibilities based on empathic understanding of stakeholders and synthesis of behavioral theory \cite{zimmerman2014research,lc_designing_2021,song_drizzle_2022,song_narrating_2022,song_climate_2021}. Recent work by Giaccardi et al. extends this understanding by reconceptualizing \textit{prototypes} as \textit{agentive}, probabilistic entities that actively shape the design process, rather than mere preliminary solutions to fixed problems \cite{giaccardi2024prototyping}.

The emergence-based approach in design research manifests through various methods, from systematic accumulative explorations to opportunistic probing that enables rapid emergence of new ideas \cite{gaver2022emergence}. These methods share a fundamental goal: helping both designers and participants simultaneously imagine new social worlds and their implications, rather than merely gathering information or testing predetermined theories \cite{hendriks2024undertable}. 
This aligns with the "divergent" phase of design thinking, which intentionally embraces uncertainty to surface latent needs and possibilities before pursuing situated solutions \cite{giaccardi2024prototyping}.
This approach particularly values seeking idiosyncratic examples and emphasizing design in settings, allowing for unexpected insights to emerge through practice.

\subsection{Collaborative Methods in Virtual Environments}
Following emergence-based design principles, our research approaches social isolation not as a predetermined problem, but as a journey of discovery where understanding emerges through practice \cite{gaver2022emergence}. This aligns with the emergence strategy of treating research as a journey rather than a linear quest.

Collaborative methods, particularly storytelling and community engagement, exemplify the emergence-friendly approach to design research. Studies on COVID-19 resilience \cite{sakurai_resilience_2020} and health-related contexts \cite{Shen_2023, mulvale_applying_2016} demonstrate how collective activities can reveal unexpected insights into adaptive behaviors - a key aspect of emergence-based design.

The recent applications of collaborative storytelling further illustrate how design possibilities aid in conceptual understanding. These applications embody an emergence strategy for gaining new perspectives, ranging from r
ole-playing and scenario building \cite{diaz-kommonen_role_2009} to gamified storytelling for post-COVID scenarios \cite{troiano_are_2021} and cultural explorations \cite{sharma_mild_2021,linder_characterizing_2022}.


Virtual technologies provide an ideal setting for this emergence-based approach, enabling users to alter contexts and roles in ways that support both structured and emergent interactions \cite{Ortiz_Cruz-Neira_2023, stenros_nordic_2010}. While platforms like Mozilla Hub \cite{brown_employing_2022,saffo_remote_2021} offer technical capabilities, their systematic application in understanding adaptation behaviors remains limited. Moreover, though RtD has been widely adopted in HCI, its potential for investigating adaptation processes in social isolation contexts, particularly through virtual collaborative activities, remains underexplored. This presents an opportunity to examine how emergence-based design approaches in virtual environments might reveal and support natural adaptation processes.

\section{System Design}\label{sec:Methods}
Our study adopts a research-through-design (RtD) approach, with introducing and emphasizing on the emergence as a key feature in our design process \cite{gaver2022emergence}. This is aligned with the body of DIS  research that embraces emergence through iterative making and reflection, as it allows new directions and insights to surface organically \cite{Johansson_2024, Reed_2024, Epp_2024}.

We utilized several \textit{Emergence Strategies} described by \citeauthor{gaver2022emergence} \cite{gaver2022emergence} in our choice of a virtual workshop format. First, we aimed to facilitate the emergence of novel adaptation mechanisms and social interaction by creating an explorative space through workshops \cite{gaver2022emergence}. Social isolation becomes increasingly prevalent, traditional adaptation strategies may not fully address emerging scenarios. 
Second, since examining unique contexts can yield broader, more innovative insights compared to typical settings \cite{gaver2022emergence}, we opted for the virtual space exploration setting, which incorporated core elements common to various isolation experiences while enabling imaginative and unexpected interactions. 

To implement this vision, we constructed our virtual 3D environments using the Mozilla Hubs platform \footnote{https://hubs.mozilla.com/}, which enables direct browser-based access without specialized hardware requirements. Through this platform, we conducted workshops that simulated an isolated environment with a futuristic theme, allowing participants to engage in a pre-designed interaction process. This setup enabled us to observe both their adaptive behaviors and creative outputs related to future isolation experiences.

The following sections detail our design process, including the rationale for the virtual environment, the features incorporated into the space, and the procedural flow of workshop activities.

\subsection{Virtual Environment Design}
\subsubsection{Conceptual Framework} Simulated environments that project real-world situations have been shown to effectively encourage independent thinking and enhance creativity in uncertain conditions\cite{sanders_probes_2014, pidel_collaboration_2020}. Our virtual environment design drew inspiration from various real-world isolation contexts, including space travel \cite{Chouker_2020, Arquilla_Webb_Anderson_2022,harrison_antarctica_2012}, submarines \cite{palinkas_psychosocial_2021}, and polar research stations \cite{harrison_antarctica_2012}. While these environments all offer authentic isolation experiences, we specifically chose a space exploration setting for its unique advantages. The space setting provides valuable insights for future commercial space travel and long-distance interstellar exploration. Moreover, recent research has revealed significant parallels between space isolation and other forms of social isolation, particularly during the pandemic\cite{Chouker_2020, Arquilla_Webb_Anderson_2022}. Additionally, unique elements of the space environment, such as altered gravity, can stimulate imagination while maintaining core isolation characteristics \cite{linder_characterizing_2022,oluwafemi_review_2021}.

\begin{figure*}[t]
    \centering
    \includegraphics[width=1\linewidth]{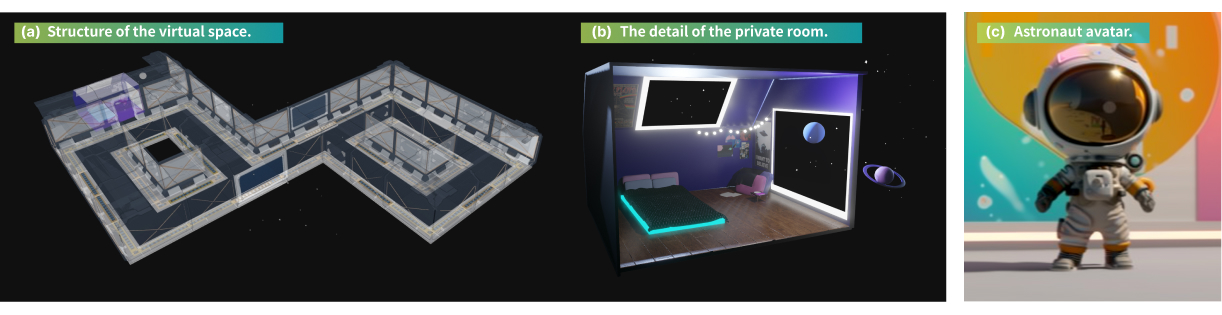}
    \caption{The figures illustrate the virtual immersive experience environment.\textbf{ (a)} The space corridor for participant gatherings, discussions, and other activities. \textbf{(b) }The structure of the private space isolation rooms where participants were asked to design their virtual isolation experience.\textbf{ (c)} The astronaut avatar used by participants to conduct all virtual activities. 3D model: Interior Spaceship + Planet by MatthewHicks, CC BY 4.0, via Sketchfab. Space Room by FutakuNightOwl, CC BY 4.0, via Sketchfab.}
    \Description{The figures illustrate the virtual immersive experience environment used in our study. \textbf{(a)} The space corridor layout shows the common areas designed for participant gatherings, discussions, and group activities. The L-shaped corridor structure provides both communal spaces and access to private rooms. \textbf{(b)} The detailed view of a private isolation room demonstrates the intimate space where participants were asked to design their virtual isolation experience, featuring customizable elements and a window view of space. \textbf{(c)} The astronaut avatar designed for participants to use during all virtual activities, showing the detailed spacesuit design and helmet reflection effects.}
    \label{fig:VirutualSpace}
\end{figure*}

Building on these foundations, we designed a minimalist yet immersive virtual space station environment (See Fig.\ref{fig:VirutualSpace}). The environment featured a cosmic backdrop with orbiting planets and stars to create an engaging atmosphere. At its core was a central public corridor with a communal clock, serving as a shared gathering space. Each participant had access to a private cabin equipped with basic furnishings for personal activities. Participants explored this environment as astronaut avatars, enhancing their sense of immersion in the space exploration scenario.

This design balanced realistic isolation conditions with imaginative elements, allowing us to observe both practical adaptation strategies and creative solutions that might inform future isolation scenarios.

\subsubsection{Support object selection for workshops} To support participant activities while maintaining experimental control, we developed a three-tiered resource system comprising essential supplies, optional items, and personal objects.

The Basic Stock and Equipment category included items essential for simulating space travel: tablets, intercoms, space-specific toiletries, water, and packaged space food. These items established a baseline environment that all participants could access (detailed descriptions in Appendix A).

\begin{figure*}[t]
    \centering
    \includegraphics[width=1\linewidth]{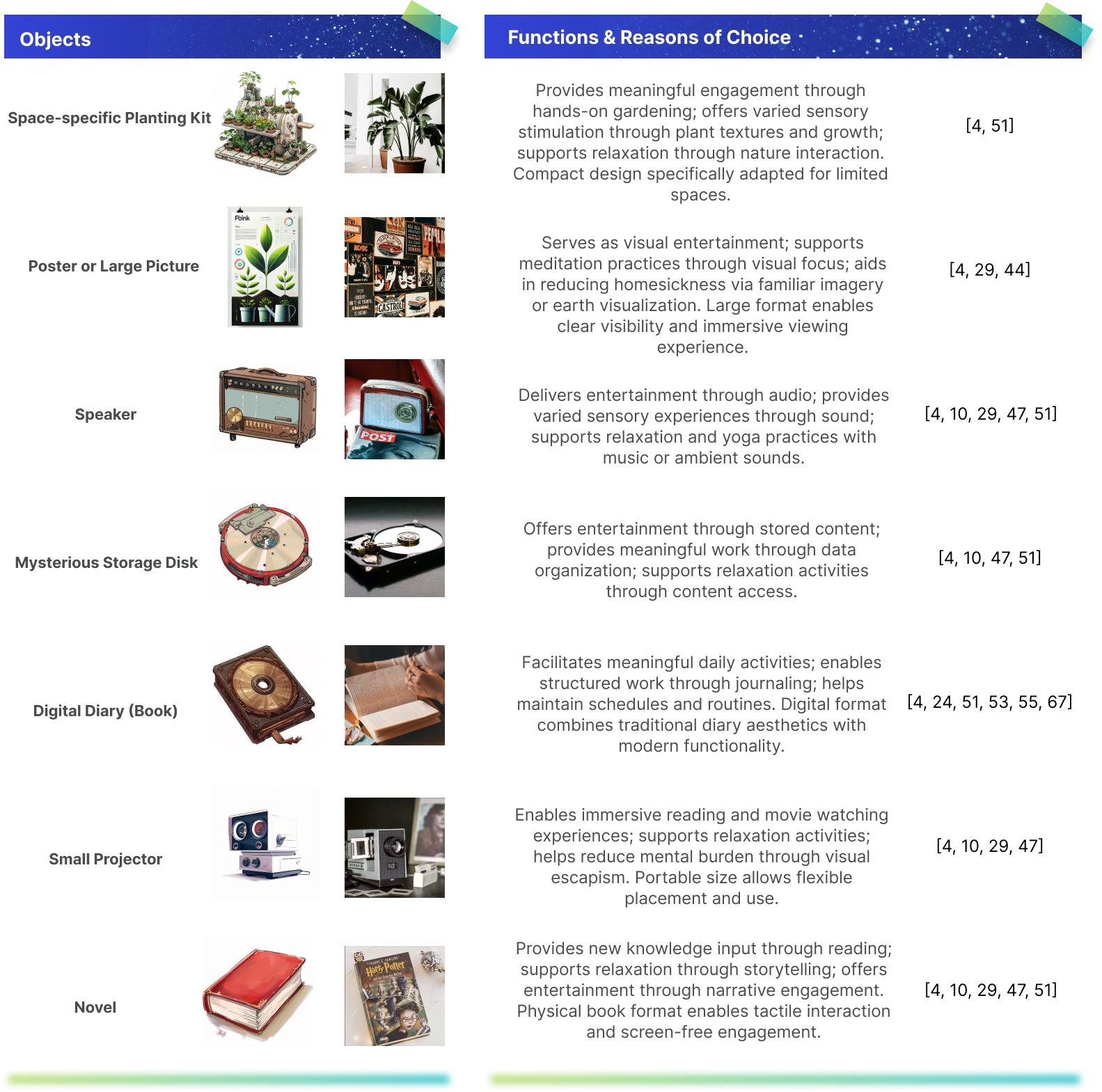}
    \Description{Reasons of Objects Choice. Objects are presented with their core functions and selection rationale. The left column shows object categories with illustrations (Space Planting Kit, Poster, Speaker, Storage Disk, Digital Diary, Projector, and Novel), while the right column presents detailed functions and justifications. Bracketed numbers indicate relevant literature references.}
    \caption{Reasons of Objects Choice. Objects are presented with their core functions and selection rationale, accompanied by supporting citations. Left column shows object categories with illustrations, while right column presents functions and justifications. Bracketed numbers indicate relevant literature references.}
    \label{fig:Selection}
\end{figure*}

Optional Items were curated based on isolation research \cite{oluwafemi_review_2021,peldszus_perfect_2014,baloran_knowledge_2020,farhang_impact_2022}. Each participant was presented with a selection of seven distinct objects: Space-specific Planting Kits, Posters or Large Pictures, Speakers, Mysterious Storage Disks, Digital Diaries (Books), Small Projectors, and Novels. The selection rationale and detailed functional descriptions for each item are illustrated in Fig.\ref{fig:Selection}. To maintain methodological rigor and prevent bias, objects were presented to participants using only neutral physical descriptions, without disclosure of their historical context or intended applications.

To balance experimental control with personal relevance, we introduced Self-selected Objects as the third category. Specifically, we allowed each participant to bring one additional item from their daily belongings that they felt was necessary for their space isolation experience. This addition served two purposes: it provided insight into individual coping preferences and revealed how participants adapted familiar objects to unusual contexts.

While our pre-selected items might appear to constrain participant choices, our research focused not on what items were chosen, but on how participants utilized these resources in varied and creative ways during isolation. The combination of standardized resources and personal objects enabled us to observe both structured and emergent adaptation strategies while keeping the possibility of comparative analysis.

\subsection{Task Design}
Our workshop structure comprised two complementary tasks designed to explore both individual and collaborative adaptation strategies in isolation contexts. Each task was carefully crafted to elicit different aspects of isolation experience while maintaining engagement through the space journey metaphor.

\subsubsection{Task 1: Personal Space Design} 
The first task focused on individual adaptation through environmental customization. Each participant was assigned an identical private cabin furnished with basic amenities and given 30 minutes to personalize their space using available resources. This individual design phase concluded with a group sharing session where participants presented their designs and discussed their spatial preferences.

 This task served two research objectives:
\begin{itemize}
\item Understanding how individuals construct personalized spaces under resource constraints.
\item Examining strategies for supporting physical and mental well-being through environmental modification.
\end{itemize}

\subsubsection{Task 2: Collaborative Storytelling} 
The second task shifted focus to collective adaptation through shared narrative creation. Participants gathered in the public corridor to collaboratively develop stories about their space journey and engaged in speculative discussions about the future, focusing on ways to maintain social connections and manage isolation-related challenges. To facilitate this process, we provided structured prompts at key points throughout the session (detailed in Appendix B).

This task was designed to:
\begin{itemize}
\item Investigate how collaborative storytelling might foster creative solutions to isolation challenges.
\item Explore adaptation strategies that emerged through both individual and group activities.
\end{itemize}

Through this two-part structure, we could observe how participants navigated both personal and shared spaces, and how they balanced individual coping mechanisms with collaborative problem-solving approaches.

\section{User Study}\label{sec:Methods}
\subsection{Participants and Recruitment}
We recruited participants who were experiencing one of three types of Covid-19-related isolation: hotel quarantine (7–14 days, alone in a hotel room), at-home isolation (remaining at home with family or alone, unable to meet others outside the home), or at-campus isolation. All participants were actively in isolation when the study began, providing authentic context for our research.
        
Recruitment was conducted through social media platforms (WeChat, Xiaohongshu, Instagram, and Facebook) using posts that included concept images of the virtual environment. We successfully recruited 18 participants (4 male, 14 female; M = 24.22 years, SD = 2.41) from diverse professional backgrounds: Management (5), Design (3), Computer Science (2), Engineering (2), Journalism (2), Social Studies (1), and Translation Studies (1). Detailed demographic information is provided in Appendix C.

The participants represented various isolation contexts at the time of the study. Four participants were in hotel quarantine, while five were isolating at home, two of whom had previously experienced hotel quarantine. Seven participants had completed hotel quarantine and were continuing isolation at home, and four were isolated on campus, with two having previously isolated at home. All participants had been in isolation for 1-5 days before the study and remained isolated for at least seven days. Each participant had reliable access to digital platforms and completed the required training before participation. 

This study received ethical approval from the University Institutional Review Board (Project Reference No:11607623). The recruitment advertisement outlined the research goals and provided participants with a brief overview of the research and experimental content before they applied for registration. The recruited personnel were over 18 years old. Prior to the study, each participant signed an informed consent and agreed to record or transcribe the experiment into text for subsequent research analysis. In the discussion presented in this paper, all participants' names were replaced with fixed numerical identifiers to protect their privacy rights.

\subsection{Process}
\subsubsection{Orientation}
The 18 participants were divided into three groups for manageable observation. Three days before each session, participants received a comprehensive email package. This package contained a detailed workshop process overview, necessary access links to both Mozilla Hub and Tencent Meeting platforms\footnote{https://meeting.tencent.com/}, step-by-step platform familiarization instructions, and dedicated technical support contact information for addressing any pre-workshop concerns.

Participants were asked to test the platforms and report any technical issues before the workshop. All participants received instruction sheets and provided consent for data collection. While the research team provided instructions in English, Chinese clarification was available when needed, as all participants were native Chinese speakers.

\begin{figure*}
        \centering
        \includegraphics[width=1\linewidth]{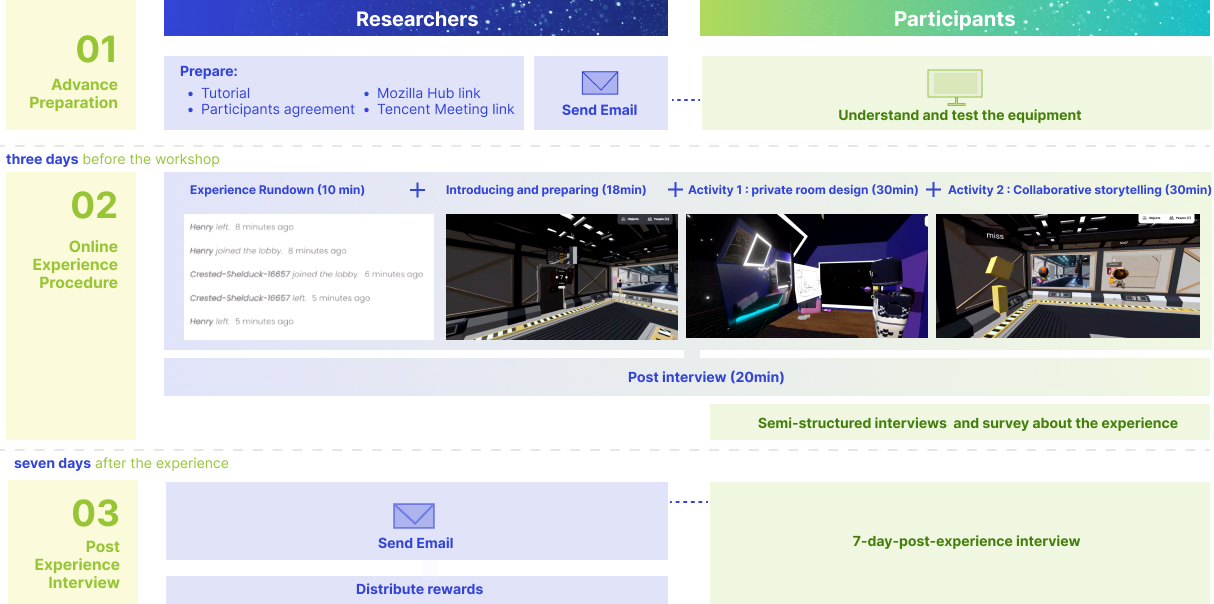}
        \caption{The complete process of the research, which consists of three main parts: preparation, online experience and assessment, and post-experience follow-up. The flowchart indicates the duration of each major experimental stage and outlines the key activities conducted by both researchers and participants.}
        \Description{The complete process of the research, which consists of three main parts (1-3): (1) Advance Preparation: Three days before the workshop, researchers prepare materials and agreements while participants familiarize themselves with equipment. (2) Online Experience Procedure: The core workshop including experience briefing (10min), introduction (10min), Activity 1: private room design (30min), Activity 2: collaborative storytelling (30min), and post-interview (30min). (3) Post Experience Interview: Seven days after the experience, conducting follow-up interviews and distributing rewards.The flowchart indicates the duration of each major experimental step and outlines the key activities conducted by both researchers and participants.}
        \label{fig:procedure}
    \end{figure*}        
    
\subsubsection{Workshop Implementation}
The workshop proceeded in two main phases (see Figure \ref{fig:procedure}). First, participants joined both the Mozilla Hub environment and Tencent Meeting, completed consent procedures, and received their astronaut roles. After selecting their Optional Items and Self-selected Objects, they began the private cabin design task in their virtual spaces. We collected their design sketches, screen recordings, written descriptions, and presentation recordings for analysis.

The second phase involved collaborative storytelling in the virtual Public Corridor. All discussions were recorded and transcribed for subsequent analysis. Throughout both phases, researchers maintained continuous observation of participant interactions and engagement. 

\subsubsection{Post-workshop Data Collection}
Our post-workshop assessment consisted of two phases. Immediately following the workshop tasks, we conducted a 30-minute focus group discussion where participants explored their perspectives on the space isolation metaphor, shared newly discovered adaptation strategies through peer interaction, and reflected on their overall workshop experience. Participants also completed a detailed questionnaire about their workshop experience.

To track the longer-term impact, we conducted a follow-up assessment seven days after the workshop. Participants responded to structured email questions addressing their implementation of workshop-discussed coping strategies, the effectiveness of these strategies in managing isolation-related emotions, and any changes in their emotional state over the week. To ensure accurate recall and timely feedback, participants were required to respond within 24 hours of receiving the follow-up questions.

\subsection{Data  Analysis} 
All collected data and field notes were transcribed and anonymized for analysis. We employed thematic analysis with open coding procedures \cite{braun_using_2006,braun_what_2014} to gain an in-depth understanding of participants' adaptation to social isolation. The analysis, conducted using NVivo software, proceeded in multiple phases.

In the initial phase, the first author reviewed the transcripts, labeling significant segments that revealed adaptation strategies and insights. For visual data such as sketches, we applied open coding to identify prominent features and patterns. This process yielded 336 initial codes encompassing design concepts, emotional responses, and collaborative behaviors. Through iterative analysis, we identified 4 major themes with 12 related subthemes. These themes emerged from participants' creative adaptation strategies, emotional coping mechanisms, and collaborative engagement patterns and so on. The analysis also revealed potential directions for extended discussion.

In the second phase, an additional author joined the analysis process, conducting a detailed sentence-by-sentence review of the transcriptions. Both coders then collaboratively reviewed the data to ensure consistency and establish inter-coder reliability. Any coding discrepancies were discussed and resolved through consensus, strengthening the validity of our findings.

\section{Results}\label{sec:Results}
Our thematic analysis results revealed three distinct dimensions of adaptation to social isolation: environmental adaptation (how participants modified and utilized their physical spaces), practical adaptation (the development of new routines and activities), and mental adaptation (psychological strategies for coping with isolation), see Table \ref{tab:fig5}. Below, we examine how participants employed these different types of adaptation strategies to maintain wellbeing during isolation.

\begin{table}[htbp]
    \caption{Strategies mentioned in the findings.}\label{tab:fig5}
    \centering
    \begin{tabular}{|p{2cm}|p{6cm}|} 
    \hline
    Category &  Strategy \\
    \hline
    Environmental Adaptations& 1. Supporting Essential Tasks During Isolation and Create Normalcy. \\
                               & 2. Enhancing Atmosphere Through Creative Decorations. \\
                               & 3. Adapting Flexibly to Limited Physical Resources. \\
    \hline
    Practical& 1. Setting and Pursuing Meaningful Goals. \\
    Adaptations  & 2. Creating Personalized Activities for Wellbeing. \\
                               & 3. Fostering Connection Through Collaborative Activities. \\
    \hline
    Mental and Emotional & 1. Building Emotional Connections Through Meditation and Reflection. \\
    Strategies & 2. Imagining Non-Human Companions for Support. \\
                                     & 3. Reframing Isolation as an Opportunity. \\
    \hline
    \end{tabular}
\end{table}
        
\subsection{Environment Adaptation Strategies} 
Our analysis revealed how participants actively modified their physical environments to maintain wellbeing during isolation. Their adaptations demonstrated a sophisticated range of strategies across four main categories: supporting essential tasks and responsibilities, creating normalcy through daily activities, enhancing atmospheric comfort through creative decorations, and demonstrating flexible adaptation to resource constraints. These strategies reflected both practical needs and psychological considerations, showing how participants balanced immediate necessities with quality-of-life improvements despite limitations.

\subsubsection{Supporting Essential Tasks During Isolation and Create Normalcy}
Despite the various reasons for social isolation, time doesn't stop during these periods. During real-world isolation, participants still faced pressing deadlines and responsibilities, including thesis writing (P13), job hunting (P8), and entrance exam preparation (P18). The disruption of normal routines significantly impacted these crucial tasks, leading to emotional distress that participants clearly expressed during both the room design and storytelling sessions of our workshop.

For example, P8 and P18 reported that isolation had severely disrupted their original plans, resulting in irregular schedules, reversed day-night cycles, and insomnia - all of which contributed to significant negative emotions. These real-world challenges emerged metaphorically during the workshop's storytelling sessions. P13 voiced concerns through the space metaphor: "We might lack the expertise and could get trapped during the space journey," later explaining in the post-workshop interview that including thesis work in their narrative reflected their current "tense state of mind."

In response to these challenges, participants' physical environment adaptation strategies specifically focused on supporting their most critical tasks (P13, P18) and ensuring quality sleep (P8). As P8 explained, "I wish to bring my teddy bear and scented candles. Even if the candles cannot be lit, just having them there makes me feel relaxed." While addressing these essential needs was primary, participants also sought ways to maintain quality of life beyond basic necessities.

\begin{figure*}[]
        \centering
        \includegraphics[width=1\linewidth]{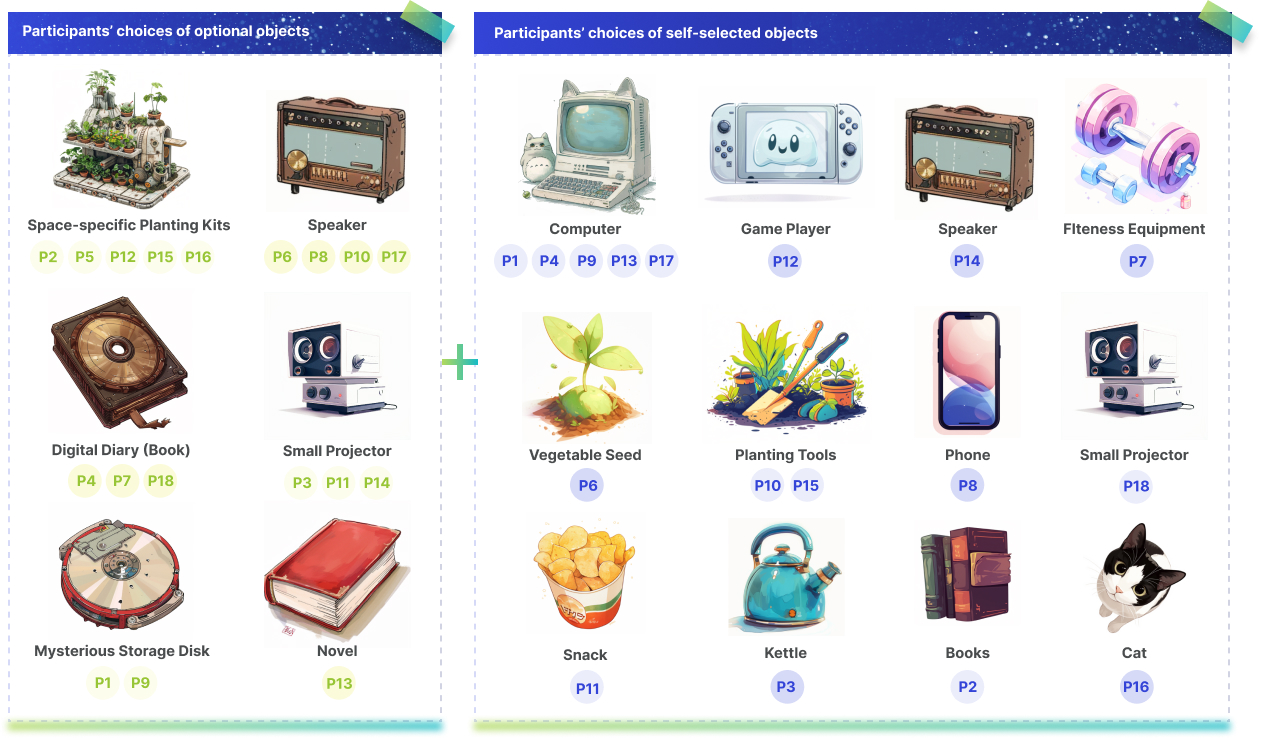}
        \caption{The figures show the items participants chose to bring into the immersive experience environment. \textbf{(Left) }The selected optional items.\textbf{ (Right)} The self-selected objects. Each item was labeled with the participant's identification number.}
        \Description{The figures show the items participants chose to bring into the immersive experience environment. The layout is divided into two panels: (Left) The selected optional items: Space-specific Planting Kits, Speaker, Digital Diary (Book), Small Projector, Mysterious Storage Disk, and Novel. These were pre-defined options provided to participants. (Right) The self-selected objects: Computer, Game Player, Speaker, Fitness Equipment, Vegetable Seed, Planting Tools, Phone, Small Projector, Snack, Kettle, Books, and Cat. These represent participants' personal choices. Each item is labeled with participant identification numbers (P1-P18) to track selection patterns.}
        \label{fig:optionalitems}
        \end{figure*}

In addition to supporting critical tasks, participants worked to transform their spaces through both daily activities and creative endeavors. When designing their private spaces in the virtual spaceship, participants most frequently chose everyday items such as plants, sofas, beds, audio equipment, yoga mats, and computers (Fig. \ref{fig:optionalitems}). Participants selected items that would support specific hobbies and routines - growing plants (P5), listening to music (P9), and exercising (P14), for maintaining normalcy through familiar activities.

When faced with resource limitations, participants demonstrated remarkable creativity in generating alternatives. For instance, P10 proposed creating meaningful environmental modifications: "I hope to obtain some tools to create a photo wall, as things of commemorative and decorative significance can bring motivation to people's lives" (Figure \ref{fig:Sketches}). Others, like P5, suggested establishing systems for sharing and trading items among residents - a strategy that mirrors the informal trading networks that emerged between neighbors during real-world building-wide quarantines.

\subsubsection{Enhancing Atmosphere Through Creative Decorations}
When describing their isolation spaces, participants consistently used terms that emphasized psychological discomfort: "empty" (P15), "sealed" (P3), "depressing" (P3), "disheartening" (P3), "lonely" (P2), and "desolate" (P2). These descriptions revealed not just physical constraints but the emotional impact of their environments. In response, participants developed deliberate strategies to transform their spaces' atmosphere.

Artistic expression emerged as one key approach to environmental modification. P15 utilized their painting kit to create artwork depicting both real landscapes from their surroundings and imaginary scenes, using these to decorate otherwise bare walls. Similarly, P1 devised a technological solution, planning to use their computer to display landscape images, making the space feel less monotonous.

Natural elements played a crucial role in atmospheric enhancement. P3 specifically chose a planting kit, noting that incorporating greenery provided a "visually relaxing and pleasing experience." This idea resonated with P12, who found the concept of maintaining plants in space particularly novel and incorporated them as architectural elements in their space design. Taking a different approach to creating comfort, P2 focused on tactile and visual warmth, requesting carpets and blankets to counter feelings of loneliness and desolation.

Despite the limited resources available, participants demonstrated ingenuity in using accessible items to modify their environments, addressing the oppressive atmosphere from multiple angles - visual, natural, and textural. These adaptations reveal how even simple modifications can impact the psychological experience of isolated spaces.
        
\subsubsection{Adapting Flexibly to Limited Physical Resources}
An important aspect of participants' adaptation strategies was their ability to creatively downgrade or substitute resources when faced with constraints. This flexibility manifested in both physical items and spatial perceptions. For example, during the workshop's private room sketching activity, P1 demonstrated remarkable adaptability when faced with a lack of drawing paper in their actual isolation space - they ingeniously used napkins to complete their sketch (Figure \ref{fig:P1sketch}).

This concept of "downgrade substitution" extended beyond physical objects to spatial adaptations. Participants showed particular creativity in addressing spatial limitations through both physical and virtual solutions. P3 articulated the psychological impact of confined spaces: "The room for isolation is too small and enclosed, it is easy to feel depressed. If there are windows to see the scenery outside, it can make people feel happy." Building on this desire for expanded space, participants proposed technological solutions - P5 suggested using drones, while P6 recommended VR goggles to create a sense of spatial expansion. Beyond these, strategies in Creative Decoration also demonstrated similar principles - P15 created artworks depicting both surrounding real landscapes and imaginary scenes, while P1 used their computer to display landscape images, making the space less monotonous.

These adaptations reveal how participants reconceptualized their environment, treating even inaccessible external spaces as potential extensions of their physical environment. Their creative solutions demonstrate how people in special circumstances can discover novel uses for objects and spaces, expanding their utility to meet personal needs. This flexibility in adaptation suggests that environmental constraints can actually stimulate innovative thinking about resource utilization.

\subsection{Adapting to Isolation Through New Practices} 
Our survey prior to the workshop revealed that 78 percent (14/18) of participants worried about social isolation becoming an inescapable part of daily life. In response, participants expressed the need to normalize these circumstances and readjust their daily routines for long-term adaptation. Throughout the workshop, activity-based adaptation emerged as the most frequently mentioned coping strategy.

\subsubsection{Setting and Pursuing Meaningful Goals} Under long-term isolation, participants reported experiencing significant psychological challenges, describing feelings of being "lost" (P8, P9), "bored and anxious" (P4, P15), "depressed" (P14), and struggling with time management (P4, P18). These negative emotions stemmed from two primary sources: physiological disruptions to biological rhythms due to confined spaces, and the practical dissolution of established routines without new structured patterns to replace them.

Participants recognized the critical importance of maintaining mental wellness through purposeful goal-setting. As P9 insightfully noted: "If there are no thoughts and goals for the future during isolation, after a long time, you will lose perception of time and fall into a state of doing nothing." This observation was echoed by P7, who emphasized: "Once a person is alone for a long time, they may become loose in state, and both mental and physical states deteriorate."

In response, participants developed a two-tiered approach to goal-setting. Long-term goals often centered around collective purposes, such as "exploring ways to save the collective future" or "discovering new harmonious ways of interaction." P4 suggested establishing regular community meetings: "We should gather weekly to adjust our division of labor. When it's necessary to complete a major goal, we need democratic voting." Short-term goals focused on maintaining daily structure and breaking down larger objectives into manageable tasks. P5 advocated for weekly planning: "There should be a rough plan every week, with clear main directions for what to accomplish." P8 and P17 emphasized the importance of incremental progress, noting that "important things can't be done overnight" (P8) and "small goals should be split from larger ones, step by step" (P17).

To enhance their sense of time and progress, participants developed various documentation strategies. Some preferred visual methods, such as P8's proposal to track trajectory through daily planet observations, or P10 and P15's approach of creating picture walls to chronicle their experiences. Others favored audio documentation, with P15 suggesting wall-mounted voice notebooks for spontaneous recording of thoughts during long stays in isolation. P4 proposed maintaining digital "flight logs," explaining: "I will voice-record new things seen and experienced every day with the digital diary, and then edit them with the computer." These recording practices served not only as progress markers but also, as P15 noted, as means of creating unique memories of their experience.

\subsubsection{Creating Personalized Activities for Wellbeing}
Participants developed diverse activities to combat isolation's psychological effects, ranging from everyday leisure pursuits to professionally-informed creative endeavors.

These activities were deliberately chosen for their emotional benefits. P11 expressed a desire for astronomical observation: "If possible, I really want to put a telescope by the window, then I can look at our Earth and gaze at the stars, which makes me feel calm." Music emerged as a particularly powerful tool for emotional regulation. P2 specifically requested Nordic folk songs, explaining that "the elegant melody evokes the open feeling of Northern Europe, creating a contrast with the constraints of isolation." Others used music for different purposes - P17 for emotional release through singing, and P14 for managing insomnia: "I hope music can accompany me through sleepless times."

Participants particularly excelled at leveraging their professional backgrounds to create meaningful activities suited to their interests and expertise. For instance, a participant studying Electronic Engineering (P9) found purpose in repairing mysterious disks, while a Journalism major (P10) proposed collaborative novel writing as a social activity. Those with design backgrounds, like P15, channeled their creativity into visual documentation, drawing "beautiful memories, stories of past interactions with family and friends, even the take-out food I miss, and the activities happening outside the window." Management students approached adaptation through systematic thinking - P6 focused on creating structured social interactions: "If we have extra paper, we can design new card games for everyone to play together."

These personalized activities served multiple purposes: maintaining professional identity, creating structure, fostering creativity, and providing emotional outlets. The diversity of approaches demonstrates how participants actively transformed their professional knowledge and personal interests into coping mechanisms, making isolation not just bearable but potentially enriching.

\subsubsection{Fostering Connection Through Collaborative Activities}
Despite physical restrictions, participants developed innovative ways to maintain social connections through four distinct types of collaborative activities, ranging from minimal to intensive interaction: relay activities, exchange systems, resource sharing, and joint participation. These approaches varied in their applicability depending on isolation conditions, from complete individual isolation to collective quarantine scenarios.

Relay activities emerged as an effective method for maintaining connection without requiring simultaneous participation. These activities allowed creativity and emotional experiences to interweave across time and space. Participants proposed collaborative storytelling through "writing a novel together" (P9), "drawing a storybook collectively" (P10), and taking turns caring for shared plants, which P12 noted would create "an emotional connection between participants through the shared responsibility of nurturing life."

For situations allowing limited physical interaction, participants developed structured exchange systems. P5 proposed organizing regular trading events: "We could establish a marketplace, similar to traditional rural Chinese fairs, setting fixed weekly times when people can briefly leave their spaces to trade." This approach not only facilitated resource distribution but also created opportunities for social interaction within safety constraints.

Resource sharing represented a more communal approach to collaboration. Several participants proposed sharing both physical and digital resources to enhance collective wellbeing. For instance, P11 offered to share their projector for community use, while others suggested shared pet care as a way to build connections. When P16 mentioned having a cat, P17 and P18 immediately proposed collective pet care arrangements, demonstrating how shared responsibilities could foster community bonds.

The most intensive form of collaboration emerged through joint participation activities, particularly relevant in collective isolation settings. These included both physical and virtual collective experiences: communal breakfast preparation (P2), scheduled group labor periods (P5), and the creation of shared spaces for community gardening (P11). Digital technology enabled remote participation in activities such as group exercise sessions via video calls (P7), multiplayer games (P14, P15, P18), and synchronized movie watching with discussion sessions (P4).

Beyond conventional activities, participants envisioned creative collaborative possibilities inspired by their unique context. P12 imagined conducting shared experiments with space-mutated plants, while P16 proposed innovative uses for radiation-altered vegetation to create communal spaces. These imaginative proposals demonstrated how isolation constraints could spark creative thinking about new forms of social connection.

This spectrum of collaborative activities reveals how participants strategically adapted their social interactions to maintain connections despite isolation restrictions, from individual relay activities to fully collaborative projects.

\subsection{Adapting to Isolation Through Mental Measures}
During the workshop, participants revealed sophisticated mental adaptation strategies that fell into three main categories: meditative practices, imagining non-human companions, and reframing isolation experiences. These strategies demonstrated how participants actively shaped their mental landscape to cope with isolation.

\subsubsection{Building Emotional Connections Through Meditation and Reflection} Participants engaged in both inward and outward meditation practices. Inward meditation focused on personal reflection and meaning-making. As P11 explained: "I want to put a yoga mat by the window of my room, so when I feel bored or in a bad mood, I can practice yoga and then meditate." P16 elaborated on this contemplative practice: "There is no need to work under the current situation, so I can abandon some social rules, as well as the logic of social operation. The concept of time also 'does not exist', so I can meditate more according to my physical and mental feelings. For example, I think about my existence in the world without social relations." 

Outward meditation involved creating connections with the external world through imagination and memory.Participants viewed hometown photos and videos (P1) on their computers, or used telescopes to gaze at Earth and stars (P11) . While participants acknowledged they couldn't immediately return to normal "Earth life," these distant connections through memories and stargazing provided emotional healing and hope for the future. Participants also used fictional works like movies to maintain emotional connections. P15 noted that watching movies helped "recall beautiful experiences with friends or family," while P18 used them to "experience life in fictional stories and better feel the passage of time."

\begin{figure*}[]
        \centering
        \includegraphics[width=1\linewidth]{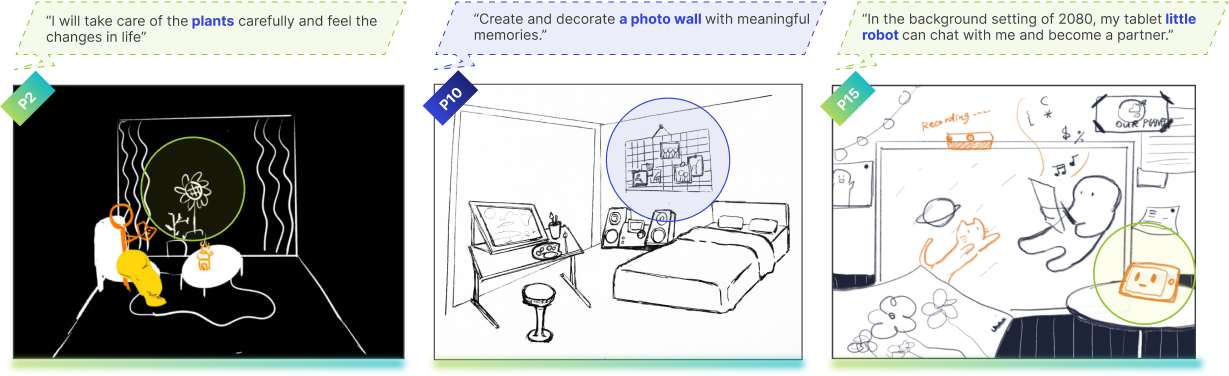}
        \caption{Three representative sketches that show participants' private room designs and envisioned adaptation activities. \textbf{(Left) }P2: "There are sunflowers and various succulent plants on the windowsill." P2 also described activities related to the plants. \textbf{(Middle)} P10: "There are some clips, thumbtacks, and wires for making a photo collage. Maybe I can make a photo collage on one side of the wall to decorate the room with some precious photos and pictures I am satisfied with." \textbf{(Right) }P15: "My tablet looks like a little robot, accompanying me."}
        \Description{Three representative sketches that demonstrate participants' envisioned private room designs and adaptation activities in their virtual isolation spaces. (Left) P2's design features a contemplative space with plants by a window, emphasizing botanical care as a means of tracking time and life changes. (Middle) P10's concept incorporates a photo wall collage area, suggesting the use of personal memories and images for emotional connection. (Right) P15's futuristic room design includes an AI companion tablet, illustrating how technology might provide social interaction in isolation. Each sketch is accompanied by the participant's own description of their design rationale, showing different approaches to creating meaningful spaces for isolation adaptation.}
        \label{fig:Sketches}
        \end{figure*}

\subsubsection{Imagining Non-Human Companions for Support} When traditional social interactions were limited, participants creatively expanded their social world by developing relationships with non-human companions. P15 anthropomorphized technology, imagining their tablet as a "little robot" that could "chat with me and become a partner" in their envisioned 2080 setting (See Fig. \ref{fig:Sketches}). Plants emerged as particularly significant companions, serving multiple emotional functions. P3 expressed this connection: "If conditions permit, I want two broad-leaved green plants from Earth to accompany me. They symbolize a kind of companionship from Earth, joining me on this new journey in space." P2 similarly noted the emotional sustenance provided by plant care: "During the long space journey, I will carefully tend to the plants and feel the changes in life. At least spiritually, there will be a feeling of being accompanied."

\subsubsection{Reframing Isolation as an Opportunity} Participants demonstrated remarkable cognitive flexibility in reframing isolation from a constraint to an opportunity. P9 discovered unexpected benefits: "Although you can't go out during home isolation, you rarely have time to spend with your family. During this time, basking in the sun and chatting together, I feel another kind of happiness." P10 saw isolation as a chance for self-development: "As I grow older, I have less and less time to seriously do what I like. Whether in isolation or in a space capsule, such a life will be a quiet opportunity to try new things."

Some participants drew inspiration from virtual experiences to create motivating narratives. P15 referenced Keep (a Chinese running app\footnote{https://www.gotokeep.com/}) that transforms exercise into storytelling through audio from "Empresses in the Palace"\footnote{https://en.wikipedia.org/wiki/Empresses in the Palace}: "The runner becomes a character avoiding pursuit... These narratives like 'Little Master, run slower, you're not recovered yet' help adjust pace. Creating such stories for isolated life can provide motivation and courage to face challenges."

This reframing extended to reimagining one's role in a larger narrative. P4 embraced an explorer mindset: "I had two choices - stay on Earth or venture into space for adventure. Even if I might not return, I would choose to explore new life in space." P17 found meaning in collective achievement: "The ending of this space journey may be that we complete an experiment together and find ways to relieve Earth's pressures. Although the process is hard, when we return to normal life, we will face the future with a more optimistic attitude." However, acknowledging the challenges of isolation, P5 expressed a poignant wish for normalcy: "When I think about all this, the alarm clock rings, and my mother asks why I haven't gotten up yet. 'Don't think about these things anymore,' she says, 'I've made breakfast for you, remember to go to work after eating.' And then everything disappears. Everything is as usual, and isolation was just a dream." This contrast between aspirational reframing and longing for normality illustrates the complex psychological dynamics of adapting to isolation.

\subsection{Virtual Environment Workshops as a Facilitator of Emergent Design Practices}
The virtual workshop environment served as both a simulation platform and an ideation space, allowing participants to experience isolation conditions while developing and testing adaptation strategies. Through careful design of environmental elements and social interactions, the workshop created a structured yet flexible space for exploring isolation coping mechanisms.

\subsubsection{Simulating Real-Life Isolation} 
The virtual workshop environment successfully replicated key aspects of real-world isolation, creating an authentic yet less pressured space for participants to engage with isolation-related challenges. Participants identified several parallels between the virtual environment and their actual isolation experiences. P17 noted environmental similarities: "The dim environment in space is very similar to my experience of closing the curtains in hotel isolation, making day and night indistinguishable." The workshop's structural elements also mirrored real isolation conditions, with P18 observing that "the limitation on the number of items we could bring reminded me of life in isolation."

The social dynamics within the workshop particularly resonated with participants' real experiences. P11 highlighted how the scenario of mutual assistance among strangers was "very similar to the cooperation during real isolation." P9 and P10 observed that the entire workshop process authentically reflected the social progression in collective isolation environments - from initial individual isolation in small rooms to gradual relationship building through ice-breaking activities, ultimately fostering what they described as a special camaraderie of "shared hardships."

The virtual nature of the workshop created a unique balance between authenticity and psychological safety. As P5 explained, "Compared to the lack of choices during isolation, the space exploration context in the workshop provided us with more options, allowing me to think beyond the details and consider ways to adapt to the isolation environment." P4 found that the space capsule setting offered an imaginative framework for approaching isolation, enabling them to "cope with difficulties in real isolation with a similar mindset." This combination of authentic simulation and creative freedom allowed participants to explore adaptation strategies more openly than they might have in actual isolation conditions.

\subsubsection{Catalyzing Adaptation Strategies}
The workshop proved effective in not only generating adaptation strategies but also motivating their real-world implementation. In the immediate aftermath of the workshop, participants identified various promising approaches, ranging from technological solutions like "using VR/AR technology to enter broader scenes" (P1, P3) and "creating virtual environments" (P12) to practical activities such as "having breakfast by the window" (P2) and "participating in collaborative activities" (P10). Some found inspiration in gaming mechanics, with P6 noting how they adopted self-motivation techniques "similar to Rimworld game."

Follow-up questionnaires after seven days revealed comprehensive adoption of workshop-inspired strategies. All participants reported implementing methods discussed during the workshop, with more than half (P5, P7, P8, P9, P13, P14, P15, P16, P17) developing structured daily schedules. These schedules incorporated various meaningful activities:

\begin{itemize}
\item Physical and outdoor engagement: Exercise routines (P2, P3, P18) and nature interaction (P3).
\item Creative pursuits: Reading (P11), drawing (P16), baking (P11), and meditation (P16).
\item Social connections: Regular communication with family and friends (P8, P9, P11, P15) and collaborative gaming (P2, P4, P7, P9, P10, P12, P14, P15, P18).
\end{itemize}

Notably, technology-focused solutions like VR/AR or virtual environments were less frequently mentioned in follow-up responses, likely due to limited access to necessary technology and suitable platforms. Instead, participants focused on adapting readily available resources and activities into meaningful experiences. Participants demonstrated particular creativity in combining different strategies. For instance, P18 integrated social gaming with exercise: "We played online board games. If we lose, we have to do push-ups." 

Some participants developed longer-term perspectives, with P15 reporting, "I tried to make a life plan," and P5 describing tangible progress: "decided on the priorities of the to-do list. In addition, I changed my resume and took part in two online interviews, which enriched my isolated life." 

The implemented strategies primarily helped participants regain a sense of control over their daily lives. P2 and P15 both emphasized that "those methods enhanced the sense of control over my life," while P9 found that "planning my life... quantified my goals and made me not worry or often feeling busy." This increased sense of control led to improved psychological adaptation, as evidenced by P8's observation: "These methods helped me adjust my mental states relatively quickly, without spending too much energy on bad emotions." P7 similarly reported, "After having some demands on my own life, my overall mental health condition is much better." Ultimately, these improvements enabled participants to better cope with isolation. As P6 reflected, "Although the date of lifting the isolation has not been determined, I have gradually adapted to the isolated days... when I complete my daily tasks, I have a sense of achievement.

\subsubsection{Measuring Workshop Impact}

Follow-up data revealing both sustained implementation of strategies and notable impacts on participants' emotional resilience. 
One example was participants' ability to maintain structured routines when facing unexpected changes in isolation conditions. As P11 described: "During hotel isolation, I felt relaxed due to sufficient preparation and careful planning. When stricter community controls were suddenly implemented and isolation time became uncertain, I was initially anxious. However, I was able to use psychological self-adjustment to revise my plans and calm myself down. These methods helped me digest negative emotions. 

The implementation of workshop-derived strategies led to observable improvements in daily life quality. P17 noted that "my routine is much more regular than before," while P6 described: "My life these 7 days was not as idle as it was before when I was just forced to be isolated at home. I think if I find something that I like to do, every day will be full of fun." This sentiment was echoed by P18, who found that combining social gaming with exercise "adjusted and eased my negative emotions to a great extent, because the games that afford communication with people are very interesting, and with the addition of sports, they are very fun and healthy."

However, the follow-up data also revealed important limitations and ongoing challenges. Some participants reported persistent difficulties that the workshop-inspired strategies couldn't fully address. P3 acknowledged that while "exercise can temporarily make me happy through dopamine release, more effective medical solutions are needed to address psychological issues at their root. I still feel depression and pressure in the fixed space and isolated environment." Similarly, P16 described emotional fluctuations as "feeling pressure and uncertainty about the future remains an underlying tone".

These findings demonstrate that virtual workshops can serve as both a platform for observing adaptation strategies and intervention tool for fostering adaptation behaviors.

The workshop's success in catalyzing positive behavioral changes suggests its value extends beyond a research methodology to become a promising intervention framework. The identified limitations point to opportunities for future refinements in designing virtual environments that can both study and support isolation adaptation.

\section{Discussion}\label{sec:Discussion}
\subsection{Unveiling Adaptation Strategies Through Virtual Narrative Engagements}

Our research demonstrates how virtual narrative engagements can serve as an effective tool for understanding adaptation strategies during social isolation. Traditional research methods often struggle to capture the dynamic and creative nature of adaptation, particularly in constrained environments where direct observation is challenging \cite{gaver2022emergence}. This includes the use of fixed prototypes, simple observations, or diary studies \cite{Park_Yoo_Choe_Park_Song_2012, peldszus_perfect_2014}. While these methods are valuable in documenting personal trajectories and reflecting contextual limitations, they limit participants' imagination of possible solutions and lack the interactive exchange necessary to reveal collaboratively emergent strategies. 
In contrast, virtual narrative engagements enabled the exploration of asynchronous cooperation models, such as relay-style activities or distributed caregiving. These forms of indirect collaboration sparked participants' imagination about feasible future practices, beyond what they had personally experienced.

Adaptation is a dynamic process where individuals construct coping strategies through personal experimentation and lived experiences, while virtual narrative engagements can create a collaborative space for this process, by allowing participants to actively share their prior knowledge, explore and articulate adaptation approaches \cite{gresalfi2009virtual}. Through this approach, we observed a condensed version of the adaptation process over seven days, capturing the journey from initial divergent thinking to practical application. 

The effectiveness of our virtual workshop design rested on three key methodological elements. First, the narrative framing provided relatable yet transformative scenarios that encouraged participants to reflect deeply on their adaptation experiences. While imaginative (e.g., space missions, isolated cabins), these scenarios maintained sufficient parallels to real-world isolation, enabling participants to practically connect fictional explorations with actual coping strategies \cite{peldszus_perfect_2014}.

Second, our design balanced structure with openness through a combination of guided activities and free exploration. Structured prompts guided participants through collaborative storytelling and virtual environment customization, while open-ended tasks encouraged creative exploration. This balance proved crucial in uncovering both anticipated and unexpected adaptation strategies. For instance, when tasked with adapting to resource limitations in their virtual environment, participants revealed creative approaches to repurposing available materials and reimagining space usage that might not have emerged in traditional interviews.

Third, the immersive virtual environment facilitated experimentation with spatial and social configurations, enabling participants to demonstrate rather than merely describe their adaptation strategies. The virtual format allowed participants to image their environment, discuss different scenarios, and engage in collaborative problem-solving, providing rich behavioral data that complemented verbal responses.

As people's adaptation strategies in social isolation form a dynamic process of continuous negotiation with changing environments \cite{Kimhi_2011, Bjorklund_2015}, our approach proved particularly valuable in eliciting tacit knowledge. Through collaborative storytelling, participants moved beyond describing known coping mechanisms to uncover strategies they had not previously articulated or consciously considered within the safety of the virtual environment.

This methodological approach demonstrated how virtual narratives can bridge the gap between traditional user studies and the complex reality of adaptation in constrained environments. The combination of structured guidance and emergent exploration provides a robust framework for understanding not just what strategies people use, but how these strategies evolve and interact in dynamic situations \cite{gaver2022emergence}. This opens opportunities for exploring adaptation in other constrained environments, such as healthcare settings or space exploration, with the potential for future research to incorporate real-time adaptation tracking and investigate long-term engagement impacts.

 Virtual narrative approach can facilitate social isolation \cite{Udapola_2022, Bahng_2020}. As people's adaptation strategies in social isolation form a dynamic process of continuous negotiation with changing environments \cite{Kimhi_2011, Bjorklund_2015}, through collaborative storytelling, participants naturally emerged and revealed strategies they hadn't previously articulated or consciously considered. The workshop format encouraged them to move beyond describing known coping mechanisms to actively discovering new approaches within the safety of the virtual environment. This methodological approach demonstrated how virtual narratives can bridge the gap between traditional user studies and the complex reality of adaptation in constrained environments. The combination of structured guidance and emergent exploration provides a robust framework for understanding not just what strategies people use, but how these strategies evolve and interact in dynamic situations \cite{gaver2022emergence}. While participants primarily implemented more accessible strategies due to practical constraints, their attempts also demonstrated both the methodology's potential for pattern discovery and its potential positive impact on participants. This opens opportunities for exploring adaptation in other constrained environments, such as healthcare settings, space exploration and marine. Future research could refine virtual environments to include more advanced tools for real-time adaptation tracking or investigate how long-term engagement with such tools impacts adaptation strategies over time.

\subsection{Patterns of Adaptation Strategies in Social Isolation} 
Through our RtD process, we identified four interrelated modes of adaptation that emerged from participants' interactions within the virtual workshop: Support, Modification, Substitution, and Connection. 

\textbf{Support} represents the most straightforward adaptation pattern, involving the direct use of tools and resources to address specific needs during isolation. This includes engaging with entertainment media, maintaining exercise routines, or using digital platforms for basic tasks, whose effectiveness often translates well from other contexts to isolation scenarios. 

\textbf{Modification} involves more personalization than support, emphasizing how design can enable users to customize their environment or create new norms. This includes physical modifications (e.g., drawing decorations, plant cultivation) and non-physical adaptations (e.g., schedule adjustments). As prior research suggests that enabling creative possibilities can enhance adaptive capacity, highlighting the relationship between creativity and resilience \cite{xu2021depression}.

\textbf{Substitution} emerged as perhaps the most flexible and innovative pattern, encompassing both physical and imaginative substitutions. The virtual workshop revealed how participants developed creative alternatives when preferred options were unavailable, such as using everyday objects in novel ways or creating virtual equivalents of physical experiences. More abstractly, imaginative substitution allowed participants to reframe their isolation experience through narrative engagement, demonstrating how cognitive reframing can support adaptation during extended periods of isolation.

\textbf{Connection} patterns extended beyond traditional social interactions to encompass various forms of engagement. While existing research has focused on maintaining fixed relationships through remote IoT interactions or online gaming platforms \cite{gaver_yo_2022, Nardi_Harris_2006}, our findings suggest that connection needs are more fluid and multifaceted. The virtual narrative approach revealed nuanced connection strategies that included interactions with virtual communities, engagement with AI companions, and even relationships with physical objects in one's environment. Particularly notable was how collaborative creative activities, such as relayed storytelling or virtual space design, facilitated multiple forms of connection simultaneously.

\begin{figure*}[]
    \centering
    \includegraphics[width=1\linewidth]{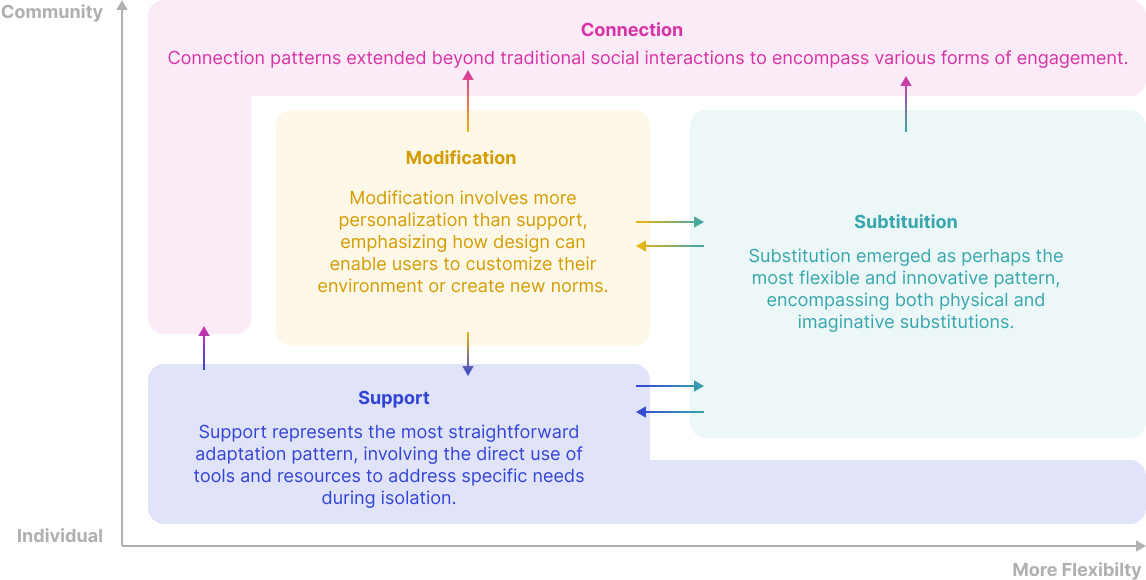}
    \caption{Interaction Patterns Among Four Adaptation Modes. This visualization maps the dynamic relationships between adaptation modes along individual-community and flexibility dimensions, showing their interconnections and distinct characteristics through spatial positioning and directional flows.}
    \Description{Interaction Patterns Among Four Adaptation Modes. This visualization maps the dynamic relationships between adaptation strategies (Support, Modification, Substitution, and Connection) along two dimensions: individual-community (vertical axis) and flexibility (horizontal axis). The diagram shows how these modes interact and flow into each other.}
    \label{fig:Patterns}
\end{figure*}

These four modes frequently operate in concert rather than in isolation. Effective design interventions for social isolation should consider how to enable and encourage the flexible combination of these modes rather than focusing on any single approach, as shown in Fig.\ref{fig:Patterns}.

This framework extends beyond immediate isolation scenarios, offering insights for designing adaptive systems in various constrained environments, from remote work situations to future space travel. It suggests that HCI research should focus not just on providing direct support but on creating platforms and catalysts that enable users to discover and combine different adaptation strategies in personally meaningful ways.

\subsection{Virtual narrative engagement as a potential inventention}

While our study's duration, limited by the quarantine condition, could not capture the long-term impact of virtual narrative engagement, the follow-up focus groups and post-workshop questionnaires revealed intriguing findings. For example, derived from its original purpose of understanding adaptation, the process itself influenced participants' real-world behaviors. This interaction between participants and the system exemplified the productive potential of embracing uncertainty \cite{giaccardi2024prototyping}.

Consistent with prior findings that online interactions can mitigate negative effects of social isolation \cite{baloran_knowledge_2020, norton_depression_2007}, our study revealed that virtual narrative engagement offered emotional benefits to participants. Participants reported reduced feelings of loneliness during focus group discussions, because the virtual space allowed them to express adaptation ideas, share existing strategies, and envision potential approaches collectively. 

Furthermore, our approach introduced new perspectives as participants incorporated their prior experiences into their narratives. This cross-pollination of ideas through virtual narratives provided experiential learning \cite{Nardi_Harris_2006}. The emergence of diverse adaptation strategies convinced participants that quarantine life could be changed by shifting perspectives. 
Participants' exploration of strategies aligned with our definition of adaptation \cite{Bjorklund_2015} and yielded positive outcomes. While some participants found certain strategies had limited effectiveness, they found solutions that might better address their own needs, such as medical support or methods to enhance future predictability. These insights can inform future exploration of adaptation mechanisms for people in isolation.

In particular, catalyzing participants' behavioral changes suggests virtual narrative engagement's value extends beyond a research methodology to become a potential intervention framework. While initially designed as a research probe, our approach potentially triggered lasting behavioral and emotional shifts that warrant further investigation. 
This methodological duality—where research tools become transformative interventions—opens new opportunities for designing virtual environments that can both study and support isolation adaptation. Future research should examine how such narrative-driven engagements shape not just immediate reflections but lasting adaptations, potentially informing the design of interventions that are both participatory and transformative\cite{agcal_bricolage_2025,lc_designing_2022}. Of particular interest are settings of prolonged isolation such as Antarctic expeditions or space station residencies \cite{baloran_knowledge_2020,norton_depression_2007,palinkas_psychosocial_2021,peldszus_perfect_2014,yevdokimova_coping_2021}. Understanding how speculative, narrative-driven engagement shapes not just immediate reflections but lasting behavior informs the design of participatory narratives for positive behavioral adaptation.

\subsection{Design Implications for Supporting Adaptation in Social Isolation} 
\subsubsection{Design Implications}
Drawing from the patterns we identified, we propose design recommendations across three levels: pattern-specific support, integrated adaptation, and long-term sustainability.

For support-oriented adaptation, systems should provide modular resources that users can organize and access based on their needs. This could include customizable virtual workspace templates, adaptable routine planners, and flexible resource management tools. The key is to enable users to curate their support infrastructure rather than imposing fixed solutions.

For modification-based adaptation, designs should facilitate gradual environment transformation and enable meaningful creative outputs that foster user satisfaction. This includes tools for virtual-physical space modification, where users can experiment with changes in virtual environments before implementing them physically. Systems could offer visualization tools for space reconfiguration and provide suggestions for creative space utilization based on successful adaptations from other users.

To enable substitution strategies, designs should incorporate mechanisms that inspire creative alternatives. This might include AI-powered systems that suggest novel uses for available resources or community-driven databases of substitution ideas. The goal is to help users discover unexpected possibilities within their constraints rather than focusing on what's missing. When online communities are unavailable, toolkits that stimulate imaginative alternatives can serve as effective substitutes. Additionally, embedding meaningful narrative scenarios within supportive tools and applications could enhance this process, allowing users to explore adaptation strategies through contextually relevant storytelling.

For connection-focused adaptation, systems should support multi-layered social interaction. This involves integrating synchronous and asynchronous communication channels, collaborative creation spaces, and narrative-sharing platforms. Importantly, these features should allow users to modulate their social engagement levels, recognizing that connection needs vary across individuals and situations.
The key is to avoid treating isolation adaptation as fixed solutions, instead designing systems that support flexible combinations of strategies and organic evolution of coping mechanisms.

\subsubsection{Supporting Long-term and Sustainable Adaptation}
Our findings highlight adaptation as an ongoing process rather than a one-time adjustment. To support this temporal dimension, we propose three key design considerations: First, systems should implement stage-aware support mechanisms that evolve with users' isolation experience - from establishing basic routines in early stages to supporting complex adaptations later. Second, designs should incorporate reflection tools like interactive journals and visualization features that help users track and understand their adaptation patterns. Third, platforms should facilitate community knowledge sharing, enabling users to contribute and learn from others' successful strategies while maintaining contextual relevance.

Beyond individual-focused benefits, our approach promotes sustainability by leveraging people's inherent adaptation capabilities and encouraging the flexible use of existing tools. By reducing reliance on physical resources and avoiding redundant resource creation, digital platforms ensure accessibility for diverse populations while minimizing environmental impact. The scalability of adaptation allows for cost-effective applications across varied contexts, from supporting aging populations to addressing challenges in extreme work environments and space exploration.

While our study focused on short-term pandemic-related isolation, we see potential for applying this approach in other prolonged or self-imposed isolation contexts. For instance, individuals living alone with chronic illness, or those who voluntarily withdraw from social interaction, may benefit from similar forms of virtual narrative engagement. When the narrative environment is thoughtfully designed, the emotional safety, imaginative flexibility, and low-pressure collaboration make this observational method potentially more acceptable than traditional user research approaches, enabling us to better reach and understand marginalized populations.

For individuals with social interaction challenges, virtual narrative engagement could serve as a gentle pathway toward self-expression, similar to exposure therapy principles. This approach allows them to gradually improve their social interaction abilities or develop better coping mechanisms. For those dealing with chronic illnesses or facing new pandemic situations - conditions that involve ongoing challenges and changing circumstances - virtual narrative engagement offers a sustainable and accessible support system. Through virtual workshops, we can observe their adaptation processes, gather strategy feedback, and develop responsive support mechanisms. This iterative cycle of implementation and rapid prototyping through narrative engagement enables participants to continuously refine their adaptation strategies while helping researchers develop more effective support systems.

As society prepares for future health crises or navigates long-term shifts in social behavior, virtual narrative engagements provide a scalable methodology to study and support adaptation across various contexts. By emphasizing emotional depth and shared understanding over transactional interactions, our approach fosters empathy and reduces communication barriers in diverse communities. This framework not only addresses immediate challenges in isolation but also contributes to building collective resilience while adhering to environmental consciousness—an essential consideration for the future of adaptation support systems.

\subsection{Limitations and Future Work}
Our study has several limitations that should be considered when interpreting the results. 

\subsubsection{Research Method and Its Scope of Applicability}
While we chose the space metaphor for its imaginative potential in simulating virtual isolation, this specific context may not fully represent other isolation scenarios. Our study was situated in the context of COVID-19 quarantine, where the length of isolation was relatively standardized and finite. 
As a result, the findings derived from this study may not be fully generalizable to extreme or open-ended isolation contexts, such as policy-driven lockdowns with uncertain endpoints. Nevertheless, we believe our approach still offers valuable insights into how individuals adapt under uncertainty, and speculative narrative engagement might help shape emotional responses in such ambiguous conditions.

In addition, for populations facing isolation due to internal or structural factors, such as older adults, individuals with limited mobility or reduced social networks, the engagement we used may require further adaptation. These groups may exhibit different adaptation patterns, and future studies should carefully consider the population and reasons behind the isolation. Given the use of narrative elements in our method, it is also important to ensure that the story-building process does not inadvertently trigger secondary trauma, especially among vulnerable participants.

\subsubsection{Verification and Data Analysis}
Our analytical approach using thematic analysis, while common in HCI research, has inherent limitations including subjectivity and lack of standardization \cite{Kiger_Varpio_2020}.

Our focus on using the experience primarily as a probing method for adaptation strategies, combined with the relatively short isolation period studied, limited our ability to conduct a comprehensive evaluation. Future studies would benefit from incorporating both immediate quantitative measures and long-term validation methods, such as qualitative assessments or daily journaling.
Furthermore, the inclusion of control groups and quantitative measurement tools would help better evaluate the specific effects of virtual narrative engagement on participants' actual adaptation processes. 

\subsubsection{Demographic Considerations} 
Our participant pool was exclusively from China, potentially biasing the isolation mitigation strategies toward Chinese cultural perspectives, which tend to be more conservative compared to Western approaches \cite{Shaw_1997, Vaus_2018}. While this cultural specificity presents a limitation, it also offers unique insights into adaptation patterns within Eastern cultural contexts. The limited age range of participants also leaves questions about the approach's suitability for vulnerable populations such as the elderly and children. Furthermore, our sampling method resulted in an unbalanced gender distribution, which may affect our findings given that gender influences research participation patterns \cite{Otufowora_Liu_Young_Egan_Varma_Striley_Cottler_2021}.

\subsubsection{Participant Engagement and Potential Bias} The collaborative nature of our storytelling activities revealed varying levels of participant engagement, influenced by individual factors such as personality, professional background, and current circumstances. This variation in contribution levels should be considered as not all participants contributed equally to the activity. In contrast, participants who showed higher levels of engagement often demonstrated stronger technical literacy and self-adaptation capabilities, potentially resulting in more flexible adaptation patterns. This engagement bias should be carefully considered when developing design implications, as the resulting recommendations may better serve users who are naturally more proactive in exploring adaptation strategies.

\subsubsection{Applications of GenAI} Recent work shows that GenAI can support creative processes like creative design\cite{ling_sketchar_2024}, fiction writing\cite{yang_ai_2022}, and narrative expression\cite{he_i_2025,fu_being_2024}. 
Participatory workflows with GenAI have potential to encourage users to take positive behaviors by natural dialogue, but they could also lead to negative consequences of negative comparisons and biased perspectives\cite{han_when_2024}. Future research may explore how participatory strategies involving an empowering use of GenAI may promote wellness\cite{cacioppo2008loneliness} in isolation context while avoiding its potential pitfalls. 

\section{Conclusion}\label{sec:Conclusion}
We investigated how virtual narrative engagements, set in space-themed scenarios, can facilitate understanding and support for adaptation strategies during social isolation. Our findings highlight the potential of combining immersive virtual experiences with collaborative storytelling and design activities to uncover participants' adaptation patterns while also fostering their coping processes. 

While our study was conducted within a specific cultural context, the methodological framework and insights we developed are adaptable to other contexts and isolation scenarios, including remote work, healthcare, and extreme environments. Future research can extend this approach to refine its application across diverse populations, leveraging virtual narratives to enhance social well-being and resilience during periods of isolation.

\bibliographystyle{ACM-Reference-Format}
\bibliography{references}


\begin{thebibliography}{85}


\ifx \showCODEN    \undefined \def \showCODEN     #1{\unskip}     \fi
\ifx \showDOI      \undefined \def \showDOI       #1{#1}\fi
\ifx \showISBNx    \undefined \def \showISBNx     #1{\unskip}     \fi
\ifx \showISBNxiii \undefined \def \showISBNxiii  #1{\unskip}     \fi
\ifx \showISSN     \undefined \def \showISSN      #1{\unskip}     \fi
\ifx \showLCCN     \undefined \def \showLCCN      #1{\unskip}     \fi
\ifx \shownote     \undefined \def \shownote      #1{#1}          \fi
\ifx \showarticletitle \undefined \def \showarticletitle #1{#1}   \fi
\ifx \showURL      \undefined \def \showURL       {\relax}        \fi
\providecommand\bibfield[2]{#2}
\providecommand\bibinfo[2]{#2}
\providecommand\natexlab[1]{#1}
\providecommand\showeprint[2][]{arXiv:#2}

\bibitem[Arq(2022)]%
        {Arquilla_Webb_Anderson_2022}
 \bibinfo{year}{2022}\natexlab{}.
\newblock \showarticletitle{Isolation and confinement due to the COVID-19 pandemic: Lessons for human spaceflight}.
\newblock   \bibinfo{volume}{196} (\bibinfo{date}{July} \bibinfo{year}{2022}), \bibinfo{pages}{282–289}.
\newblock
\showISSN{0094-5765}
\urldef\tempurl%
\url{https://doi.org/10.1016/j.actaastro.2022.04.026}
\showDOI{\tempurl}


\bibitem[Agcal et~al\mbox{.}(2025)]%
        {agcal_bricolage_2025}
\bibfield{author}{\bibinfo{person}{Bengi Agcal}, \bibinfo{person}{Ines~Ziyou Yin}, \bibinfo{person}{Marty Miller}, {and} \bibinfo{person}{RAY LC}.} \bibinfo{year}{2025}\natexlab{}.
\newblock \showarticletitle{Bricolage: aligning with climate action through playful participatory design in speculative scenarios}.
\newblock \bibinfo{journal}{\emph{International Journal of Play}} (\bibinfo{date}{Jan.} \bibinfo{year}{2025}).
\newblock
\showISSN{2159-4937}
\urldef\tempurl%
\url{https://www.tandfonline.com/doi/full/10.1080/21594937.2025.2464324}
\showURL{%
\tempurl}
\newblock
\shownote{Publisher: Routledge}.


\bibitem[Arone et~al\mbox{.}({[n.\,d.]})]%
        {Arone_2021}
\bibfield{author}{\bibinfo{person}{Alessandro Arone}, \bibinfo{person}{Tea Ivaldi}, \bibinfo{person}{Konstantin Loganovsky}, \bibinfo{person}{Stefania Palermo}, \bibinfo{person}{Elisabetta Parra}, \bibinfo{person}{Walter Flamini}, {and} \bibinfo{person}{Donatella Marazziti}.} \bibinfo{year}{[n.\,d.]}\natexlab{}.
\newblock \showarticletitle{The Burden of Space Exploration on the Mental Health of Astronauts: A Narrative Review}.
\newblock \bibinfo{journal}{\emph{Clinical Neuropsychiatry}} \bibinfo{volume}{18}, \bibinfo{number}{5} (\bibinfo{year}{[n.\,d.]}), \bibinfo{pages}{237–246}.
\newblock
\showISSN{1724-4935}
\urldef\tempurl%
\url{https://doi.org/10.36131/cnfioritieditore20210502}
\showDOI{\tempurl}


\bibitem[Baecker et~al\mbox{.}(2014)]%
        {baecker_technology_2014}
\bibfield{author}{\bibinfo{person}{Ron Baecker}, \bibinfo{person}{Kate Sellen}, \bibinfo{person}{Sarah Crosskey}, \bibinfo{person}{Veronique Boscart}, {and} \bibinfo{person}{Barbara Barbosa~Neves}.} \bibinfo{year}{2014}\natexlab{}.
\newblock \showarticletitle{Technology to Reduce Social Isolation and Loneliness}. In \bibinfo{booktitle}{\emph{Proceedings of the 16th International ACM SIGACCESS Conference on Computers \& Accessibility}} \emph{(\bibinfo{series}{ASSETS '14})}. \bibinfo{publisher}{Association for Computing Machinery}, \bibinfo{address}{New York, NY, USA}, \bibinfo{pages}{27--34}.
\newblock
\showISBNx{978-1-4503-2720-6}
\urldef\tempurl%
\url{https://doi.org/10.1145/2661334.2661375}
\showDOI{\tempurl}


\bibitem[Bahng et~al\mbox{.}(2020)]%
        {Bahng_2020}
\bibfield{author}{\bibinfo{person}{Sojung Bahng}, \bibinfo{person}{Ryan~M. Kelly}, {and} \bibinfo{person}{Jon McCormack}.} \bibinfo{year}{2020}\natexlab{}.
\newblock \showarticletitle{Reflexive VR Storytelling Design Beyond Immersion: Facilitating Self-Reflection on Death and Loneliness}. In \bibinfo{booktitle}{\emph{Proceedings of the 2020 CHI Conference on Human Factors in Computing Systems}} (Honolulu, HI, USA) \emph{(\bibinfo{series}{CHI '20})}. \bibinfo{publisher}{Association for Computing Machinery}, \bibinfo{address}{New York, NY, USA}, \bibinfo{pages}{1–13}.
\newblock
\showISBNx{9781450367080}
\urldef\tempurl%
\url{https://doi.org/10.1145/3313831.3376582}
\showDOI{\tempurl}


\bibitem[Baloran(2020)]%
        {baloran_knowledge_2020}
\bibfield{author}{\bibinfo{person}{Erick~T. Baloran}.} \bibinfo{year}{2020}\natexlab{}.
\newblock \showarticletitle{Knowledge, Attitudes, Anxiety, and Coping Strategies of Students during COVID-19 Pandemic}.
\newblock \bibinfo{journal}{\emph{Journal of Loss and Trauma}} \bibinfo{volume}{25}, \bibinfo{number}{8} (\bibinfo{year}{2020}), \bibinfo{pages}{635--642}.
\newblock
\showISSN{1532-5024}
\urldef\tempurl%
\url{https://doi.org/10.1080/15325024.2020.1769300}
\showDOI{\tempurl}


\bibitem[Bjorklund(2015)]%
        {Bjorklund_2015}
\bibfield{author}{\bibinfo{person}{David~F. Bjorklund}.} \bibinfo{year}{2015}\natexlab{}.
\newblock \showarticletitle{Developing adaptations}.
\newblock \bibinfo{journal}{\emph{Developmental Review}}  \bibinfo{volume}{38} (\bibinfo{date}{Dec.} \bibinfo{year}{2015}), \bibinfo{pages}{13–35}.
\newblock
\showISSN{02732297}
\urldef\tempurl%
\url{https://doi.org/10.1016/j.dr.2015.07.002}
\showDOI{\tempurl}


\bibitem[Bonsaksen et~al\mbox{.}({[n.\,d.]})]%
        {Bonsaksen_2023}
\bibfield{author}{\bibinfo{person}{Tore Bonsaksen}, \bibinfo{person}{Mary Ruffolo}, \bibinfo{person}{Daicia Price}, \bibinfo{person}{Janni Leung}, \bibinfo{person}{Hilde Thygesen}, \bibinfo{person}{Gary Lamph}, \bibinfo{person}{Isaac Kabelenga}, {and} \bibinfo{person}{Amy~Østertun Geirdal}.} \bibinfo{year}{[n.\,d.]}\natexlab{}.
\newblock \showarticletitle{Associations between social media use and loneliness in a cross-national population: do motives for social media use matter?}
\newblock \bibinfo{journal}{\emph{Health Psychology and Behavioral Medicine}} \bibinfo{volume}{11}, \bibinfo{number}{1} (\bibinfo{year}{[n.\,d.]}), \bibinfo{pages}{2158089}.
\newblock
\showISSN{2164-2850}
\urldef\tempurl%
\url{https://doi.org/10.1080/21642850.2022.2158089}
\showDOI{\tempurl}


\bibitem[Brandt et~al\mbox{.}(2022)]%
        {Brandt_2022}
\bibfield{author}{\bibinfo{person}{Lasse Brandt}, \bibinfo{person}{Shuyan Liu}, \bibinfo{person}{Christine Heim}, {and} \bibinfo{person}{Andreas Heinz}.} \bibinfo{year}{2022}\natexlab{}.
\newblock \showarticletitle{The effects of social isolation stress and discrimination on mental health}.
\newblock \bibinfo{journal}{\emph{Translational Psychiatry}}  \bibinfo{volume}{12} (\bibinfo{date}{Sept.} \bibinfo{year}{2022}), \bibinfo{pages}{398}.
\newblock
\showISSN{2158-3188}
\urldef\tempurl%
\url{https://doi.org/10.1038/s41398-022-02178-4}
\showDOI{\tempurl}


\bibitem[Braun and Clarke(2006)]%
        {braun_using_2006}
\bibfield{author}{\bibinfo{person}{Virginia Braun} {and} \bibinfo{person}{Victoria Clarke}.} \bibinfo{year}{2006}\natexlab{}.
\newblock \showarticletitle{Using Thematic Analysis in Psychology}.
\newblock \bibinfo{journal}{\emph{Qualitative Research in Psychology}} \bibinfo{volume}{3}, \bibinfo{number}{2} (\bibinfo{year}{2006}), \bibinfo{pages}{77--101}.
\newblock
\showISSN{1478-0887}
\urldef\tempurl%
\url{https://doi.org/10.1191/1478088706qp063oa}
\showDOI{\tempurl}


\bibitem[Braun and Clarke(2014)]%
        {braun_what_2014}
\bibfield{author}{\bibinfo{person}{Virginia Braun} {and} \bibinfo{person}{Victoria Clarke}.} \bibinfo{year}{2014}\natexlab{}.
\newblock \showarticletitle{What Can ``Thematic Analysis'' Offer Health and Wellbeing Researchers?}
\newblock \bibinfo{journal}{\emph{International Journal of Qualitative Studies on Health and Well-being}} \bibinfo{volume}{9}, \bibinfo{number}{1} (\bibinfo{year}{2014}), \bibinfo{pages}{26152}.
\newblock
\showISSN{null}
\urldef\tempurl%
\url{https://doi.org/10.3402/qhw.v9.26152}
\showDOI{\tempurl}


\bibitem[Brooks et~al\mbox{.}(2020)]%
        {brooks_psychological_2020}
\bibfield{author}{\bibinfo{person}{Samantha~K Brooks}, \bibinfo{person}{Rebecca~K Webster}, \bibinfo{person}{Louise~E Smith}, \bibinfo{person}{Lisa Woodland}, \bibinfo{person}{Simon Wessely}, \bibinfo{person}{Neil Greenberg}, {and} \bibinfo{person}{Gideon~James Rubin}.} \bibinfo{year}{2020}\natexlab{}.
\newblock \showarticletitle{The Psychological Impact of Quarantine and How to Reduce It: Rapid Review of the Evidence}.
\newblock \bibinfo{journal}{\emph{The Lancet}} \bibinfo{volume}{395}, \bibinfo{number}{10227} (\bibinfo{year}{2020}), \bibinfo{pages}{912--920}.
\newblock
\showISSN{0140-6736}
\urldef\tempurl%
\url{https://doi.org/10.1016/S0140-6736(20)30460-8}
\showDOI{\tempurl}


\bibitem[Brown et~al\mbox{.}(2022)]%
        {brown_employing_2022}
\bibfield{author}{\bibinfo{person}{Ryan Brown}, \bibinfo{person}{Samin {Habibi-Luevano}}, \bibinfo{person}{Gil Robern}, \bibinfo{person}{Kody Wood}, \bibinfo{person}{Sharman Perera}, \bibinfo{person}{Alvaro {Uribe-Quevedo}}, \bibinfo{person}{Callan Brown}, \bibinfo{person}{Khalid Rizk}, \bibinfo{person}{Filippo Genco}, \bibinfo{person}{Jennifer McKellar}, \bibinfo{person}{Kirk Atkinson}, {and} \bibinfo{person}{Akira Tokuhiro}.} \bibinfo{year}{2022}\natexlab{}.
\newblock \showarticletitle{Employing Mozilla Hubs as an Alternative Tool for Student Outreach: A Design Challenge Use Case}. In \bibinfo{booktitle}{\emph{New Realities, Mobile Systems and Applications}} \emph{(\bibinfo{series}{Lecture Notes in Networks and Systems})}, \bibfield{editor}{\bibinfo{person}{Michael~E. Auer} {and} \bibinfo{person}{Thrasyvoulos Tsiatsos}} (Eds.). \bibinfo{publisher}{Springer International Publishing}, \bibinfo{address}{Cham}, \bibinfo{pages}{213--222}.
\newblock
\showISBNx{978-3-030-96296-8}
\urldef\tempurl%
\url{https://doi.org/10.1007/978-3-030-96296-8-20}
\showDOI{\tempurl}


\bibitem[Cacioppo and Hawkley(2003)]%
        {Cacioppo_Hawkley_2003}
\bibfield{author}{\bibinfo{person}{John~T. Cacioppo} {and} \bibinfo{person}{Louise~C. Hawkley}.} \bibinfo{year}{2003}\natexlab{}.
\newblock \showarticletitle{Social isolation and health, with an emphasis on underlying mechanisms}.
\newblock \bibinfo{journal}{\emph{Perspectives in Biology and Medicine}} \bibinfo{volume}{46}, \bibinfo{number}{3 Suppl} (\bibinfo{year}{2003}), \bibinfo{pages}{S39--52}.
\newblock
\showISSN{0031-5982}


\bibitem[Cacioppo et~al\mbox{.}(2002)]%
        {Cacioppo_2002}
\bibfield{author}{\bibinfo{person}{John~T. Cacioppo}, \bibinfo{person}{Louise~C. Hawkley}, \bibinfo{person}{Gary~G. Berntson}, \bibinfo{person}{John~M. Ernst}, \bibinfo{person}{Amber~C. Gibbs}, \bibinfo{person}{Robert Stickgold}, {and} \bibinfo{person}{J.~Allan Hobson}.} \bibinfo{year}{2002}\natexlab{}.
\newblock \showarticletitle{Do lonely days invade the nights? Potential social modulation of sleep efficiency}.
\newblock \bibinfo{journal}{\emph{Psychological Science}} \bibinfo{volume}{13}, \bibinfo{number}{4} (\bibinfo{date}{July} \bibinfo{year}{2002}), \bibinfo{pages}{384–387}.
\newblock
\showISSN{0956-7976}
\urldef\tempurl%
\url{https://doi.org/10.1111/1467-9280.00469}
\showDOI{\tempurl}


\bibitem[Cacioppo and Patrick(2008)]%
        {cacioppo2008loneliness}
\bibfield{author}{\bibinfo{person}{John~T Cacioppo} {and} \bibinfo{person}{William Patrick}.} \bibinfo{year}{2008}\natexlab{}.
\newblock \bibinfo{booktitle}{\emph{Loneliness: Human nature and the need for social connection}}.
\newblock \bibinfo{publisher}{WW Norton \& Company}.
\newblock


\bibitem[Chancellor et~al\mbox{.}(2016a)]%
        {Chancellor_Lin_Goodman_Zerwas_Choudhury_2016}
\bibfield{author}{\bibinfo{person}{Stevie Chancellor}, \bibinfo{person}{Zhiyuan Lin}, \bibinfo{person}{Erica~L. Goodman}, \bibinfo{person}{Stephanie Zerwas}, {and} \bibinfo{person}{Munmun De~Choudhury}.} \bibinfo{year}{2016}\natexlab{a}.
\newblock \showarticletitle{Quantifying and Predicting Mental Illness Severity in Online Pro-Eating Disorder Communities}. In \bibinfo{booktitle}{\emph{Proceedings of the 19th ACM Conference on Computer-Supported Cooperative Work \& Social Computing}} \emph{(\bibinfo{series}{CSCW ’16})}. \bibinfo{publisher}{Association for Computing Machinery}, \bibinfo{address}{New York, NY, USA}, \bibinfo{pages}{1171–1184}.
\newblock
\showISBNx{978-1-4503-3592-8}
\urldef\tempurl%
\url{https://doi.org/10.1145/2818048.2819973}
\showDOI{\tempurl}


\bibitem[Chancellor et~al\mbox{.}(2016b)]%
        {Chancellor_Pater_Clear_Choudhury_2016}
\bibfield{author}{\bibinfo{person}{Stevie Chancellor}, \bibinfo{person}{Jessica~Annette Pater}, \bibinfo{person}{Trustin Clear}, \bibinfo{person}{Eric Gilbert}, {and} \bibinfo{person}{Munmun De~Choudhury}.} \bibinfo{year}{2016}\natexlab{b}.
\newblock \showarticletitle{\#thyghgapp: Instagram Content Moderation and Lexical Variation in Pro-Eating Disorder Communities}. In \bibinfo{booktitle}{\emph{Proceedings of the 19th ACM Conference on Computer-Supported Cooperative Work \& Social Computing}} \emph{(\bibinfo{series}{CSCW ’16})}. \bibinfo{publisher}{Association for Computing Machinery}, \bibinfo{address}{New York, NY, USA}, \bibinfo{pages}{1201–1213}.
\newblock
\showISBNx{978-1-4503-3592-8}
\urldef\tempurl%
\url{https://doi.org/10.1145/2818048.2819963}
\showDOI{\tempurl}


\bibitem[Choudhury et~al\mbox{.}(2013)]%
        {Choudhury_Gamon_Counts_Horvitz_2013}
\bibfield{author}{\bibinfo{person}{Munmun~De Choudhury}, \bibinfo{person}{Michael Gamon}, \bibinfo{person}{Scott Counts}, {and} \bibinfo{person}{Eric Horvitz}.} \bibinfo{year}{2013}\natexlab{}.
\newblock \showarticletitle{Predicting Depression via Social Media}.
\newblock \bibinfo{journal}{\emph{Proceedings of the International AAAI Conference on Web and Social Media}} \bibinfo{volume}{7}, \bibinfo{number}{11} (\bibinfo{year}{2013}), \bibinfo{pages}{128–137}.
\newblock
\showISSN{2334-0770}
\urldef\tempurl%
\url{https://doi.org/10.1609/icwsm.v7i1.14432}
\showDOI{\tempurl}


\bibitem[Choukér and Stahn(2020)]%
        {Chouker_2020}
\bibfield{author}{\bibinfo{person}{Alexander Choukér} {and} \bibinfo{person}{Alexander~C. Stahn}.} \bibinfo{year}{2020}\natexlab{}.
\newblock \showarticletitle{COVID-19-The largest isolation study in history: the value of shared learnings from spaceflight analogs}.
\newblock \bibinfo{journal}{\emph{NPJ microgravity}}  \bibinfo{volume}{6} (\bibinfo{year}{2020}), \bibinfo{pages}{32}.
\newblock
\showISSN{2373-8065}
\urldef\tempurl%
\url{https://doi.org/10.1038/s41526-020-00122-8}
\showDOI{\tempurl}


\bibitem[De~Vaus et~al\mbox{.}(2018)]%
        {Vaus_2018}
\bibfield{author}{\bibinfo{person}{June De~Vaus}, \bibinfo{person}{Matthew~J. Hornsey}, \bibinfo{person}{Peter Kuppens}, {and} \bibinfo{person}{Brock Bastian}.} \bibinfo{year}{2018}\natexlab{}.
\newblock \showarticletitle{Exploring the East-West Divide in Prevalence of Affective Disorder: A Case for Cultural Differences in Coping With Negative Emotion}.
\newblock \bibinfo{journal}{\emph{Personality and Social Psychology Review}} \bibinfo{volume}{22}, \bibinfo{number}{3} (\bibinfo{date}{Aug.} \bibinfo{year}{2018}), \bibinfo{pages}{285–304}.
\newblock
\showISSN{1088-8683}
\urldef\tempurl%
\url{https://doi.org/10.1177/1088868317736222}
\showDOI{\tempurl}


\bibitem[Desjardins and Key(2020)]%
        {Desjardins_2020}
\bibfield{author}{\bibinfo{person}{Audrey Desjardins} {and} \bibinfo{person}{Cayla Key}.} \bibinfo{year}{2020}\natexlab{}.
\newblock \showarticletitle{Parallels, Tangents, and Loops: Reflections on the 'Through' Part of RtD}. In \bibinfo{booktitle}{\emph{Proceedings of the 2020 ACM Designing Interactive Systems Conference}} (Eindhoven, Netherlands) \emph{(\bibinfo{series}{DIS '20})}. \bibinfo{publisher}{Association for Computing Machinery}, \bibinfo{address}{New York, NY, USA}, \bibinfo{pages}{2133–2147}.
\newblock
\showISBNx{9781450369749}
\urldef\tempurl%
\url{https://doi.org/10.1145/3357236.3395586}
\showDOI{\tempurl}


\bibitem[{D{\'i}az-Kommonen} et~al\mbox{.}(2009)]%
        {diaz-kommonen_role_2009}
\bibfield{author}{\bibinfo{person}{Lily {D{\'i}az-Kommonen}}, \bibinfo{person}{Markku Reunanen}, {and} \bibinfo{person}{Anna Salmi}.} \bibinfo{year}{2009}\natexlab{}.
\newblock \showarticletitle{Role Playing and Collaborative Scenario Design Development}.
\newblock \bibinfo{journal}{\emph{ICED 2009, 17th International Conference on Engineering Design, August 24-27, Center for Design Research, Stanford University}} (\bibinfo{year}{2009}), \bibinfo{pages}{79--86}.
\newblock


\bibitem[Einav and Margalit(2023)]%
        {Einav_Margalit_2023}
\bibfield{author}{\bibinfo{person}{Michal Einav} {and} \bibinfo{person}{Malka Margalit}.} \bibinfo{year}{2023}\natexlab{}.
\newblock \showarticletitle{Loneliness before and after COVID-19: Sense of Coherence and Hope as Coping Mechanisms}.
\newblock \bibinfo{journal}{\emph{International Journal of Environmental Research and Public Health}} \bibinfo{volume}{20}, \bibinfo{number}{10} (\bibinfo{date}{May} \bibinfo{year}{2023}), \bibinfo{pages}{5840}.
\newblock
\showISSN{1661-7827}
\urldef\tempurl%
\url{https://doi.org/10.3390/ijerph20105840}
\showDOI{\tempurl}


\bibitem[Epp et~al\mbox{.}(2024)]%
        {Epp_2024}
\bibfield{author}{\bibinfo{person}{Felix~Anand Epp}, \bibinfo{person}{Anton~Poikolainen Ros\'{e}n}, \bibinfo{person}{Antti Salovaara}, {and} \bibinfo{person}{Camilo Sanchez}.} \bibinfo{year}{2024}\natexlab{}.
\newblock \showarticletitle{Uncertainties as Generative Resources in Research through Design: Three Dynamics for Moving in a Design Space}.
\newblock \bibinfo{journal}{\emph{ACM Trans. Comput.-Hum. Interact.}} \bibinfo{volume}{31}, \bibinfo{number}{6}, Article \bibinfo{articleno}{70} (\bibinfo{date}{Dec.} \bibinfo{year}{2024}), \bibinfo{numpages}{31}~pages.
\newblock
\showISSN{1073-0516}
\urldef\tempurl%
\url{https://doi.org/10.1145/3689041}
\showDOI{\tempurl}


\bibitem[Fancourt et~al\mbox{.}(2021)]%
        {fancourt2021trajectories}
\bibfield{author}{\bibinfo{person}{Daisy Fancourt}, \bibinfo{person}{Andrew Steptoe}, {and} \bibinfo{person}{Feifei Bu}.} \bibinfo{year}{2021}\natexlab{}.
\newblock \showarticletitle{Trajectories of anxiety and depressive symptoms during enforced isolation due to COVID-19 in England: a longitudinal observational study}.
\newblock \bibinfo{journal}{\emph{The Lancet Psychiatry}} \bibinfo{volume}{8}, \bibinfo{number}{2} (\bibinfo{year}{2021}), \bibinfo{pages}{141--149}.
\newblock


\bibitem[Farhang et~al\mbox{.}(2022)]%
        {farhang_impact_2022}
\bibfield{author}{\bibinfo{person}{Maryam Farhang}, \bibinfo{person}{Claudia {Miranda-Castillo}}, \bibinfo{person}{Maria~Isabel Behrens}, \bibinfo{person}{Eduardo Castillo}, \bibinfo{person}{Sandra Mosquera~Amar}, {and} \bibinfo{person}{Graciela Rojas}.} \bibinfo{year}{2022}\natexlab{}.
\newblock \showarticletitle{Impact of Social Isolation and Coping Strategies in Older Adults with Mild Cognitive Impairment during the Covid-19 Pandemic: A Qualitative Study}.
\newblock \bibinfo{journal}{\emph{Aging \& Mental Health}} \bibinfo{volume}{26}, \bibinfo{number}{7} (\bibinfo{year}{2022}), \bibinfo{pages}{1395--1416}.
\newblock
\showISSN{1360-7863}
\urldef\tempurl%
\url{https://doi.org/10.1080/13607863.2021.1958145}
\showDOI{\tempurl}


\bibitem[Fu et~al\mbox{.}(2024)]%
        {fu_being_2024}
\bibfield{author}{\bibinfo{person}{Kexue Fu}, \bibinfo{person}{Ruishan Wu}, \bibinfo{person}{Yuying Tang}, \bibinfo{person}{Yixin Chen}, \bibinfo{person}{Bowen Liu}, {and} \bibinfo{person}{RAY LC}.} \bibinfo{year}{2024}\natexlab{}.
\newblock \showarticletitle{"{Being} {Eroded}, {Piece} by {Piece}": {Enhancing} {Engagement} and {Storytelling} in {Cultural} {Heritage} {Dissemination} by {Exhibiting} {GenAI} {Co}-{Creation} {Artifacts}}. In \bibinfo{booktitle}{\emph{Proceedings of the 2024 {ACM} {Designing} {Interactive} {Systems} {Conference}}} \emph{(\bibinfo{series}{{DIS} '24})}. \bibinfo{publisher}{Association for Computing Machinery}, \bibinfo{address}{New York, NY, USA}, \bibinfo{pages}{2833--2850}.
\newblock
\showISBNx{9798400705830}
\urldef\tempurl%
\url{https://doi.org/10.1145/3643834.3660711}
\showDOI{\tempurl}


\bibitem[Gaver et~al\mbox{.}(2022a)]%
        {gaver_yo_2022}
\bibfield{author}{\bibinfo{person}{William Gaver}, \bibinfo{person}{Andy Boucher}, \bibinfo{person}{Dean Brown}, \bibinfo{person}{David Chatting}, \bibinfo{person}{Naho Matsuda}, \bibinfo{person}{Liliana Ovalle}, \bibinfo{person}{Andy Sheen}, {and} \bibinfo{person}{Michail Vanis}.} \bibinfo{year}{2022}\natexlab{a}.
\newblock \showarticletitle{Yo\textendash Yo Machines: Self-Build Devices That Support Social Connections During the Pandemic}. In \bibinfo{booktitle}{\emph{CHI Conference on Human Factors in Computing Systems}}. \bibinfo{publisher}{ACM}, \bibinfo{address}{New Orleans LA USA}, \bibinfo{pages}{1--17}.
\newblock
\showISBNx{978-1-4503-9157-3}
\urldef\tempurl%
\url{https://doi.org/10.1145/3491102.3517547}
\showDOI{\tempurl}


\bibitem[Gaver et~al\mbox{.}(2022b)]%
        {gaver2022emergence}
\bibfield{author}{\bibinfo{person}{William Gaver}, \bibinfo{person}{Peter~Gall Krogh}, \bibinfo{person}{Andy Boucher}, {and} \bibinfo{person}{David Chatting}.} \bibinfo{year}{2022}\natexlab{b}.
\newblock \showarticletitle{Emergence as a feature of practice-based design research}. In \bibinfo{booktitle}{\emph{Proceedings of the 2022 ACM designing interactive systems conference}}. \bibinfo{pages}{517--526}.
\newblock


\bibitem[Giaccardi et~al\mbox{.}(2024)]%
        {giaccardi2024prototyping}
\bibfield{author}{\bibinfo{person}{Elisa Giaccardi}, \bibinfo{person}{Dave Murray-Rust}, \bibinfo{person}{Johan Redstr{\"o}m}, {and} \bibinfo{person}{Baptiste Caramiaux}.} \bibinfo{year}{2024}\natexlab{}.
\newblock \bibinfo{title}{Prototyping with uncertainties: Data, algorithms, and research through design}.
\newblock , \bibinfo{numpages}{21}~pages.
\newblock


\bibitem[Gresalfi et~al\mbox{.}(2009)]%
        {gresalfi2009virtual}
\bibfield{author}{\bibinfo{person}{Melissa Gresalfi}, \bibinfo{person}{Sasha Barab}, \bibinfo{person}{Sinem Siyahhan}, {and} \bibinfo{person}{Tyler Christensen}.} \bibinfo{year}{2009}\natexlab{}.
\newblock \showarticletitle{Virtual worlds, conceptual understanding, and me: Designing for consequential engagement}.
\newblock \bibinfo{journal}{\emph{On the Horizon}} \bibinfo{volume}{17}, \bibinfo{number}{1} (\bibinfo{year}{2009}), \bibinfo{pages}{21--34}.
\newblock


\bibitem[Gunderson(1973)]%
        {gunderson_individual_1973}
\bibfield{author}{\bibinfo{person}{E.~K.~Eric Gunderson}.} \bibinfo{year}{1973}\natexlab{}.
\newblock \showarticletitle{Individual Behavior in Confined or Isolated Groups}.
\newblock In \bibinfo{booktitle}{\emph{Man in Isolation \& Confinement}}. \bibinfo{publisher}{Routledge}.
\newblock
\showISBNx{978-0-203-78657-4}


\bibitem[Han et~al\mbox{.}(2024)]%
        {han_when_2024}
\bibfield{author}{\bibinfo{person}{Yuanning Han}, \bibinfo{person}{Ziyi Qiu}, \bibinfo{person}{Jiale Cheng}, {and} \bibinfo{person}{RAY LC}.} \bibinfo{year}{2024}\natexlab{}.
\newblock \showarticletitle{When {Teams} {Embrace} {AI}: {Human} {Collaboration} {Strategies} in {Generative} {Prompting} in a {Creative} {Design} {Task}}. In \bibinfo{booktitle}{\emph{Proceedings of the {CHI} {Conference} on {Human} {Factors} in {Computing} {Systems}}} \emph{(\bibinfo{series}{{CHI} '24})}. \bibinfo{publisher}{Association for Computing Machinery}, \bibinfo{address}{New York, NY, USA}, \bibinfo{pages}{1--14}.
\newblock
\showISBNx{9798400703300}
\urldef\tempurl%
\url{https://doi.org/10.1145/3613904.3642133}
\showDOI{\tempurl}


\bibitem[Harrison et~al\mbox{.}(2012)]%
        {harrison_antarctica_2012}
\bibfield{author}{\bibinfo{person}{Albert~A. Harrison}, \bibinfo{person}{Yvonne~A. Clearwater}, {and} \bibinfo{person}{Christopher~P. McKay}.} \bibinfo{year}{2012}\natexlab{}.
\newblock \bibinfo{booktitle}{\emph{From Antarctica to Outer Space: Life in Isolation and Confinement}}.
\newblock \bibinfo{publisher}{Springer Science \& Business Media}.
\newblock
\showISBNx{978-1-4612-3012-0}


\bibitem[He et~al\mbox{.}(2025)]%
        {he_i_2025}
\bibfield{author}{\bibinfo{person}{Zhiting He}, \bibinfo{person}{Jiayi Su}, \bibinfo{person}{Li Chen}, \bibinfo{person}{Tianqi Wang}, {and} \bibinfo{person}{RAY LC}.} \bibinfo{year}{2025}\natexlab{}.
\newblock \showarticletitle{"{I} {Recall} the {Past}": {Exploring} {How} {People} {Collaborate} with {Generative} {AI} to {Create} {Cultural} {Heritage} {Narratives}}.
\newblock \bibinfo{journal}{\emph{Proceedings of the ACM on Human-Computer Interaction}} \bibinfo{volume}{9}, \bibinfo{number}{CSCW 108} (\bibinfo{date}{April} \bibinfo{year}{2025}), \bibinfo{pages}{30}.
\newblock
\urldef\tempurl%
\url{https://doi.org/10.1145/3711006}
\showDOI{\tempurl}


\bibitem[Hendriks et~al\mbox{.}(2024)]%
        {hendriks2024undertable}
\bibfield{author}{\bibinfo{person}{Sjoerd Hendriks}, \bibinfo{person}{Mafalda Gamboa}, {and} \bibinfo{person}{Mohammad Obaid}.} \bibinfo{year}{2024}\natexlab{}.
\newblock \showarticletitle{The Undertable: A Design Remake of the Mediated Body}. In \bibinfo{booktitle}{\emph{Proceedings of the 2024 ACM Designing Interactive Systems Conference}}. \bibinfo{pages}{2591--2610}.
\newblock


\bibitem[Homan et~al\mbox{.}(2014)]%
        {Homan_2014}
\bibfield{author}{\bibinfo{person}{Christopher~M. Homan}, \bibinfo{person}{Naiji Lu}, \bibinfo{person}{Xin Tu}, \bibinfo{person}{Megan~C. Lytle}, {and} \bibinfo{person}{Vincent~M.B. Silenzio}.} \bibinfo{year}{2014}\natexlab{}.
\newblock \showarticletitle{Social structure and depression in TrevorSpace}. In \bibinfo{booktitle}{\emph{Proceedings of the 17th ACM conference on Computer supported cooperative work \& social computing}} \emph{(\bibinfo{series}{CSCW ’14})}. \bibinfo{publisher}{Association for Computing Machinery}, \bibinfo{address}{New York, NY, USA}, \bibinfo{pages}{615–625}.
\newblock
\showISBNx{978-1-4503-2540-0}
\urldef\tempurl%
\url{https://doi.org/10.1145/2531602.2531704}
\showDOI{\tempurl}


\bibitem[Johansson et~al\mbox{.}(2024)]%
        {Johansson_2024}
\bibfield{author}{\bibinfo{person}{Karin Johansson}, \bibinfo{person}{Raquel Robinson}, \bibinfo{person}{Jon Back}, \bibinfo{person}{Sarah~Lynne Bowman}, \bibinfo{person}{James Fey}, \bibinfo{person}{Elena M\'{a}rquez~Segura}, \bibinfo{person}{Annika Waern}, {and} \bibinfo{person}{Katherine Isbister}.} \bibinfo{year}{2024}\natexlab{}.
\newblock \showarticletitle{Why Larp? A Synthesis Article on Live Action Roleplay in Relation to HCI Research and Practice}.
\newblock \bibinfo{journal}{\emph{ACM Trans. Comput.-Hum. Interact.}} \bibinfo{volume}{31}, \bibinfo{number}{5}, Article \bibinfo{articleno}{64} (\bibinfo{date}{Nov.} \bibinfo{year}{2024}), \bibinfo{numpages}{35}~pages.
\newblock
\showISSN{1073-0516}
\urldef\tempurl%
\url{https://doi.org/10.1145/3689045}
\showDOI{\tempurl}


\bibitem[Karimi and Neustaedter(2012)]%
        {Karimi_Neustaedter_2012}
\bibfield{author}{\bibinfo{person}{Azmina Karimi} {and} \bibinfo{person}{Carman Neustaedter}.} \bibinfo{year}{2012}\natexlab{}.
\newblock \showarticletitle{From high connectivity to social isolation: communication practices of older adults in the digital age}. In \bibinfo{booktitle}{\emph{Proceedings of the ACM 2012 conference on Computer Supported Cooperative Work Companion}} \emph{(\bibinfo{series}{CSCW ’12})}. \bibinfo{publisher}{Association for Computing Machinery}, \bibinfo{address}{New York, NY, USA}, \bibinfo{pages}{127–130}.
\newblock
\showISBNx{978-1-4503-1051-2}
\urldef\tempurl%
\url{https://doi.org/10.1145/2141512.2141559}
\showDOI{\tempurl}


\bibitem[Kiger and Varpio(2020)]%
        {Kiger_Varpio_2020}
\bibfield{author}{\bibinfo{person}{Michelle~E. Kiger} {and} \bibinfo{person}{Lara Varpio}.} \bibinfo{year}{2020}\natexlab{}.
\newblock \showarticletitle{Thematic analysis of qualitative data: AMEE Guide No. 131}.
\newblock \bibinfo{journal}{\emph{Medical Teacher}} \bibinfo{volume}{42}, \bibinfo{number}{8} (\bibinfo{date}{Aug.} \bibinfo{year}{2020}), \bibinfo{pages}{846–854}.
\newblock
\showISSN{0142159X}
\urldef\tempurl%
\url{https://doi.org/10.1080/0142159X.2020.1755030}
\showDOI{\tempurl}


\bibitem[Kimhi(2011)]%
        {Kimhi_2011}
\bibfield{author}{\bibinfo{person}{Shaul Kimhi}.} \bibinfo{year}{2011}\natexlab{}.
\newblock \showarticletitle{Understanding Good Coping: A Submarine Crew Coping with Extreme Environmental Conditions}.
\newblock \bibinfo{journal}{\emph{Psychology}} \bibinfo{volume}{02}, \bibinfo{number}{09} (\bibinfo{year}{2011}), \bibinfo{pages}{961–968}.
\newblock
\showISSN{2152-7180, 2152-7199}
\urldef\tempurl%
\url{https://doi.org/10.4236/psych.2011.29145}
\showDOI{\tempurl}


\bibitem[{Korpilahti-Leino} et~al\mbox{.}(2022)]%
        {korpilahti-leino_resilience_2022}
\bibfield{author}{\bibinfo{person}{Tarja {Korpilahti-Leino}}, \bibinfo{person}{Terhi Luntamo}, \bibinfo{person}{Terja Ristkari}, \bibinfo{person}{Susanna {Hinkka-Yli-Salom{\"a}ki}}, \bibinfo{person}{Laura {Pulkki-R{\aa}back}}, \bibinfo{person}{Otto Waris}, \bibinfo{person}{Hanna-Maria Matinolli}, \bibinfo{person}{Atte Sinokki}, \bibinfo{person}{Yuko Mori}, \bibinfo{person}{Mami Fukaya}, \bibinfo{person}{Yuko Yamada}, {and} \bibinfo{person}{Andre Sourander}.} \bibinfo{year}{2022}\natexlab{}.
\newblock \showarticletitle{Resilience against Crises: COVID-19 and Lessons from Natural Disasters}.
\newblock \bibinfo{journal}{\emph{Journal of Medical Internet Research}} \bibinfo{volume}{24}, \bibinfo{number}{4} (\bibinfo{year}{2022}), \bibinfo{pages}{e26438}.
\newblock
\urldef\tempurl%
\url{https://doi.org/10.2196/26438}
\showDOI{\tempurl}


\bibitem[LC and Mizuno(2021)]%
        {lc_designing_2021}
\bibfield{author}{\bibinfo{person}{RAY LC} {and} \bibinfo{person}{Daijiro Mizuno}.} \bibinfo{year}{2021}\natexlab{}.
\newblock \showarticletitle{Designing for {Narrative} {Influence}: {Speculative} {Storytelling} for {Social} {Good} in {Times} of {Public} {Health} and {Climate} {Crises}}.
\newblock In \bibinfo{booktitle}{\emph{Extended {Abstracts} of the 2021 {CHI} {Conference} on {Human} {Factors} in {Computing} {Systems}}}. Number~29. \bibinfo{publisher}{Association for Computing Machinery}, \bibinfo{address}{New York, NY, USA}, \bibinfo{pages}{1--13}.
\newblock
\showISBNx{978-1-4503-8095-9}
\urldef\tempurl%
\url{https://doi.org/10.1145/3411763.3450373}
\showURL{%
\tempurl}


\bibitem[LC et~al\mbox{.}(2022)]%
        {lc_designing_2022}
\bibfield{author}{\bibinfo{person}{RAY LC}, \bibinfo{person}{Zijing Song}, \bibinfo{person}{Yating Sun}, {and} \bibinfo{person}{Cheng Yang}.} \bibinfo{year}{2022}\natexlab{}.
\newblock \showarticletitle{Designing narratives and data visuals in comic form for social influence in climate action}.
\newblock \bibinfo{journal}{\emph{Frontiers in Psychology}}  \bibinfo{volume}{13} (\bibinfo{year}{2022}).
\newblock
\showISSN{1664-1078}
\urldef\tempurl%
\url{https://www.frontiersin.org/articles/10.3389/fpsyg.2022.893181}
\showURL{%
\tempurl}


\bibitem[Linder et~al\mbox{.}(2022)]%
        {linder_characterizing_2022}
\bibfield{author}{\bibinfo{person}{Rhema Linder}, \bibinfo{person}{Chase Hunter}, \bibinfo{person}{Jacob McLemore}, \bibinfo{person}{Senjuti Dutta}, \bibinfo{person}{Fatema Akbar}, \bibinfo{person}{Ted Grover}, \bibinfo{person}{Thomas Breideband}, \bibinfo{person}{Judith~W. Borghouts}, \bibinfo{person}{Yuwen Lu}, \bibinfo{person}{Gloria Mark}, \bibinfo{person}{Austin~Z. Henley}, {and} \bibinfo{person}{Alex~C. Williams}.} \bibinfo{year}{2022}\natexlab{}.
\newblock \showarticletitle{Characterizing Work-Life for Information Work on Mars: A Design Fiction for the New Future of Work on Earth}.
\newblock \bibinfo{journal}{\emph{Proceedings of the ACM on Human-Computer Interaction}} \bibinfo{volume}{6}, \bibinfo{number}{GROUP} (\bibinfo{year}{2022}), \bibinfo{pages}{40:1--40:27}.
\newblock
\urldef\tempurl%
\url{https://doi.org/10.1145/3492859}
\showDOI{\tempurl}


\bibitem[Ling et~al\mbox{.}(2024)]%
        {ling_sketchar_2024}
\bibfield{author}{\bibinfo{person}{Long Ling}, \bibinfo{person}{Xinyi Chen}, \bibinfo{person}{Ruoyu Wen}, \bibinfo{person}{Toby Jia-Jun Li}, {and} \bibinfo{person}{RAY LC}.} \bibinfo{year}{2024}\natexlab{}.
\newblock \showarticletitle{Sketchar: {Supporting} {Character} {Design} and {Illustration} {Prototyping} {Using} {Generative} {AI}}.
\newblock \bibinfo{journal}{\emph{Proc. ACM Hum.-Comput. Interact.}} \bibinfo{volume}{8}, \bibinfo{number}{CHI PLAY} (\bibinfo{date}{Oct.} \bibinfo{year}{2024}), \bibinfo{pages}{337:1--337:28}.
\newblock
\urldef\tempurl%
\url{https://doi.org/10.1145/3677102}
\showDOI{\tempurl}


\bibitem[McPhail et~al\mbox{.}(2024)]%
        {McPhail_2024}
\bibfield{author}{\bibinfo{person}{Ruth McPhail}, \bibinfo{person}{Xi~Wen~(Carys) Chan}, \bibinfo{person}{Robyn May}, {and} \bibinfo{person}{Adrian Wilkinson}.} \bibinfo{year}{2024}\natexlab{}.
\newblock \showarticletitle{Post-COVID remote working and its impact on people, productivity, and the planet: an exploratory scoping review}.
\newblock \bibinfo{journal}{\emph{The International Journal of Human Resource Management}} \bibinfo{volume}{35}, \bibinfo{number}{1} (\bibinfo{date}{Jan.} \bibinfo{year}{2024}), \bibinfo{pages}{154–182}.
\newblock
\showISSN{0958-5192}
\urldef\tempurl%
\url{https://doi.org/10.1080/09585192.2023.2221385}
\showDOI{\tempurl}


\bibitem[Mulvale et~al\mbox{.}(2016)]%
        {mulvale_applying_2016}
\bibfield{author}{\bibinfo{person}{Alison Mulvale}, \bibinfo{person}{Ashleigh Miatello}, \bibinfo{person}{Christina Hackett}, {and} \bibinfo{person}{Gillian Mulvale}.} \bibinfo{year}{2016}\natexlab{}.
\newblock \showarticletitle{Applying Experience-Based Co-Design with Vulnerable Populations: Lessons from a Systematic Review of Methods to Involve Patients, Families and Service Providers in Child and Youth Mental Health Service Improvement}.
\newblock \bibinfo{journal}{\emph{Patient Experience Journal}} \bibinfo{volume}{3}, \bibinfo{number}{1} (\bibinfo{year}{2016}), \bibinfo{pages}{117--129}.
\newblock
\showISSN{2372-0247}
\urldef\tempurl%
\url{https://doi.org/10.35680/2372-0247.1104}
\showDOI{\tempurl}


\bibitem[Nardi and Harris(2006)]%
        {Nardi_Harris_2006}
\bibfield{author}{\bibinfo{person}{Bonnie Nardi} {and} \bibinfo{person}{Justin Harris}.} \bibinfo{year}{2006}\natexlab{}.
\newblock \showarticletitle{Strangers and friends: collaborative play in world of warcraft}. In \bibinfo{booktitle}{\emph{Proceedings of the 2006 20th anniversary conference on Computer supported cooperative work}} \emph{(\bibinfo{series}{CSCW ’06})}. \bibinfo{publisher}{Association for Computing Machinery}, \bibinfo{address}{New York, NY, USA}, \bibinfo{pages}{149–158}.
\newblock
\showISBNx{978-1-59593-249-5}
\urldef\tempurl%
\url{https://doi.org/10.1145/1180875.1180898}
\showDOI{\tempurl}


\bibitem[Nelson et~al\mbox{.}(2007)]%
        {nelson2007adaptation}
\bibfield{author}{\bibinfo{person}{Donald~R Nelson}, \bibinfo{person}{W~Neil Adger}, {and} \bibinfo{person}{Katrina Brown}.} \bibinfo{year}{2007}\natexlab{}.
\newblock \showarticletitle{Adaptation to environmental change: contributions of a resilience framework}.
\newblock \bibinfo{journal}{\emph{Annu. Rev. Environ. Resour.}} \bibinfo{volume}{32}, \bibinfo{number}{1} (\bibinfo{year}{2007}), \bibinfo{pages}{395--419}.
\newblock


\bibitem[Newman et~al\mbox{.}(2011)]%
        {Newman_Lauterbach_Munson_Resnick_Morris_2011}
\bibfield{author}{\bibinfo{person}{Mark~W. Newman}, \bibinfo{person}{Debra Lauterbach}, \bibinfo{person}{Sean~A. Munson}, \bibinfo{person}{Paul Resnick}, {and} \bibinfo{person}{Margaret~E. Morris}.} \bibinfo{year}{2011}\natexlab{}.
\newblock \showarticletitle{It’s not that i don’t have problems, i’m just not putting them on facebook: challenges and opportunities in using online social networks for health}. In \bibinfo{booktitle}{\emph{Proceedings of the ACM 2011 conference on Computer supported cooperative work}} \emph{(\bibinfo{series}{CSCW ’11})}. \bibinfo{publisher}{Association for Computing Machinery}, \bibinfo{address}{New York, NY, USA}, \bibinfo{pages}{341–350}.
\newblock
\showISBNx{978-1-4503-0556-3}
\urldef\tempurl%
\url{https://doi.org/10.1145/1958824.1958876}
\showDOI{\tempurl}


\bibitem[Norton(2007)]%
        {norton_depression_2007}
\bibfield{author}{\bibinfo{person}{Peter~J. Norton}.} \bibinfo{year}{2007}\natexlab{}.
\newblock \showarticletitle{Depression Anxiety and Stress Scales (DASS-21): Psychometric Analysis across Four Racial Groups}.
\newblock \bibinfo{journal}{\emph{Anxiety, Stress, \& Coping}} \bibinfo{volume}{20}, \bibinfo{number}{3} (\bibinfo{year}{2007}), \bibinfo{pages}{253--265}.
\newblock
\showISSN{1061-5806}
\urldef\tempurl%
\url{https://doi.org/10.1080/10615800701309279}
\showDOI{\tempurl}


\bibitem[Oluwafemi et~al\mbox{.}(2021)]%
        {oluwafemi_review_2021}
\bibfield{author}{\bibinfo{person}{Funmilola~A. Oluwafemi}, \bibinfo{person}{Rayan Abdelbaki}, \bibinfo{person}{James C.~Y. Lai}, \bibinfo{person}{Jose~G. {Mora-Almanza}}, {and} \bibinfo{person}{Esther~M. Afolayan}.} \bibinfo{year}{2021}\natexlab{}.
\newblock \showarticletitle{A Review of Astronaut Mental Health in Manned Missions: Potential Interventions for Cognitive and Mental Health Challenges}.
\newblock \bibinfo{journal}{\emph{Life Sciences in Space Research}}  \bibinfo{volume}{28} (\bibinfo{year}{2021}), \bibinfo{pages}{26--31}.
\newblock
\showISSN{2214-5524}
\urldef\tempurl%
\url{https://doi.org/10.1016/j.lssr.2020.12.002}
\showDOI{\tempurl}


\bibitem[Ortiz and Cruz-Neira(2023)]%
        {Ortiz_Cruz-Neira_2023}
\bibfield{author}{\bibinfo{person}{Jason~A. Ortiz} {and} \bibinfo{person}{Carolina Cruz-Neira}.} \bibinfo{year}{2023}\natexlab{}.
\newblock \showarticletitle{Workspace VR: A Social and Collaborative Telework Virtual Reality Application}. In \bibinfo{booktitle}{\emph{Companion Publication of the 2023 Conference on Computer Supported Cooperative Work and Social Computing}} \emph{(\bibinfo{series}{CSCW ’23 Companion})}. \bibinfo{publisher}{Association for Computing Machinery}, \bibinfo{address}{New York, NY, USA}, \bibinfo{pages}{381–383}.
\newblock
\showISBNx{9798400701290}
\urldef\tempurl%
\url{https://doi.org/10.1145/3584931.3607502}
\showDOI{\tempurl}


\bibitem[Otufowora et~al\mbox{.}(2021)]%
        {Otufowora_Liu_Young_Egan_Varma_Striley_Cottler_2021}
\bibfield{author}{\bibinfo{person}{Ayodeji Otufowora}, \bibinfo{person}{Yiyang Liu}, \bibinfo{person}{Henry Young}, \bibinfo{person}{Kathleen~L. Egan}, \bibinfo{person}{Deepthi~S. Varma}, \bibinfo{person}{Catherine~W. Striley}, {and} \bibinfo{person}{Linda~B. Cottler}.} \bibinfo{year}{2021}\natexlab{}.
\newblock \showarticletitle{Sex Differences in Willingness to Participate in Research Based on Study Risk Level Among a Community Sample of African Americans in North Central Florida}.
\newblock \bibinfo{journal}{\emph{Journal of immigrant and minority health}} \bibinfo{volume}{23}, \bibinfo{number}{1} (\bibinfo{date}{Feb.} \bibinfo{year}{2021}), \bibinfo{pages}{19–25}.
\newblock
\showISSN{1557-1912}
\urldef\tempurl%
\url{https://doi.org/10.1007/s10903-020-01015-4}
\showDOI{\tempurl}


\bibitem[Palinkas and Suedfeld(2021)]%
        {palinkas_psychosocial_2021}
\bibfield{author}{\bibinfo{person}{Lawrence~A. Palinkas} {and} \bibinfo{person}{Peter Suedfeld}.} \bibinfo{year}{2021}\natexlab{}.
\newblock \showarticletitle{Psychosocial Issues in Isolated and Confined Extreme Environments}.
\newblock \bibinfo{journal}{\emph{Neuroscience \& Biobehavioral Reviews}}  \bibinfo{volume}{126} (\bibinfo{year}{2021}), \bibinfo{pages}{413--429}.
\newblock
\showISSN{0149-7634}
\urldef\tempurl%
\url{https://doi.org/10.1016/j.neubiorev.2021.03.032}
\showDOI{\tempurl}


\bibitem[Pan et~al\mbox{.}(2013)]%
        {Pan_2013}
\bibfield{author}{\bibinfo{person}{Lu Pan}, \bibinfo{person}{Feng Tian}, \bibinfo{person}{Fei Lu}, \bibinfo{person}{Xiaolong~(Luke) Zhang}, \bibinfo{person}{Ying Liu}, \bibinfo{person}{Wenxin Feng}, \bibinfo{person}{Guozhong Dai}, {and} \bibinfo{person}{Hongan Wang}.} \bibinfo{year}{2013}\natexlab{}.
\newblock \showarticletitle{An exploration on long-distance communications between left-behind children and their parents in China}. In \bibinfo{booktitle}{\emph{Proceedings of the 2013 conference on Computer supported cooperative work}} \emph{(\bibinfo{series}{CSCW ’13})}. \bibinfo{publisher}{Association for Computing Machinery}, \bibinfo{address}{New York, NY, USA}, \bibinfo{pages}{1147–1156}.
\newblock
\showISBNx{978-1-4503-1331-5}
\urldef\tempurl%
\url{https://doi.org/10.1145/2441776.2441906}
\showDOI{\tempurl}


\bibitem[Park et~al\mbox{.}(2015)]%
        {Park_2015}
\bibfield{author}{\bibinfo{person}{Sungkyu Park}, \bibinfo{person}{Inyeop Kim}, \bibinfo{person}{Sang~Won Lee}, \bibinfo{person}{Jaehyun Yoo}, \bibinfo{person}{Bumseok Jeong}, {and} \bibinfo{person}{Meeyoung Cha}.} \bibinfo{year}{2015}\natexlab{}.
\newblock \showarticletitle{Manifestation of Depression and Loneliness on Social Networks: A Case Study of Young Adults on Facebook}. In \bibinfo{booktitle}{\emph{Proceedings of the 18th ACM Conference on Computer Supported Cooperative Work \& Social Computing}} \emph{(\bibinfo{series}{CSCW ’15})}. \bibinfo{publisher}{Association for Computing Machinery}, \bibinfo{address}{New York, NY, USA}, \bibinfo{pages}{557–570}.
\newblock
\showISBNx{978-1-4503-2922-4}
\urldef\tempurl%
\url{https://doi.org/10.1145/2675133.2675139}
\showDOI{\tempurl}


\bibitem[Park et~al\mbox{.}(2012)]%
        {Park_Yoo_Choe_Park_Song_2012}
\bibfield{author}{\bibinfo{person}{Taiwoo Park}, \bibinfo{person}{Chungkuk Yoo}, \bibinfo{person}{Sungwon~Peter Choe}, \bibinfo{person}{Byunglim Park}, {and} \bibinfo{person}{Junehwa Song}.} \bibinfo{year}{2012}\natexlab{}.
\newblock \showarticletitle{Transforming solitary exercises into social exergames}. In \bibinfo{booktitle}{\emph{Proceedings of the ACM 2012 conference on Computer Supported Cooperative Work}} \emph{(\bibinfo{series}{CSCW ’12})}. \bibinfo{publisher}{Association for Computing Machinery}, \bibinfo{address}{New York, NY, USA}, \bibinfo{pages}{863–866}.
\newblock
\showISBNx{978-1-4503-1086-4}
\urldef\tempurl%
\url{https://doi.org/10.1145/2145204.2145332}
\showDOI{\tempurl}


\bibitem[Peldszus et~al\mbox{.}(2014)]%
        {peldszus_perfect_2014}
\bibfield{author}{\bibinfo{person}{Regina Peldszus}, \bibinfo{person}{Hilary Dalke}, \bibinfo{person}{Stephen Pretlove}, {and} \bibinfo{person}{Chris Welch}.} \bibinfo{year}{2014}\natexlab{}.
\newblock \showarticletitle{The Perfect Boring Situation\textemdash Addressing the Experience of Monotony during Crewed Deep Space Missions through Habitability Design}.
\newblock \bibinfo{journal}{\emph{Acta Astronautica}} \bibinfo{volume}{94}, \bibinfo{number}{1} (\bibinfo{year}{2014}), \bibinfo{pages}{262--276}.
\newblock
\showISSN{0094-5765}
\urldef\tempurl%
\url{https://doi.org/10.1016/j.actaastro.2013.04.024}
\showDOI{\tempurl}


\bibitem[Pidel and Ackermann(2020)]%
        {pidel_collaboration_2020}
\bibfield{author}{\bibinfo{person}{Catlin Pidel} {and} \bibinfo{person}{Philipp Ackermann}.} \bibinfo{year}{2020}\natexlab{}.
\newblock \showarticletitle{Collaboration in Virtual and Augmented Reality: A Systematic Overview}. In \bibinfo{booktitle}{\emph{Augmented Reality, Virtual Reality, and Computer Graphics}}, \bibfield{editor}{\bibinfo{person}{Lucio~Tommaso De~Paolis} {and} \bibinfo{person}{Patrick Bourdot}} (Eds.). \bibinfo{publisher}{Springer International Publishing}, \bibinfo{address}{Cham}, \bibinfo{pages}{141--156}.
\newblock
\showISBNx{978-3-030-58465-8}
\urldef\tempurl%
\url{https://doi.org/10.1007/978-3-030-58465-8-10}
\showDOI{\tempurl}


\bibitem[Polizzi et~al\mbox{.}(2020)]%
        {polizzi_stress_2020}
\bibfield{author}{\bibinfo{person}{Craig Polizzi}, \bibinfo{person}{Steven~Jay Lynn}, {and} \bibinfo{person}{Andrew Perry}.} \bibinfo{year}{2020}\natexlab{}.
\newblock \showarticletitle{Stress and Coping in the Time of Covid-19: Pathways to Resilience and Recovery}.
\newblock \bibinfo{journal}{\emph{Clinical Neuropsychiatry}} \bibinfo{volume}{17}, \bibinfo{number}{2} (\bibinfo{year}{2020}), \bibinfo{pages}{59--62}.
\newblock
\showISSN{1724-4935}
\urldef\tempurl%
\url{https://doi.org/10.36131/CN20200204}
\showDOI{\tempurl}


\bibitem[Reed et~al\mbox{.}(2024)]%
        {Reed_2024}
\bibfield{author}{\bibinfo{person}{Courtney~N. Reed}, \bibinfo{person}{Adan~L. Benito}, \bibinfo{person}{Franco Caspe}, {and} \bibinfo{person}{Andrew~P. McPherson}.} \bibinfo{year}{2024}\natexlab{}.
\newblock \showarticletitle{Shifting Ambiguity, Collapsing Indeterminacy: Designing with Data as Baradian Apparatus}.
\newblock \bibinfo{journal}{\emph{ACM Trans. Comput.-Hum. Interact.}} \bibinfo{volume}{31}, \bibinfo{number}{6}, Article \bibinfo{articleno}{73} (\bibinfo{date}{Dec.} \bibinfo{year}{2024}), \bibinfo{numpages}{41}~pages.
\newblock
\showISSN{1073-0516}
\urldef\tempurl%
\url{https://doi.org/10.1145/3689043}
\showDOI{\tempurl}


\bibitem[Saczuk et~al\mbox{.}(2022)]%
        {saczuk_temporomandibular_2022}
\bibfield{author}{\bibinfo{person}{Klara Saczuk}, \bibinfo{person}{Barbara Lapinska}, \bibinfo{person}{Adam Wawrzynkiewicz}, \bibinfo{person}{Alicja Witkowska}, \bibinfo{person}{Heber~Isac {Arbildo-Vega}}, \bibinfo{person}{Monika Domarecka}, {and} \bibinfo{person}{Monika {Lukomska-Szymanska}}.} \bibinfo{year}{2022}\natexlab{}.
\newblock \showarticletitle{Temporomandibular Disorders, Bruxism, Perceived Stress, and Coping Strategies among Medical University Students in Times of Social Isolation during Outbreak of COVID-19 Pandemic}.
\newblock \bibinfo{journal}{\emph{Healthcare}} \bibinfo{volume}{10}, \bibinfo{number}{4} (\bibinfo{year}{2022}), \bibinfo{pages}{740}.
\newblock
\showISSN{2227-9032}
\urldef\tempurl%
\url{https://doi.org/10.3390/healthcare10040740}
\showDOI{\tempurl}


\bibitem[Saffo et~al\mbox{.}(2021)]%
        {saffo_remote_2021}
\bibfield{author}{\bibinfo{person}{David Saffo}, \bibinfo{person}{Sara Di~Bartolomeo}, \bibinfo{person}{Caglar Yildirim}, {and} \bibinfo{person}{Cody Dunne}.} \bibinfo{year}{2021}\natexlab{}.
\newblock \showarticletitle{Remote and Collaborative Virtual Reality Experiments via Social VR Platforms}. In \bibinfo{booktitle}{\emph{Proceedings of the 2021 CHI Conference on Human Factors in Computing Systems}} \emph{(\bibinfo{series}{CHI '21})}. \bibinfo{publisher}{Association for Computing Machinery}, \bibinfo{address}{New York, NY, USA}, \bibinfo{pages}{1--15}.
\newblock
\showISBNx{978-1-4503-8096-6}
\urldef\tempurl%
\url{https://doi.org/10.1145/3411764.3445426}
\showDOI{\tempurl}


\bibitem[Sakurai and Chughtai(2020)]%
        {sakurai_resilience_2020}
\bibfield{author}{\bibinfo{person}{Mihoko Sakurai} {and} \bibinfo{person}{Hameed Chughtai}.} \bibinfo{year}{2020}\natexlab{}.
\newblock \showarticletitle{Resilience against Crises: COVID-19 and Lessons from Natural Disasters}.
\newblock \bibinfo{journal}{\emph{European Journal of Information Systems}} \bibinfo{volume}{29}, \bibinfo{number}{5} (\bibinfo{year}{2020}), \bibinfo{pages}{585--594}.
\newblock
\showISSN{0960-085X}
\urldef\tempurl%
\url{https://doi.org/10.1080/0960085X.2020.1814171}
\showDOI{\tempurl}


\bibitem[Sanders and Stappers(2014)]%
        {sanders_probes_2014}
\bibfield{author}{\bibinfo{person}{Elizabeth B.-N. Sanders} {and} \bibinfo{person}{Pieter~Jan Stappers}.} \bibinfo{year}{2014}\natexlab{}.
\newblock \showarticletitle{Probes, Toolkits and Prototypes: Three Approaches to Making in Codesigning}.
\newblock \bibinfo{journal}{\emph{CoDesign}} \bibinfo{volume}{10}, \bibinfo{number}{1} (\bibinfo{year}{2014}), \bibinfo{pages}{5--14}.
\newblock
\showISSN{1571-0882}
\urldef\tempurl%
\url{https://doi.org/10.1080/15710882.2014.888183}
\showDOI{\tempurl}


\bibitem[Shankar(2023)]%
        {Shankar_2023}
\bibfield{author}{\bibinfo{person}{Ravi Shankar}.} \bibinfo{year}{2023}\natexlab{}.
\newblock \showarticletitle{Loneliness, Social Isolation, and its Effects on Physical and Mental Health}.
\newblock \bibinfo{journal}{\emph{Missouri Medicine}} \bibinfo{volume}{120}, \bibinfo{number}{2} (\bibinfo{year}{2023}), \bibinfo{pages}{106–108}.
\newblock
\showISSN{0026-6620}


\bibitem[Sharma et~al\mbox{.}(2021)]%
        {sharma_mild_2021}
\bibfield{author}{\bibinfo{person}{Sumita Sharma}, \bibinfo{person}{Netta Iivari}, \bibinfo{person}{Marianne Kinnula}, \bibinfo{person}{Grace Eden}, \bibinfo{person}{Alipta Ballav}, \bibinfo{person}{Rocio Fatas}, \bibinfo{person}{Ritwik Kar}, \bibinfo{person}{Deepak Ranjan~Padhi}, \bibinfo{person}{Vahid Sadeghie}, \bibinfo{person}{Pratiti Sarkar}, \bibinfo{person}{Riya Sinha}, \bibinfo{person}{Rucha Tulaskar}, {and} \bibinfo{person}{Nikita Valluri}.} \bibinfo{year}{2021}\natexlab{}.
\newblock \showarticletitle{From Mild to Wild: Reimagining Friendships and Romance in the Time of Pandemic Using Design Fiction}. In \bibinfo{booktitle}{\emph{Designing Interactive Systems Conference 2021}} \emph{(\bibinfo{series}{DIS '21})}. \bibinfo{publisher}{Association for Computing Machinery}, \bibinfo{address}{New York, NY, USA}, \bibinfo{pages}{64--77}.
\newblock
\showISBNx{978-1-4503-8476-6}
\urldef\tempurl%
\url{https://doi.org/10.1145/3461778.3462110}
\showDOI{\tempurl}


\bibitem[Shaw et~al\mbox{.}(1997)]%
        {Shaw_1997}
\bibfield{author}{\bibinfo{person}{W.~S. Shaw}, \bibinfo{person}{T.~L. Patterson}, \bibinfo{person}{S.~J. Semple}, \bibinfo{person}{I. Grant}, \bibinfo{person}{E.~S. Yu}, \bibinfo{person}{M. Zhang}, \bibinfo{person}{Y.~Y. He}, {and} \bibinfo{person}{W.~Y. Wu}.} \bibinfo{year}{1997}\natexlab{}.
\newblock \showarticletitle{A cross-cultural validation of coping strategies and their associations with caregiving distress}.
\newblock \bibinfo{journal}{\emph{The Gerontologist}} \bibinfo{volume}{37}, \bibinfo{number}{4} (\bibinfo{date}{Aug.} \bibinfo{year}{1997}), \bibinfo{pages}{490–504}.
\newblock
\showISSN{0016-9013}
\urldef\tempurl%
\url{https://doi.org/10.1093/geront/37.4.490}
\showDOI{\tempurl}


\bibitem[Shen et~al\mbox{.}(2023)]%
        {Shen_2023}
\bibfield{author}{\bibinfo{person}{Ximing Shen}, \bibinfo{person}{Yun~Suen Pai}, \bibinfo{person}{Dai Kiuchi}, \bibinfo{person}{Kehan Bao}, \bibinfo{person}{Tomomi Aoki}, \bibinfo{person}{Hikari Meguro}, \bibinfo{person}{Kanoko Oishi}, \bibinfo{person}{Ziyue Wang}, \bibinfo{person}{Sohei Wakisaka}, {and} \bibinfo{person}{Kouta Minamizawa}.} \bibinfo{year}{2023}\natexlab{}.
\newblock \showarticletitle{Dementia Eyes: Co-Design and Evaluation of a Dementia Education Augmented Reality Experience for Medical Workers}. In \bibinfo{booktitle}{\emph{Proceedings of the 2023 CHI Conference on Human Factors in Computing Systems}} \emph{(\bibinfo{series}{CHI ’23})}. \bibinfo{publisher}{Association for Computing Machinery}, \bibinfo{address}{New York, NY, USA}, \bibinfo{pages}{1–18}.
\newblock
\showISBNx{978-1-4503-9421-5}
\urldef\tempurl%
\url{https://doi.org/10.1145/3544548.3581009}
\showDOI{\tempurl}


\bibitem[Song et~al\mbox{.}(2022a)]%
        {song_drizzle_2022}
\bibfield{author}{\bibinfo{person}{Zijing Song}, \bibinfo{person}{Yating Sun}, {and} \bibinfo{person}{Ray Lc}.} \bibinfo{year}{2022}\natexlab{a}.
\newblock \showarticletitle{{DRIZZLE}: {A} {Comic} for {Covert} {Climate} {Action} {Influence}}. In \bibinfo{booktitle}{\emph{[ ] {With} {Design}: {Reinventing} {Design} {Modes}}}, \bibfield{editor}{\bibinfo{person}{Gerhard Bruyns} {and} \bibinfo{person}{Huaxin Wei}} (Eds.). \bibinfo{publisher}{Springer Nature}, \bibinfo{address}{Singapore}, \bibinfo{pages}{1613--1623}.
\newblock
\showISBNx{978-981-19447-2-7}
\urldef\tempurl%
\url{https://doi.org/10.1007/978-981-19-4472-7_105}
\showDOI{\tempurl}


\bibitem[Song et~al\mbox{.}(2022b)]%
        {song_narrating_2022}
\bibfield{author}{\bibinfo{person}{Zijing Song}, \bibinfo{person}{Yating Sun}, {and} \bibinfo{person}{RAY LC}.} \bibinfo{year}{2022}\natexlab{b}.
\newblock \showarticletitle{Narrating {Climate} {Change}: {Speculative} data stories in comic form for affecting climate action}. In \bibinfo{booktitle}{\emph{Data {Art} for {Climate} {Action} ({DACA}'22)}}. \bibinfo{address}{Hong Kong}.
\newblock
\urldef\tempurl%
\url{https://www.scm.cityu.edu.hk/events/daca}
\showURL{%
\tempurl}


\bibitem[Song et~al\mbox{.}(2021)]%
        {song_climate_2021}
\bibfield{author}{\bibinfo{person}{Zijing Song}, \bibinfo{person}{Yating Sun}, \bibinfo{person}{Vincent Ruijters}, {and} \bibinfo{person}{RAY LC}.} \bibinfo{year}{2021}\natexlab{}.
\newblock \showarticletitle{Climate {Influence}: {Implicit} {Game}-{Based} {Interactive} {Storytelling} for {Climate} {Action} {Purpose}}. In \bibinfo{booktitle}{\emph{Interactive {Storytelling}}}, \bibfield{editor}{\bibinfo{person}{Alex Mitchell} {and} \bibinfo{person}{Mirjam Vosmeer}} (Eds.). \bibinfo{publisher}{Springer International Publishing}, \bibinfo{address}{Cham}, \bibinfo{pages}{425--429}.
\newblock
\showISBNx{978-3-030-92300-6}
\urldef\tempurl%
\url{https://doi.org/10.1007/978-3-030-92300-6_42}
\showDOI{\tempurl}


\bibitem[Stenros and Montola(2010)]%
        {stenros_nordic_2010}
\bibfield{editor}{\bibinfo{person}{Jaakko Stenros} {and} \bibinfo{person}{Markus Montola}} (Eds.). \bibinfo{year}{2010}\natexlab{}.
\newblock \bibinfo{booktitle}{\emph{Nordic Larp} (\bibinfo{edition}{1st print} ed.)}.
\newblock \bibinfo{publisher}{Fea Livia}, \bibinfo{address}{Stockholm}.
\newblock
\showISBNx{978-91-633-7856-0}
\showLCCN{GV1469.6 .N67 2010}


\bibitem[Troiano et~al\mbox{.}(2021)]%
        {troiano_are_2021}
\bibfield{author}{\bibinfo{person}{Giovanni~M Troiano}, \bibinfo{person}{Matthew Wood}, \bibinfo{person}{Mustafa~Feyyaz Sonbudak}, \bibinfo{person}{Riddhi~Chandan Padte}, {and} \bibinfo{person}{Casper Harteveld}.} \bibinfo{year}{2021}\natexlab{}.
\newblock \showarticletitle{``Are We Now Post-COVID?'': Exploring Post-COVID Futures Through a Gamified Story Completion Method}. In \bibinfo{booktitle}{\emph{Designing Interactive Systems Conference 2021}} \emph{(\bibinfo{series}{DIS '21})}. \bibinfo{publisher}{Association for Computing Machinery}, \bibinfo{address}{New York, NY, USA}, \bibinfo{pages}{48--63}.
\newblock
\showISBNx{978-1-4503-8476-6}
\urldef\tempurl%
\url{https://doi.org/10.1145/3461778.3462069}
\showDOI{\tempurl}


\bibitem[Uchino et~al\mbox{.}(1996)]%
        {Uchino_Cacioppo_1996}
\bibfield{author}{\bibinfo{person}{B.~N. Uchino}, \bibinfo{person}{J.~T. Cacioppo}, {and} \bibinfo{person}{J.~K. Kiecolt-Glaser}.} \bibinfo{year}{1996}\natexlab{}.
\newblock \showarticletitle{The relationship between social support and physiological processes: a review with emphasis on underlying mechanisms and implications for health}.
\newblock \bibinfo{journal}{\emph{Psychological Bulletin}} \bibinfo{volume}{119}, \bibinfo{number}{3} (\bibinfo{date}{May} \bibinfo{year}{1996}), \bibinfo{pages}{488–531}.
\newblock
\showISSN{0033-2909}
\urldef\tempurl%
\url{https://doi.org/10.1037/0033-2909.119.3.488}
\showDOI{\tempurl}


\bibitem[Udapola(2022)]%
        {Udapola_2022}
\bibfield{author}{\bibinfo{person}{Udapola Balage Hansi~Shashiprabha Udapola}.} \bibinfo{year}{2022}\natexlab{}.
\newblock \showarticletitle{Social VR for Socially Isolated Adolescents with Significant Illnesses}. In \bibinfo{booktitle}{\emph{Companion Proceedings of the 2022 Conference on Interactive Surfaces and Spaces}} (Wellington, New Zealand) \emph{(\bibinfo{series}{ISS Companion '22})}. \bibinfo{publisher}{Association for Computing Machinery}, \bibinfo{address}{New York, NY, USA}, \bibinfo{pages}{50–53}.
\newblock
\showISBNx{9781450393560}
\urldef\tempurl%
\url{https://doi.org/10.1145/3532104.3571466}
\showDOI{\tempurl}


\bibitem[Xu et~al\mbox{.}(2021)]%
        {xu2021depression}
\bibfield{author}{\bibinfo{person}{Yanhua Xu}, \bibinfo{person}{Jinlian Shao}, \bibinfo{person}{Wei Zeng}, \bibinfo{person}{Xingrou Wu}, \bibinfo{person}{Dongtao Huang}, \bibinfo{person}{Yuqing Zeng}, {and} \bibinfo{person}{Jiamin Wu}.} \bibinfo{year}{2021}\natexlab{}.
\newblock \showarticletitle{Depression and creativity during COVID-19: psychological resilience as a mediator and deliberate rumination as a moderator}.
\newblock \bibinfo{journal}{\emph{Frontiers in Psychology}}  \bibinfo{volume}{12} (\bibinfo{year}{2021}), \bibinfo{pages}{665961}.
\newblock


\bibitem[Yang et~al\mbox{.}(2022)]%
        {yang_ai_2022}
\bibfield{author}{\bibinfo{person}{Daijin Yang}, \bibinfo{person}{Yanpeng Zhou}, \bibinfo{person}{Zhiyuan Zhang}, \bibinfo{person}{Toby Jia-Jun Li}, {and} \bibinfo{person}{RAY LC}.} \bibinfo{year}{2022}\natexlab{}.
\newblock \showarticletitle{{AI} as an {Active} {Writer}: {Interaction} strategies with generated text in human-{AI} collaborative fiction writing}. In \bibinfo{booktitle}{\emph{Joint {Proceedings} of the {IUI} 2022 {Workshops}: {APEx}-{UI}, {HAI}-{GEN}, {HEALTHI}, {HUMANIZE}, {TExSS}, {SOCIALIZE}}}. \bibinfo{publisher}{CEUR-WS Team}, \bibinfo{pages}{56--65}.
\newblock
\urldef\tempurl%
\url{https://scholars.cityu.edu.hk/en/publications/publication(d901f5a2-0600-422f-b588-db5a59871961).html}
\showURL{%
\tempurl}


\bibitem[Yevdokimova and Okhrimenko(2021)]%
        {yevdokimova_coping_2021}
\bibfield{author}{\bibinfo{person}{Olena Yevdokimova} {and} \bibinfo{person}{Ivan Okhrimenko}.} \bibinfo{year}{2021}\natexlab{}.
\newblock \showarticletitle{Coping Strategies for Overcoming Stress in Atypical Situations}.
\newblock \bibinfo{journal}{\emph{BRAIN. Broad Research in Artificial Intelligence and Neuroscience}} \bibinfo{volume}{11}, \bibinfo{number}{2Sup1} (\bibinfo{year}{2021}), \bibinfo{pages}{56--63}.
\newblock
\showISSN{2067-3957}
\urldef\tempurl%
\url{https://doi.org/10.18662/brain/11.2Sup1/94}
\showDOI{\tempurl}


\bibitem[Zimmerman and Forlizzi(2014)]%
        {zimmerman2014research}
\bibfield{author}{\bibinfo{person}{John Zimmerman} {and} \bibinfo{person}{Jodi Forlizzi}.} \bibinfo{year}{2014}\natexlab{}.
\newblock \showarticletitle{Research through design in HCI}.
\newblock In \bibinfo{booktitle}{\emph{Ways of Knowing in HCI}}. \bibinfo{publisher}{Springer}, \bibinfo{pages}{167--189}.
\newblock


\bibitem[Zimmerman et~al\mbox{.}(2007)]%
        {zimmerman2007research}
\bibfield{author}{\bibinfo{person}{John Zimmerman}, \bibinfo{person}{Jodi Forlizzi}, {and} \bibinfo{person}{Shelley Evenson}.} \bibinfo{year}{2007}\natexlab{}.
\newblock \showarticletitle{Research through design as a method for interaction design research in HCI}. In \bibinfo{booktitle}{\emph{Proceedings of the SIGCHI conference on Human factors in computing systems}}. \bibinfo{pages}{493--502}.
\newblock


\bibitem[Zimmerman et~al\mbox{.}(2010)]%
        {zimmerman2010analysis}
\bibfield{author}{\bibinfo{person}{John Zimmerman}, \bibinfo{person}{Erik Stolterman}, {and} \bibinfo{person}{Jodi Forlizzi}.} \bibinfo{year}{2010}\natexlab{}.
\newblock \showarticletitle{An analysis and critique of Research through Design: towards a formalization of a research approach}. In \bibinfo{booktitle}{\emph{proceedings of the 8th ACM conference on designing interactive systems}}. \bibinfo{pages}{310--319}.
\newblock


\end{thebibliography}

\appendix\label{sec:Appendix}
\clearpage
\section{Items for social isolation during space journey}

\vspace{2mm}
\begin{center}
\begin{minipage}{\textwidth}
\begin{center}
\refstepcounter{table}
Table \thetable: Items for social isolation during space journey
\label{tab1:3}

\vspace{2mm}
\begin{tabular}{c p{0.3\textwidth} p{0.65\textwidth}}
\toprule
\multicolumn{3}{l}{\textbf{\begin{tabular}[c]{@{}l@{}}Basic Stock and Equipment\\ Everyone is given all the listed items\end{tabular}}} \\ 
\midrule
1 & Tablet & It is used for work, daily communication, and ordering basic space services like food delivery and cleaning. It can also be used to save videos, photos , and music. \\ [5ex]
2 & Intercom in Private Cabin & It can be used for direct contact with other explorers in other cabins or contact with spaceship assistants for help. It can also be used to give emergent calls to the control center.\\ [5ex]
3 & Space-specific Toiletries & They are the daily toiletries that can be used in space for self-cleaning. \\ [5ex]
4 & Water & There is enough water for each explorer during space travel. \\ [5ex]
5 & A Package of Space-specific Instant Food & Some supplements for obtaining energy in extra time outside the three main meals; Given to each of the explorers once a week. \\  [5ex]
\hline
\multicolumn{3}{ l }{\textbf{\begin{tabular}[c]{@{}l@{}}Optional Items\\ Each explorer can choose one from the list\end{tabular}}} \\ 
\hline
6 & Space-specific Planting Kits & "It was given to me by a plant lover before onboarding." It is a set of tools that can help with planting. You can define the number and types of plants yourself and how the tools are used, but once you decide, they are all set and cannot be changed.\\ [9ex]
7 & Poster or Large Picture & "Hummm...the poster(s)/picture(s) might be something of great significance." You can imagine what the poster or picture is about and what is the size of it. You can also set the numbers of the poster and/or the picture. Once you decide, they are all set and cannot be changed. \\ [9ex]
8 & Speaker & "Music brings me power and new inspiration." It is both a music player and a speaker. You can define how to use it (USB drives? Bluetooth? Disks? Others?) \\ [5ex]
9 & Mysterious Storage Disk & "The dark disk seems to have a large capacity. I hope it can help make this space journey more interesting." The contents of the disk can be read and displayed by the equipment in the spaceship. You can imagine and define the contents stored on the disk. Once you decide, they are all set and cannot be changed.\\ [9ex]
10 & Digital Diary (Book) & "Such a mystical diary... Seems never to be finished. Even though there is no paper or pen, I still want to record the life here." This is a digital diary that affords voice and texting as inputs. \\ [5ex]
11 & Small Projector & "Do you want to watch a movie?" It needs to be used with disks or drives and data cables.\\ [5ex]
12 & Novels & "Life is a hybrid of the real and the virtual, and one cannot lose imagination." You can decide the contents of the novels and the number of novels you would like. Once you decide, they are all set and cannot be changed.\\ [10ex]
\hline
\multicolumn{3}{ l }{\textbf{\begin{tabular}[c]{p{1\textwidth}}Self-selected Objects\\ Every explorer can choose one object that he/she feels necessary during space isolation\end{tabular}}} \\
\hline
\end{tabular}
\end{center}
\end{minipage}
\end{center}

\clearpage
\section{The hints of generating stories}

\vspace{2mm}
\begin{center}
\begin{minipage}{\textwidth}
\begin{center}
\refstepcounter{table}
Table \thetable: The hints of generating stories
\label{tab1:5}

\vspace{2mm}
\begin{tabular}{l p{0.85\textwidth}}
\hline
\textbf{Stage} & \textbf{Description} \\
\hline
Start & Select or suggest a participant to initiate the story, who will be responsible for envisioning the future collective space life and providing an introduction or a captivating opening to the narrative. \\
Development & Collaboratively imagine and develop activities that can be performed together to alleviate negative emotions. You can draw inspiration from the items mentioned when designing your private room. The space for these activities is not limited to the spaceship cabin; it can also include your private spaces or an entertainment area collectively planned within the cabin. \\
Climax & Collaboratively plan life in the cabin, including a weekly schedule outlining daily activities, essential tasks, and a range of entertainment and group activities. \\
Ending & Provide a conclusion to the story that encompasses elements such as the transformation of relationships among the participants, their connection with the remaining people on Earth, and the anticipated events one or several years after the journey. \\ 
\hline
\end{tabular}
\end{center}
\end{minipage}
\end{center}

\clearpage
\section{Demographic information of participants.}

\vspace{2mm}
\begin{center}
\begin{minipage}{\textwidth}
\begin{center}
\refstepcounter{table}
Table \thetable: Demographic information of participants.
\label{tab1:2}

\vspace{2mm}
\begin{tabular}{l l l l l}
\hline
P & Age & Gender* & Domain & Isolation Type (in 7 days)\\
\hline
P1 & 20 & M & CS & Hotel Quarantine+At-home Isolation\\
P2 & 24 & F & Design & Hotel Quarantine\\
P3 & 24 & F & English + Art & Hotel Quarantine\\
P4 & 29 & F & CS & At-home Isolation\\
P5 & 25 & M & Management & Hotel Quarantine+At-home Isolation\\
P6 & 27 & F & Management & At-home Isolation\\
P7 & 22 & M & Engineering & On-campus Isolation+At-home Isolation\\
P8 & 24 & F & Journalism & Hotel Quarantine+At-home Isolation\\
P9 & 22 & F & Engineering & On-campus Isolation+At-home Isolation\\
P10 & 21 & F & Journalism & On-campus Isolation\\
P11 & 28 & F & Social Studies & At-home Isolation\\
P12 & 26 & F & Design & Hotel Quarantine+At-home Isolation\\
P13 & 24 & M & Translation Studies & Hotel Quarantine\\
P14 & 24 & F & Design & Hotel Quarantine\\
P15 & 21 & F & Design & On-campus Isolation\\
P16 & 25 & F & Management & Hotel Quarantine+At-home Isolation\\
P17 & 15 & F & Management & Hotel Quarantine+At-home Isolation\\
P18 & 25 & F & Management & Hotel Quarantine+At-home Isolation\\
\hline
\end{tabular}
\end{center}
\end{minipage}
\end{center}

\clearpage  
\section{Participants' designs of their private space for space journey isolation} 

\vspace{2mm}
\begin{center}
\begin{minipage}{\textwidth}
\begin{center}
\refstepcounter{table}
Table \thetable: Participants' designs of their private space for space journey isolation (Part 1)
\label{tab:designs1}

\vspace{2mm}
\begin{tabular}{|c|m{5cm}|m{5cm}|}  
\hline
Sketch & Content & Describe \\ \hline
\begin{minipage}[t]{0.3\columnwidth}
    \centering
    \raisebox{-.5\height}{\includegraphics[width=\linewidth]{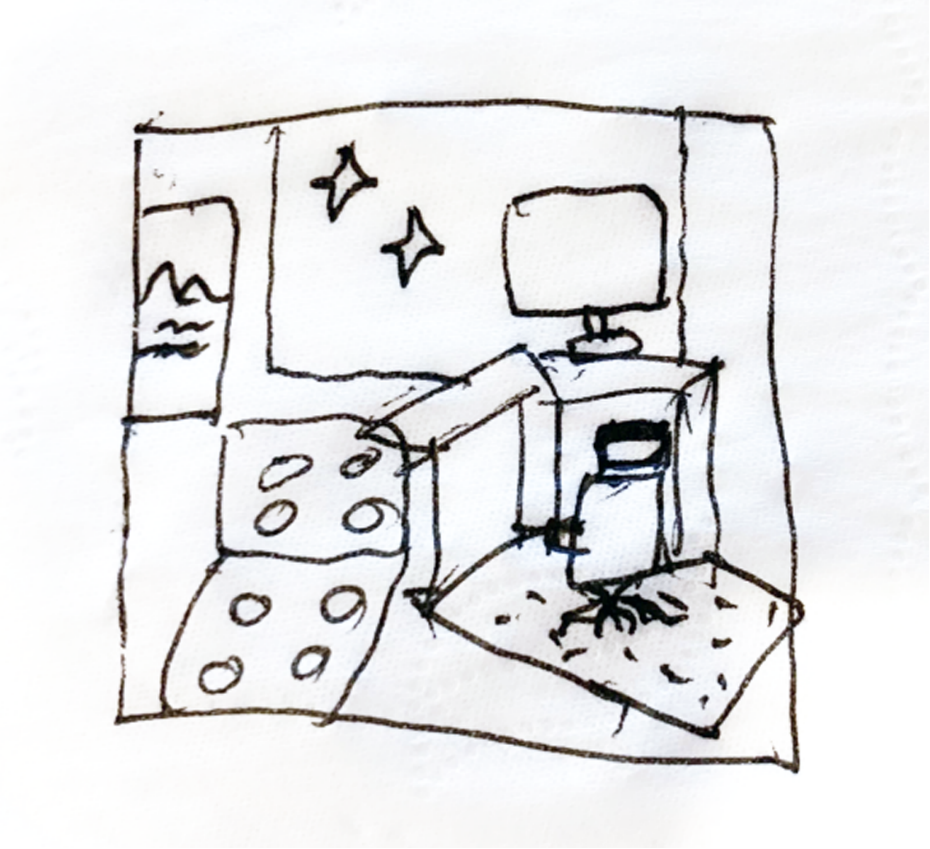}}
\end{minipage}
& A soft carpet is put on the floor, and a sofa lies on the right corner of the room.A desktop with a looping screensaver is placed on a table in front of the window. On the wall hangs a landscape painting of the Earth.
& P1:"I can lean back on the sofa and look up at the landscape painting of the Earth if I miss the view and people on Earth."
\\ \hline
\begin{minipage}[b]{0.3\columnwidth}
    \centering
    \raisebox{-.5\height}{\includegraphics[width=\linewidth]{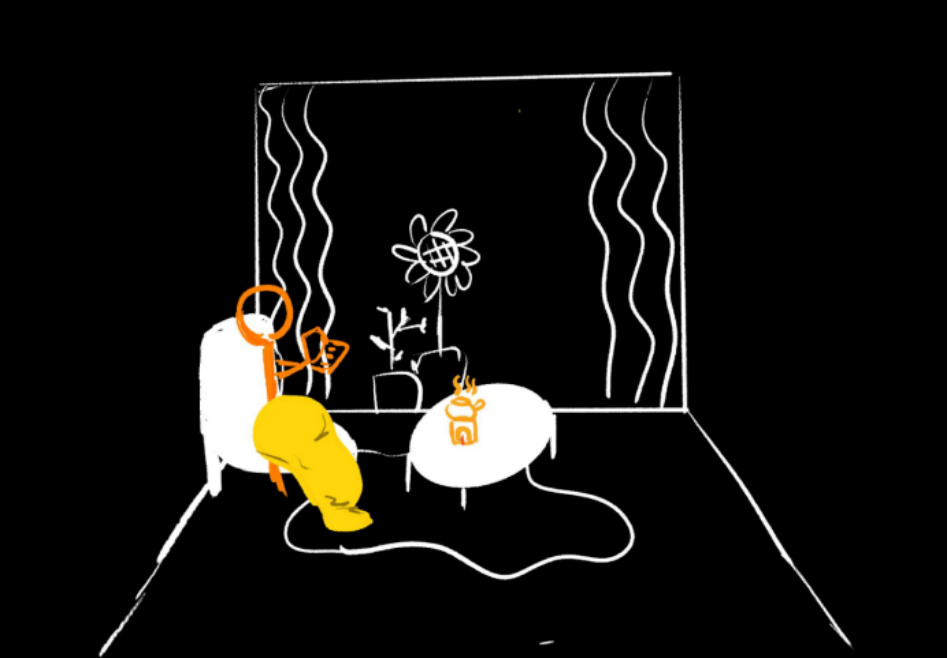}}
\end{minipage}
& There are sunflowers and various succulent plants on the windowsill. Carpets are laid on the ground. A teapot is on the tea table, and there is a slow fire boiling the tea.
& P2: "I will lie on the sofa covered with a blanket reading a novel.It will be better if there is music, Like some nordic folk. I wonder if I can bring a cat. That would be cool!"
\\ \hline
\begin{minipage}[b]{0.3\columnwidth}
    \centering
    \raisebox{-.5\height}{\includegraphics[width=\linewidth]{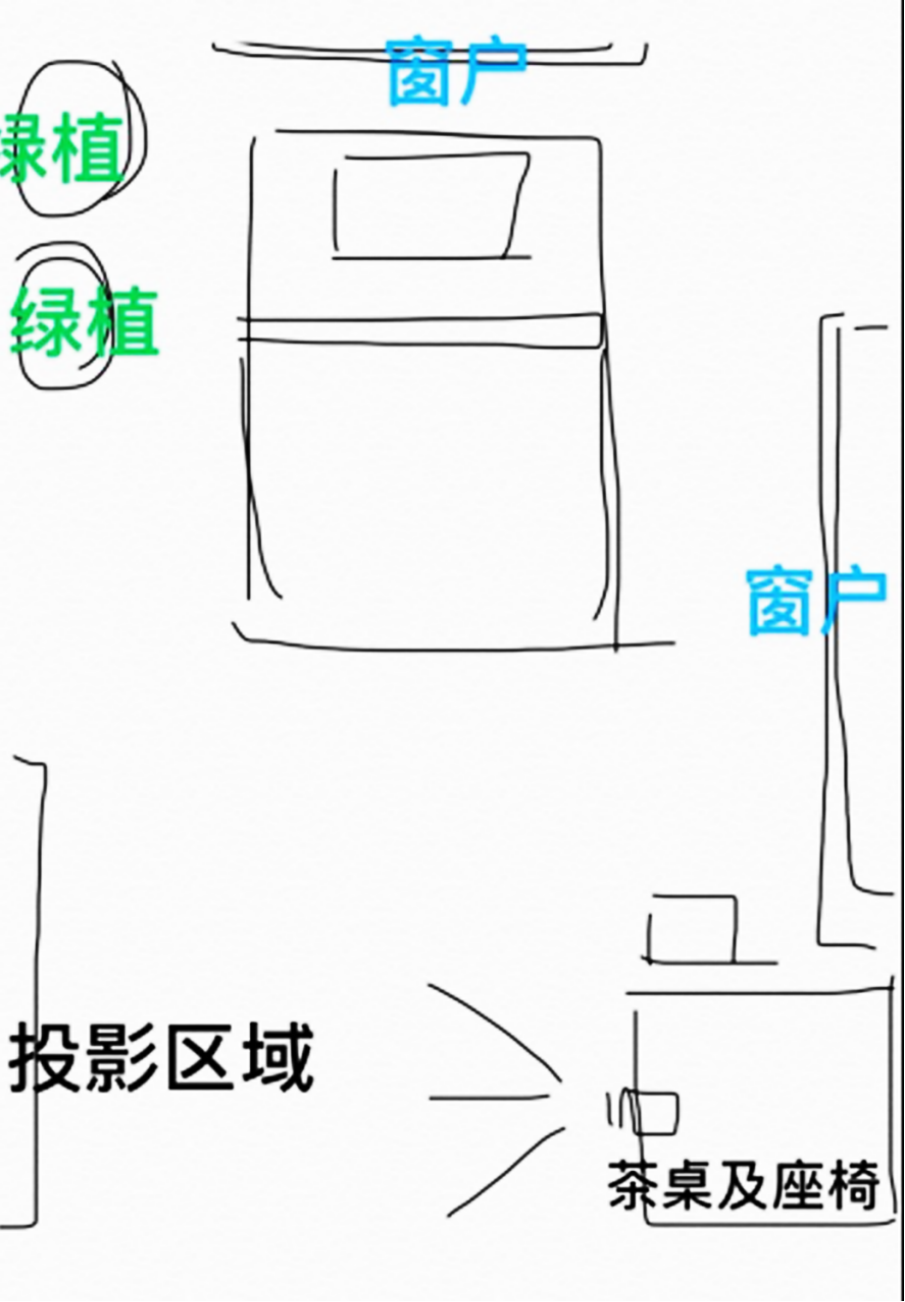}}
\end{minipage}
& There should be some broadleaf plants in the corner of the room, a tea table with a tea set in the opposite corner of the room near the window.A projector is also placed on the tea table projecting the videos on the wall.
& P3:"The broad-leaved plants will remind me of my hometown and can create refreshing atmosphere.By the window should be the place where I boil water and make tea. I can drink light tea here and enjoy my favourite movies with the projector to make me relaxed."
\\ \hline
\end{tabular}
\end{center}
\end{minipage}
\end{center}

\clearpage

\vspace{2mm}
\begin{center}
\begin{minipage}{\textwidth}
\begin{center}
\refstepcounter{table}
Table \thetable: Participants' designs of their private space for space journey isolation (Part 2)
\label{tab:designs2}

\vspace{2mm}
\begin{tabular}{|c|m{5cm}|m{5cm}|}
\hline
Sketch & Content & Describe \\ \hline
\begin{minipage}[b]{0.3\columnwidth}
    \centering
    \raisebox{-.5\height}{\includegraphics[width=\linewidth]{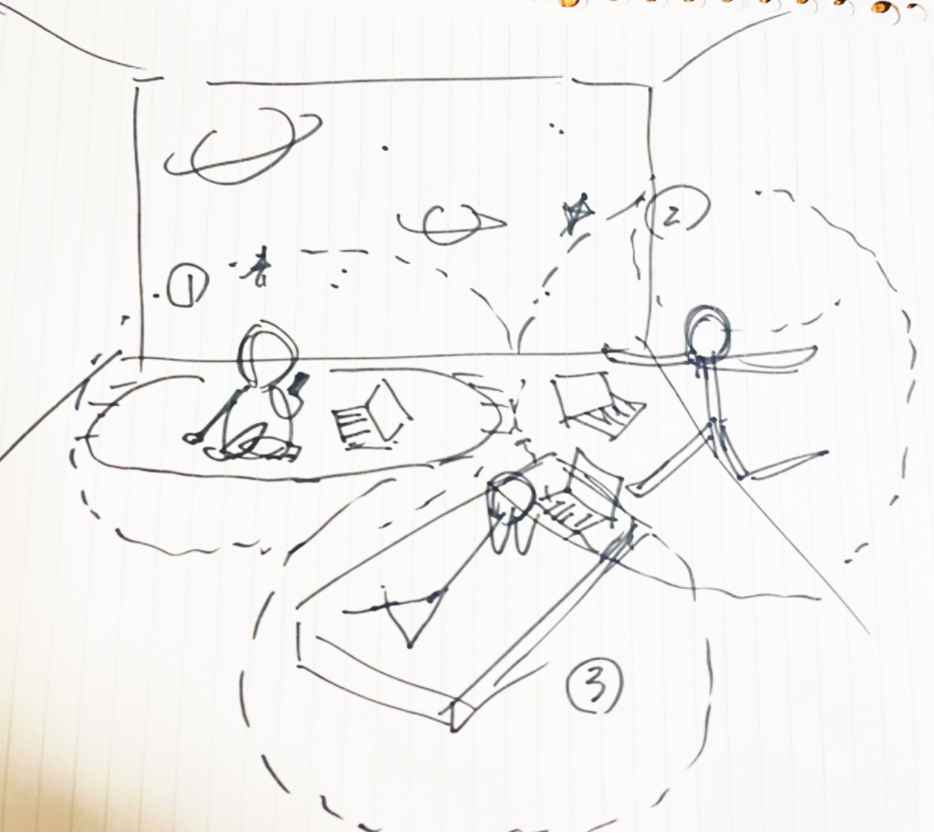}}
\end{minipage}
& The room is relatively empty with as fewer extra things as possible so that there is room to move around. There is a thick carpet put in front of the window on the floor.
& P4:"I will voice-record the new things seen and experienced every day with the digital diary, and then edit them with the computer.I may also follow the videos on the computer to exercise and keep healthy.I can watch reality shows or other entertaining videos in bed, or I may have video calls with family and friends when miss them or feel bored.I hope to move around and stare at the window for viewing the starry universe, which are relaxing."
\\ \hline
\begin{minipage}[b]{0.3\columnwidth}
    \centering
    \raisebox{-.5\height}{\includegraphics[width=\linewidth]{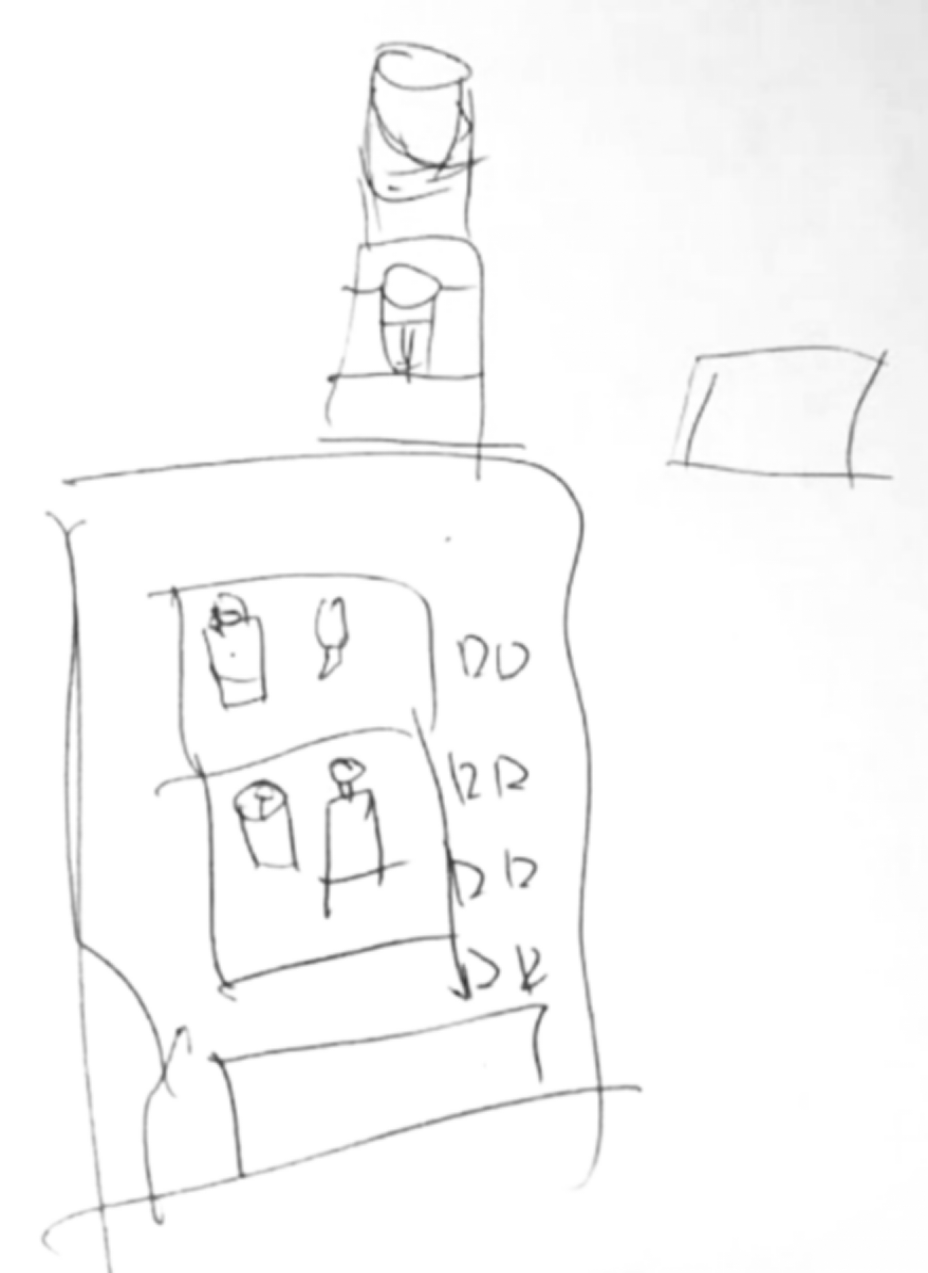}}
\end{minipage}
& There are small water dispensers and vending machines in the room.There is also a plant in the room.There is also a small carpet.
& P5: "Maybe I can plant tea with the planting kits, and so I may be able to make tea with the tealeaves.I will place all the things against the wall to make it no gaps in between, so as to create a sense of security.I can sit and relax on a small carpet."
\\ \hline
\begin{minipage}[b]{0.3\columnwidth}
    \centering
    \raisebox{-.5\height}{\includegraphics[width=\linewidth]{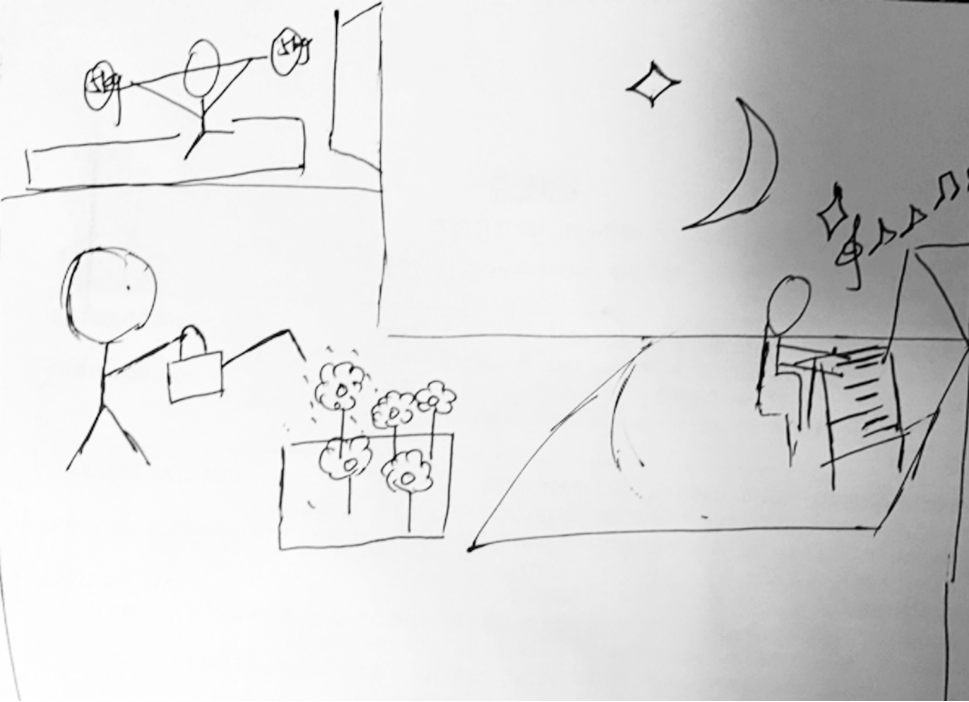}}
\end{minipage}
& There should be a floor-to-ceiling window in the room.A piano is placed in front of the window, and a yoga mat is put on the other side of the room.
& P6: "In my spare time, I can play the piano and sing. I can also plant flowers and vegetables next to the window, because I want to have fresh vegetables to eat all the time. After dinner, I can exercise in the other corner of the room."
\\ \hline
\end{tabular}
\end{center}
\end{minipage}
\end{center}

\clearpage
\vspace{2mm}
\begin{center}
\begin{minipage}{\textwidth}
\begin{center}
\refstepcounter{table}
Table \thetable: Participants' designs of their private space for space journey isolation (Part 3)
\label{tab:designs3}

\vspace{2mm}
\begin{tabular}{|c|m{5cm}|m{5cm}|}
\hline
Sketch & Content & Describe \\ \hline
\begin{minipage}[b]{0.3\columnwidth}
    \centering
    \raisebox{-.5\height}{\includegraphics[width=\linewidth]{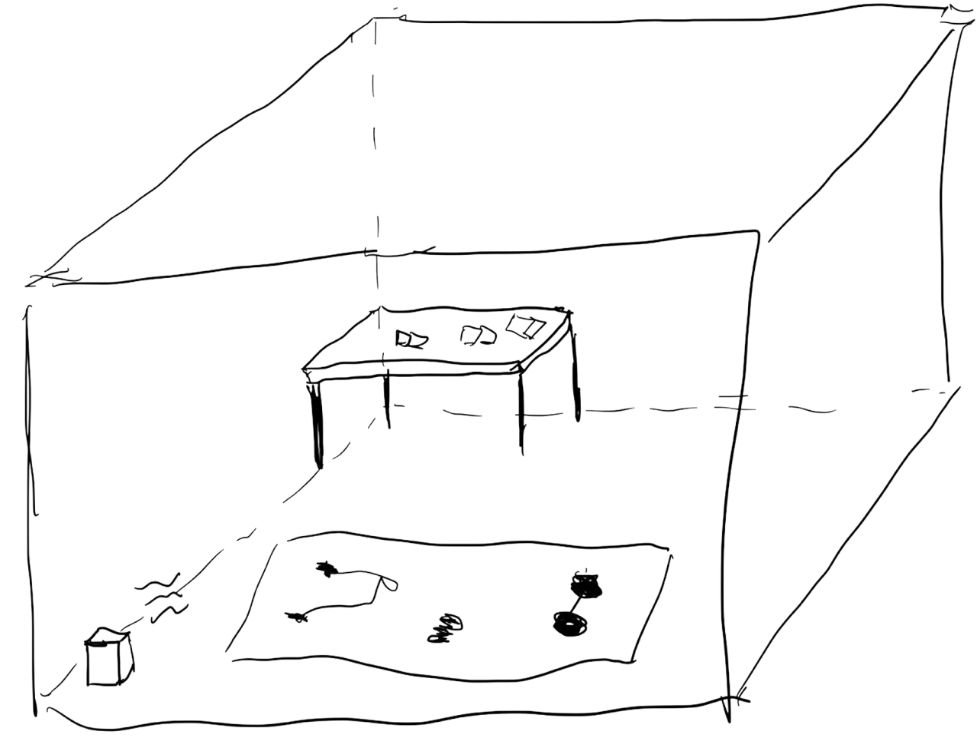}}
\end{minipage}
& The room is equipped with a desk with some novels on.There is a big bed in the middle and a carpet beside it.There is also a set of bluetooth speakers.
& P7:"I can exercise while listening to music. It would be nice to have some snacks in the room, so that I can read while eating.
\\ \hline
\begin{minipage}[b]{0.3\columnwidth}
    \centering
    \raisebox{-.5\height}{\includegraphics[width=\linewidth]{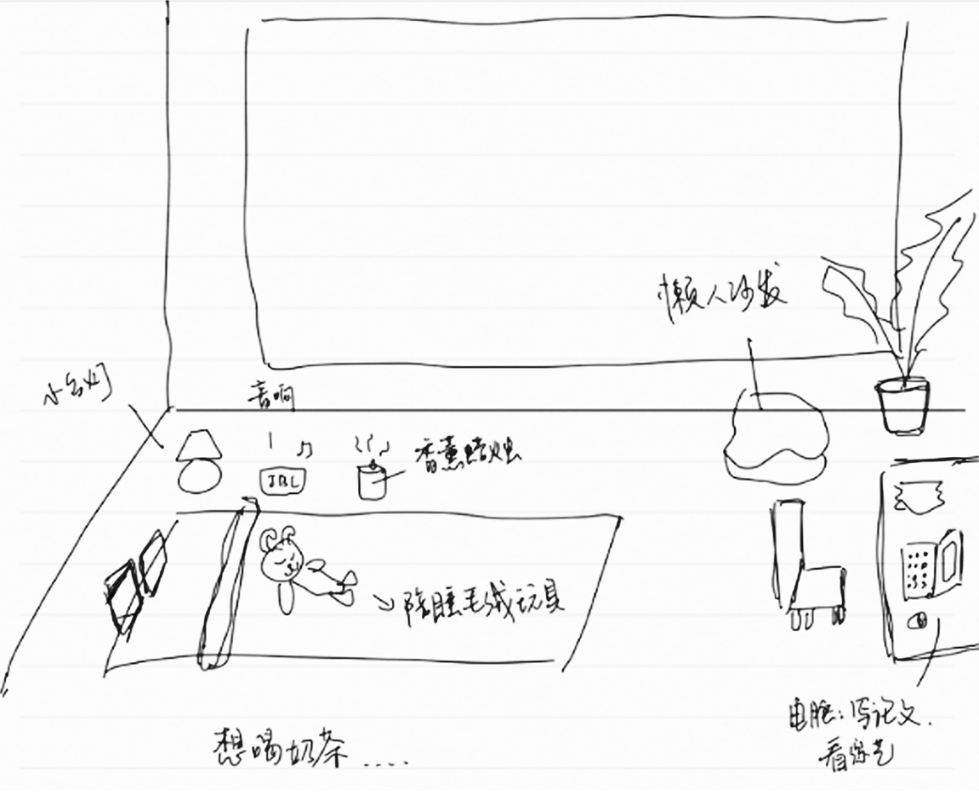}}
\end{minipage}
& Almost all the things are put directly on the floor. There is a plant next to the window, on the left of which is a lazy sofa. A desk with matching chair is on the right with a laptop on it for writing thesis. There is only a Tatami mattress, and a bear doll is placed on it. Next to the mattress is an aromatic candle, a speaker, and a small lamp.
& P8:"I often suffer from insomnia at night because of pressure and emotional fluctuations. I brought a bear doll and an aromatic candle with me since I feel like the good smell can make me relaxed and holding the teddy bear can release my bad moods. I also want to lie on the sofa to enjoy the scenery outside the window. I want to drink milk tea in my room!"
\\ \hline
\begin{minipage}[b]{0.3\columnwidth}
    \centering
    \raisebox{-.5\height}{\includegraphics[width=\linewidth]{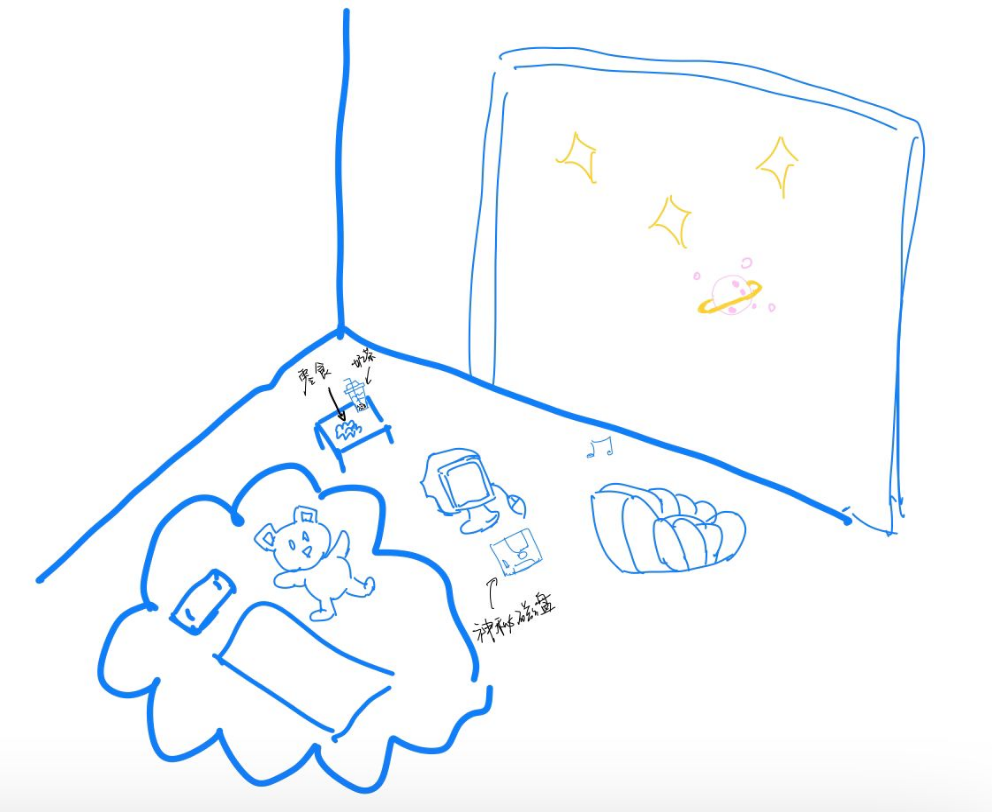}}
\end{minipage}
& There is a large soft bed with some dolls on. There are some snacks and milk tea on the bedside table. There should be a computer with a WiFi connection, and a lazy sofa placed by the floor-to-ceiling window.
& P9:"When I felt empty and lonely, I can listen to music while sitting on the sofa and staring out of the window. I will explore with my mysterious disk with the computer because there is no player. I can use the network platform to ask if there are outer space 'netizens' with relevant skills or equipment to explore a new world together."
\\ \hline
\end{tabular}
\end{center}
\end{minipage}
\end{center}

\clearpage
\vspace{2mm}
\begin{center}
\begin{minipage}{\textwidth}
\begin{center}
\refstepcounter{table}
Table \thetable: Participants' designs of their private space for space journey isolation (Part 4)
\label{tab:designs4}

\vspace{2mm}
\begin{tabular}{|c|m{5cm}|m{5cm}|}
\hline
Sketch & Content & Describe \\ \hline
\begin{minipage}[b]{0.3\columnwidth}
    \centering
    \raisebox{-.5\height}{\includegraphics[width=\linewidth]{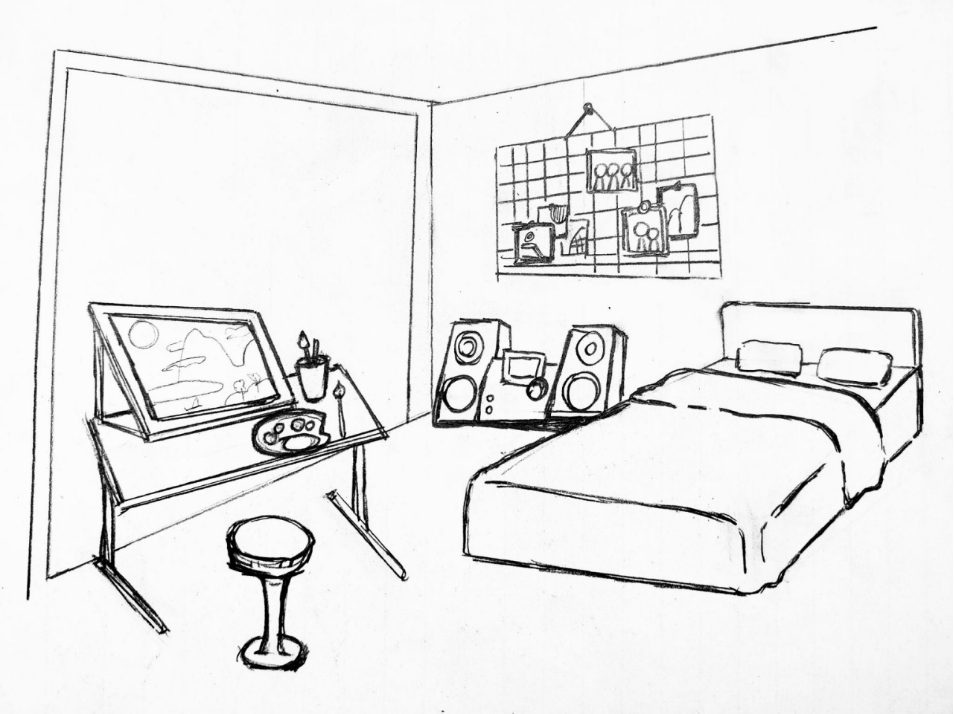}}
\end{minipage}
& There is a desk next to the window. Painting tools and plants are on the desk. A set of music stereos is next to the bed. There are some clips, thumbtacks, and wires for making a photo collage, and also some board game kits.
& P10: "I like to listen to music and paint in my spare time. Maybe I can make a photo collage on one side of the wall to decorate the room with some precious photos and pictures I am satisfied with. I also want to play board games, which may be excellent tools for socializing and relieving boredom."
\\ \hline
\begin{minipage}[b]{0.3\columnwidth}
    \centering
    \raisebox{-.5\height}{\includegraphics[width=\linewidth]{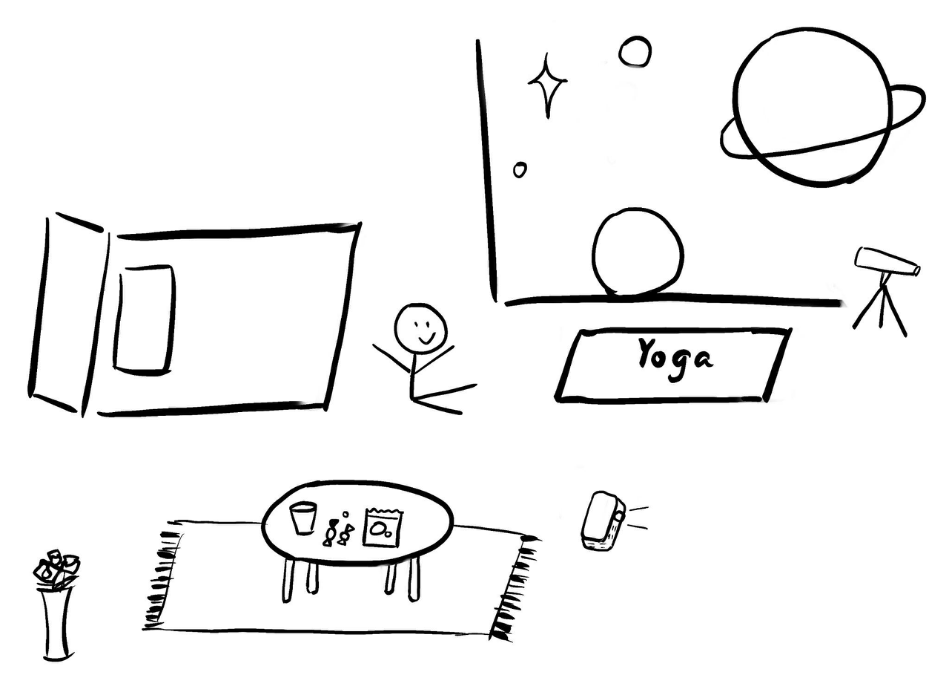}}
\end{minipage}
& There is a yoga mat placed near the window. An astronomical telescope is put next to the yoga mat facing the window. There is also a projector, a tea table and a plant in the room.
& P11: "When I'm bored, I can practice yoga near the window to keep fit and relax. I can invite friends to watch movies and eat snacks with me in my room. I wish I could look at the starry sky and be in a daze."
\\ \hline
\begin{minipage}[b]{0.3\columnwidth}
    \centering
    \raisebox{-.5\height}{\includegraphics[width=\linewidth]{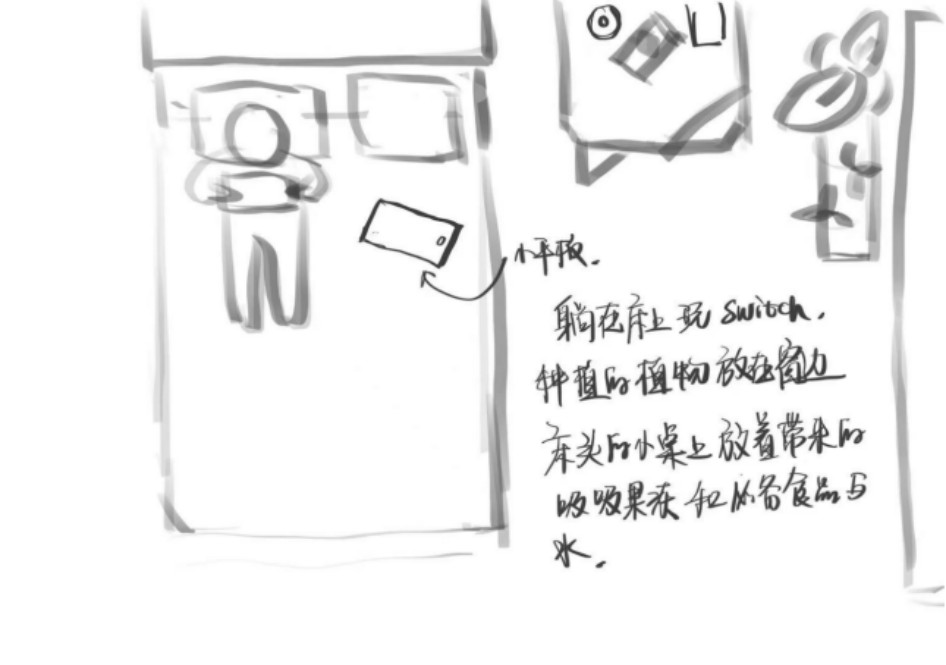}}
\end{minipage}
& A plant is placed next to the window. Near it is a table where the charger and water cup are placed. The bed is close to the table and the small tablet and a Switch are in the bed.
& P12:"I want to lie on the bed and play with the Switch console games. I like sucking jelly very much so I would like to put a lot on the table so that I can easily reach it when playing games in the bed."
\\ \hline
\end{tabular}
\end{center}
\end{minipage}
\end{center}

\clearpage
\vspace{2mm}
\begin{center}
\begin{minipage}{\textwidth}
\begin{center}
\refstepcounter{table}
Table \thetable: Participants' designs of their private space for space journey isolation (Part 5)
\label{tab:designs5}

\vspace{2mm}
\begin{tabular}{|c|m{5cm}|m{5cm}|}
\hline
Sketch & Content & Describe \\ \hline
\begin{minipage}[b]{0.3\columnwidth}
    \centering
    \raisebox{-.5\height}{\includegraphics[width=\linewidth]{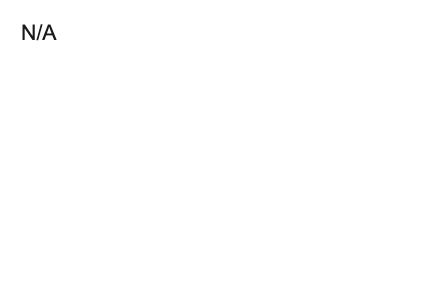}}
\end{minipage}
& There is a desk next to the window and a laptop is on the desk. The mobile hard disk is put next to the laptop. There are also some snacks around the laptop. There is also a novel on the desk near the laptop.
& P13:"I want to put a desk beside the window and put my computer on the desk so that I can ensure that I have enough space to complete my graduation thesis. There are many movies stored on my mobile hard disk. When I feel tired, I can relax by watching movies and eating snacks. If I am tired of watching movies, I can read novels for relaxation."
\\ \hline
\begin{minipage}[b]{0.3\columnwidth}
    \centering
    \raisebox{-.5\height}{\includegraphics[width=\linewidth]{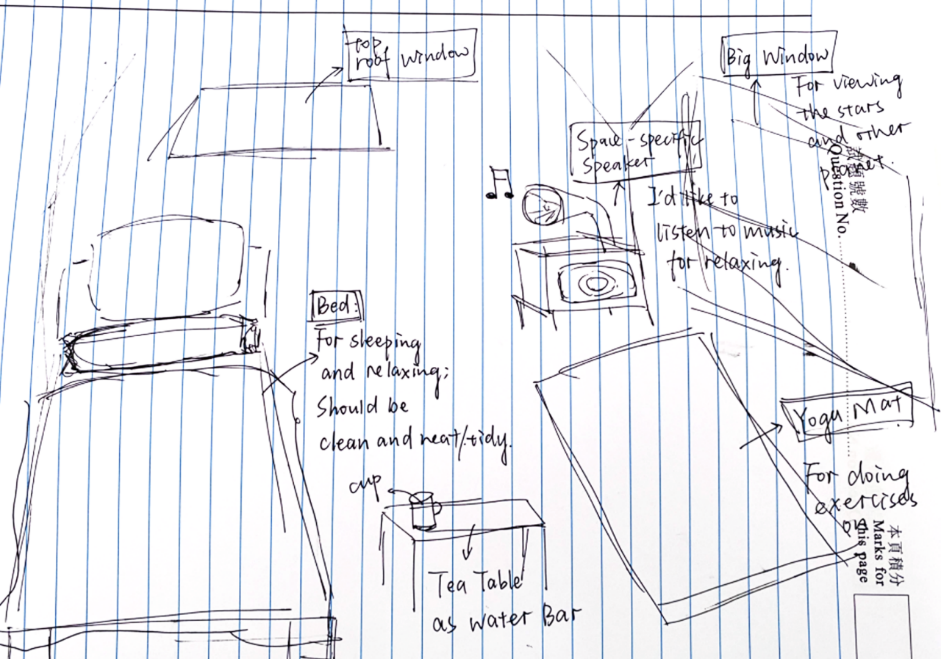}}
\end{minipage}
& The room should have artificial gravity. There is a skylight above the bed. The bed is for sleeping and relaxation, so it should be clean and tidy. A tea table as water bar should be placed near the bed with a cup on it. There is a space-specific speaker in a corner of the room. A yoga mat is placed near the big floor-to-ceiling window.
& P14:"I hope there aren't too many things in the room. I can lie on the bed and look at the starry sky outside the skylight. I will use the music player to listen to my favorite music to relax and spend the time of insomnia during space travel. To meet the needs of daily exercise, I would like to dance and do stretching on the yoga mat. The big window can be used to see the galaxy and planets outside, which must be visually relieved! I like drinking tea. I hope to have a special water bar in space. Occasionally, I can take out the projection and use it to watch the variety shows."
\\ \hline
\begin{minipage}[b]{0.3\columnwidth}
    \centering
    \raisebox{-.5\height}{\includegraphics[width=\linewidth]{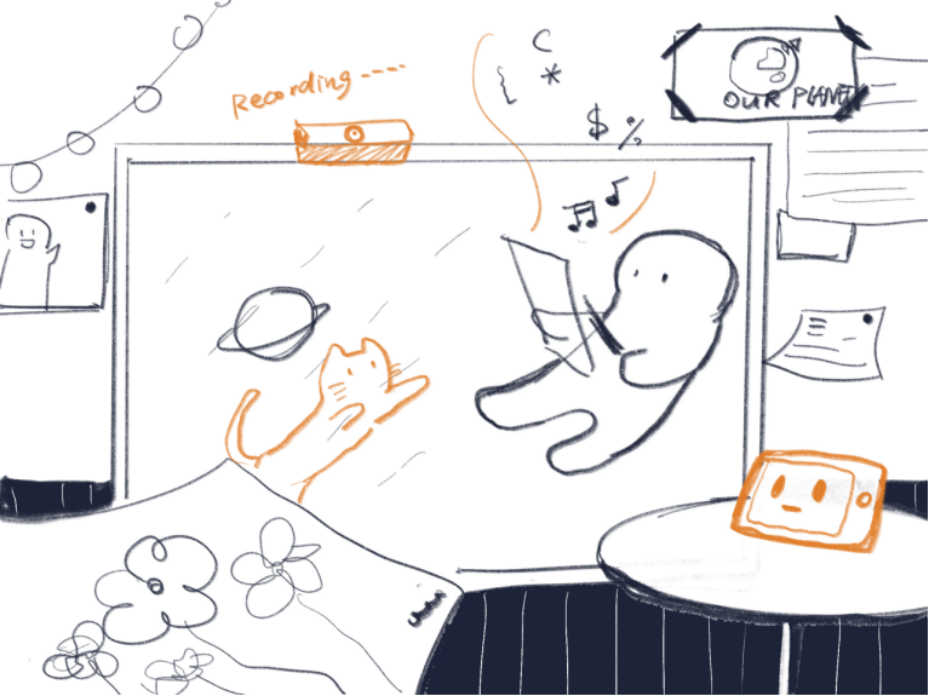}}
\end{minipage}
& Pictures and some drawings about the landscapes of Earth, like flowers, mountains or even the complete landform, are put onto the wall to make it less empty. A cat and a tobot-like tablet are in the room. A voice input notebook is hung on the wall.
& P15: "I want to paint while floating in the room, and at the same time observing the ever-changing galaxy outside the window, recording beautiful moments. That shall be time killing and can ease my loneliness. I want to play with cats in the space. My tablet looks like a little robot, accompanying me. I could directly use the notebook to record what I wanted to say, even though it may not be that deliberate."
\\ \hline
\end{tabular}
\end{center}
\end{minipage}
\end{center}

\clearpage
\vspace{2mm}
\begin{center}
\begin{minipage}{\textwidth}
\begin{center}
\refstepcounter{table}
Table \thetable: Participants' designs of their private space for space journey isolation (Part 6)
\label{tab:designs6}

\vspace{2mm}
\begin{tabular}{|c|m{5cm}|m{5cm}|}
\hline
Sketch & Content & Describe \\ \hline
\begin{minipage}[b]{0.3\columnwidth}
    \centering
    \raisebox{-.5\height}{\includegraphics[width=\linewidth]{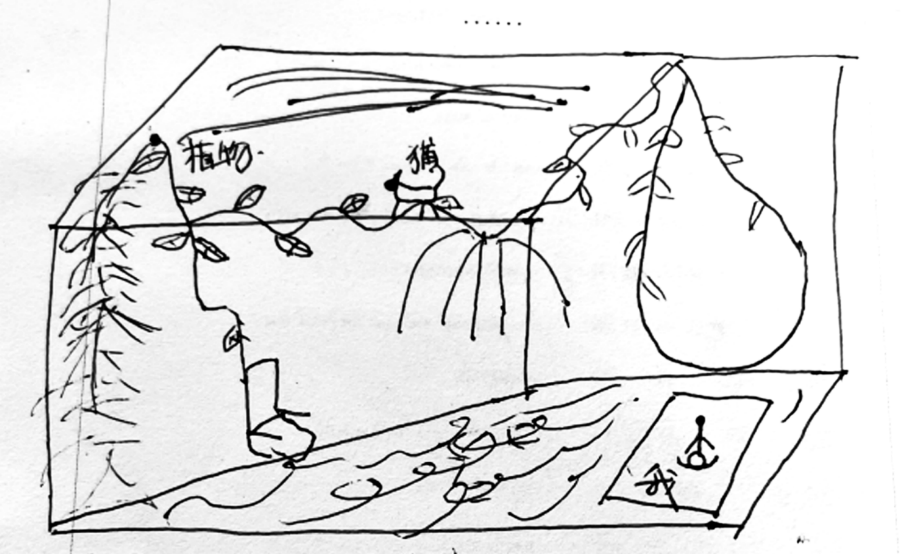}}
\end{minipage}
& The room is covered with plants, which can be made into furniture of different shapes, such as tables, chairs, hammocks and carpets.
& P16:"I want to plant vines on the top of the room to make furnitures and decorate the room. I can climb up to the ceiling and play with my cat floating in the air, because I assume there is no gravity. I will make full use the spare time for meditation and yoga, thinking about the meaning of life and the existence of the universe, and based on this, I will create works of art with space as the medium of information transmission."
\\ \hline
\begin{minipage}[b]{0.3\columnwidth}
    \centering
    \raisebox{-.5\height}{\includegraphics[width=\linewidth]{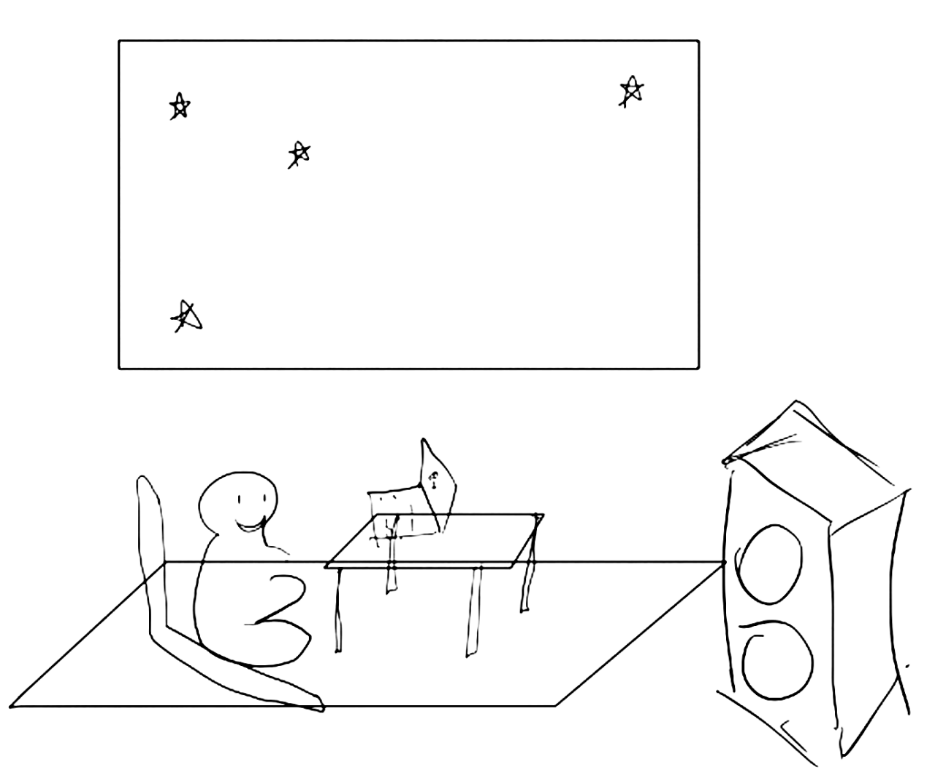}}
\end{minipage}
& A carpet is lay by the window. A small desk is on the carpet. On the small desk is a laptop computer, and next to the carpet is the music stereo.
& P17:"I would like to watch TV series and listen to music. If there is no music in this world, it will be a terrible thing! I will listen to soothing music. My favorite variety shows, TV series or novels should all be on my computer. In my spare time, I will enjoy the scenery of the universe outside the window. Perhaps, that is the so-called happiness."
\\ \hline
\begin{minipage}[b]{0.3\columnwidth}
    \centering
    \raisebox{-.5\height}{\includegraphics[width=\linewidth]{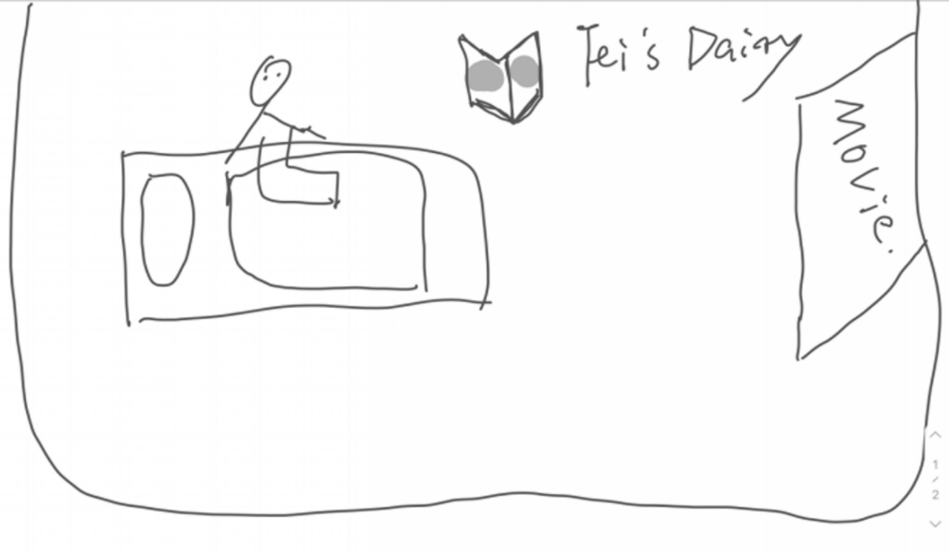}}
\end{minipage}
& There are only a soft bed, a projector, and a digital diary in the room.
& P18:"I want to wake up from the big soft bed every day. I may be sleepy, in a trance, and have migraine every day. Not awake enough, I may start to talk to my diary to record my chaotic dreams last night and share my thoughts. The time in space could be extremely long, so I will watch other people's lives through the projector. Only in this way can I feel the change of time."
\\ \hline
\end{tabular}
\end{center}
\end{minipage}
\end{center}

\end{document}